\documentclass{jpp}
\usepackage{graphicx}
\usepackage{xcolor}
\usepackage{textgreek}
\usepackage[utf8]{inputenc}
\usepackage[T1]{fontenc}
\usepackage{amsmath}

\usepackage{hyperref}
\usepackage{xcolor}
\hypersetup{
    colorlinks,
    linkcolor={red!50!black},
    citecolor={blue!50!black},
    urlcolor={blue!80!black}
}

\usepackage{multicol}

\usepackage[capitalise]{cleveref}

\usepackage[normalem]{ulem}

\shorttitle{Energetic Particle Tracing}
\shortauthor{P. A. Figueiredo, R. Jorge, J. Ferreira, P. Rodrigues}

\title{Energetic Particle Tracing in Optimized Quasisymmetric Stellarator Equilibria}

\author{P. A. Figueiredo\aff{1}
    \corresp{\email{pauloamfigueiredo@tecnico.ulisboa.pt}},
  R. Jorge\aff{1,2},
  J. Ferreira\aff{1,2},
  \and P. Rodrigues\aff{1,2}}

\affiliation{\aff{1}Departamento de Física, Instituto Superior Técnico,  Universidade de Lisboa, 1049-001 Lisboa, Portugal
\aff{2}Instituto de Plasmas e Fusão Nuclear, Instituto Superior Técnico, Universidade de Lisboa, 1049-001 Lisboa, Portugal}

\begin{document}

\maketitle

\begin{abstract}

    \par Recent developments in the design of magnetic confinement fusion devices have allowed the construction of exceptionally optimized stellarator configurations. The near-axis expansion in particular has proven to enable the construction of magnetic configurations with good confinement properties while taking only a fraction of the usual computation time to generate optimized magnetic equilibria. However, not much is known about the overall features of fast-particle orbits computed in such analytical, yet simplified, equilibria when compared to those originating from accurate equilibrium solutions. This work aims to assess and demonstrate the potential of the near-axis expansion to provide accurate information on particle orbits and to compute loss fractions in moderate to high aspect ratios. The configurations used here are all scaled to fusion-relevant parameters and approximate quasisymmetry in various degrees. This allows us to understand how deviations from quasisymmetry affect particle orbits and what are their effects on the estimation of the loss fraction. Guiding-center trajectories of fusion-born alpha particles are traced using \textit{gyronimo} and \textit{SIMPLE} codes under the \textit{NEAT} framework, showing good numerical agreement. Discrepancies between near-axis and MHD fields have minor effects on passing particles but significant effects on trapped particles, especially in quasihelically symmetric magnetic fields. Effective expressions were found for estimating orbit widths and passing-trapped separatrix in quasisymmetric near-axis fields. Loss fractions agree in the prompt losses regime but diverge afterward.

\end{abstract}

\section{Introduction}

\par Energetic alpha particles generated by fusion reactions carry a significant amount of energy and have the potential to drive a plasma towards a self-sustaining fusion state, commonly referred to as a burning plasma. To achieve this, the fraction of alpha particles that deposit their energy in the plasma before being expelled needs to be maximized to guarantee sufficient alpha heating \citep{freidberg_2007}. Additionally, energetic particles that leave the plasma can cause significant damage to the plasma-facing components of the fusion device, leading to a shorter device lifetime \citep{mauDivertorConfigurationHeat2008}. Therefore, it is crucial to confine these energetic particles and accurately predict their behavior in the plasma to achieve the desired levels of alpha heating and advance the development of fusion energy.

\par The most general condition for the confinement of orbits in stellarators is omnigenity, which requires a vanishing time-averaged radial drift, $ \langle \Delta \psi \rangle $. An important subset of omnigenity, quasisymmetry (QS), is obtained through a symmetry in the modulus of the magnetic field $\mathbf{B}$ when expressed in Boozer coordinates \citep{BoozerCoordinates}. Experiments such as the W7-X \citep{beidlerPhysicsEngineeringDesign1990} and HSX \citep{andersonHelicallySymmetricExperiment1995} intend to approximate omnigenity via quasi-isodynamic and quasisymmetric fields, respectively. Although perfectly omnigenous \citep{caryOmnigenityQuasihelicityHelical1997} and quasisymmetric \citep{garrenExistenceQuasihelicallySymmetric1991} devices have been conjectured not to exist, it has been shown that very precise approximations can be obtained \citep{landremanMagneticFieldsPrecise2022, goodmanConstructingPreciselyQuasiisodynamic2022}. 

\par While these experiments have proven to generally confine thermal plasmas, the loss of energetic alpha particles is still a key research area for stellarators \citep{bader_drevlak_2019, LeViness_2023}. Minimizing the amount of lost particles is usually performed through optimization of plasma properties such as quasisymmetry \citep{hennebergPropertiesNewQuasiaxisymmetric2019} and omnigenity \citep{goodmanConstructingPreciselyQuasiisodynamic2022} or using proxies for loss fractions, such as $\Gamma_c$, \citep{baderStellaratorEquilibriaReactor2019a} and $\Gamma_{\alpha}$ \citep{sanchezQuasiisodynamicConfigurationGood2023}. Direct optimization of loss fractions has only recently been performed, albeit at a substantial computational cost \citep{bindelDirectOptimizationFastIon2023}. 
\par Such optimization procedures require the repeated variation of the last closed flux surface and are not only hindered by the existence of multiple local minima, and thus highly dependent on the initial conditions, but are also computationally demanding and offer limited insight into the number of effective degrees of freedom of the problem, as well as the specific physical implications associated with individual coefficients or their collective groups \citep{landremanDirectConstructionOptimized2019}. Additionally, to obtain a reactor-grade stellarator, it is essential to balance the confinement of energetic particles with other physical and engineering constraints such as stability, turbulence, and coil geometry tolerances \citep{hegnaImprovingStellaratorAdvances2022}, which may increase its overall computational cost.
\par A way to circumvent the issues above is through the introduction of an analytical approximation to a magnetohydrodynamic (MHD) equilibrium. Such construction is a near-axis expansion (NAE), which is valid in the core of all stellarators and can be introduced in the early stages of optimization enabling the use of better initial conditions (warm start initial conditions) for conventional optimizations while allowing for extensive searches in the parameter space of the design \citep{landremanMappingSpaceQuasisymmetric2022}. This expansion can be obtained resorting either to the direct method \citep{mercierEquilibriumStabilityToroidal1964, solovevPlasmaConfinementClosed1970, jorgeNearAxisExpansionStellarator2020}, explicitly determining the magnetic flux surface function $\psi$ using the Mercier coordinates ($\rho$, $\theta$, $\varphi$), or the indirect method \citep{garrenExistenceQuasihelicallySymmetric1991, garrenMagneticFieldStrength1991, landremanDirectConstructionOptimized2018, landremanDirectConstructionOptimized2019}, which directly computes the position vector $\mathbf{r}$ as a function of the Boozer coordinates and is the one we will use throughout this work. Such constructions not only describe high aspect ratio devices but also the region around the axis of any stellarator, while enabling useful analytical insight due to its simplicity at lowest orders. As we show here, the estimation of loss fractions directly from orbit following codes can be computationally expedited within the near-axis framework, without significant accuracy degradation due to its simplified analytical nature for short timeframes. This speed up arises from the direct computation of the magnetic field and other related quantities as an alternative to the many Fourier coefficients needed in conventional optimization.
\par The main goal of this paper is to study the accuracy of fast particle trajectories in approximate analytical near-axis equilibria when compared with the same trajectories in accurate numerical MHD equilibria. By understanding how much these orbits deviate from each other, we are able to assess the validity of the expansion for loss fractions estimation in scenarios with a variety of aspect ratios, time scales, and perpendicular velocities. As the collisional effects have been shown negligible for particle losses in the prompt losses timescales and timescales up to a fraction of the energy deposition time \citep{lazersonSimulatingFusionAlpha2021}, we are only interested in the collisionless dynamics of the energetic particles, which were studied resorting to the integration of the particles' guiding-center orbits. The primary focus of this work is on configurations that approximate quasisymmetry in varying degrees, in order to employ the quasisymmetric version of the NAE of Refs. \cite{landremanConstructingStellaratorsQuasisymmetry2019} and \cite{landremanMappingSpaceQuasisymmetric2022}. The concept of quasisymmetry, derived from the gyroaveraged Lagrangian, implies that QS stellarators are specifically designed to confine guiding-center orbits. Therefore, our analysis focused solely on these orbits. This approach is also more computationally efficient than simulating the full orbits of particles.
\par A large number of guiding-center (GC) or full-gyromotion particle tracers, such as \textit{ANTS} \citep{drevlakFastParticleConfinement2014a}, \textit{ASCOT} \citep{hirvijokiASCOTSolvingKinetic2014}, \textit{BEAMS3D} \citep{mcmillanBEAMS3DNeutralBeam2014}, \textit{FOCUS} \citep{clauserFOCUSFullorbitCUDA2019}, \textit{LOCUST} \citep{wardVerificationValidationHighperformance2021}, \textit{OFMC} \citep{taniEffectToroidalField1981}, \textit{GNET} \citep{Masaoka2013}, \textit{SPIRAL} \citep{kramerDescriptionFullparticleorbitfollowingSPIRAL2013} and \textit{VENUS-LEVIS} \citep{pfefferleVENUSLEVISItsSplineFourier2014}, have been used for fast particle transport and loss studies. In this work we use the Euler-Lagrange equations of motion for the guiding-center \citep{littlejohn_1983} as implemented in the general-geometry particle tracing library \textit{gyronimo} \citep{Rodrigues2020}, which is a convenient tool for comparing between different geometries, and the Hamilton equations of motion as implemented in the code \textit{SIMPLE}\citep{albertSIMPLESymplecticIntegration2020}. Both particle-tracing codes are open source. In order to make the direct comparison between trajectories as straightforward as possible, the same equations and algorithms are used for particles in near-axis and MHD fields. Later specialization could improve the computational efficiency of the procedures. However, the achieved levels of speed are already relevant for near-axis equilibria optimization.
\par This paper is organized as follows. Section 2 provides an overview of the NAE and the Boozer coordinate system. In Section 3 we describe the guiding-center motion and the different approaches taken for its calculations. In Section 4 the results for single particle tracing in configurations close to quasisymmetry are shown, within a reasonable scope of initial conditions. Additionally, expressions for the estimation of the orbit radial amplitudes and the passing-trapped boundary are derived and compared with computational results. Taking this last section's knowledge into account, total loss fractions are computed and analyzed in Section 5 for two distinct configurations. In Section 6, we summarize the primary findings and delineate potential pathways for future research.
\section{The Near-Axis Expansion}
\label{sec:NAE}
\par In this section, we follow the notation of \cite{landremanConstructingStellaratorsQuasisymmetry2019} and introduce the near-axis expansion using the inverse approach first presented in \cite{garrenExistenceQuasihelicallySymmetric1991}. We begin by writing the magnetic field in Boozer coordinates $(\psi,\theta,\varphi)$, with $2\pi \psi$ the toroidal flux, and $\theta$ and $\varphi$ the poloidal and toroidal coordinates, respectively. This results in
\begin{align}
    \label{eq:BoozerCoords}
    \mathbf{B} = &\nabla\psi \times\nabla\theta + \iota \nabla\varphi \times\nabla\psi, \\
     = &\beta \nabla\psi + I \nabla\theta + G \nabla\varphi, \nonumber
\end{align}
\noindent where  $I=I(\psi)$ and $G=G(\psi)$ are flux functions, $\beta=\beta(\psi,\varphi, \theta)$ is a quantity related to the plasma pressure that usually depends on the three coordinates and $\iota=\iota(\psi)$ is the rotational transform. As we want to analyze both stellarators which exhibit a quasisymmetry in the toroidal angle, named quasiaxisymmetric (QA), and in a linear combination of poloidal and toroidal angles, termed quasihelically symmetric (QH), it is convenient to introduce an helical angle $ \vartheta = \theta - N \varphi$ where N is a constant integer equal to 0 for a QA stellarator and, usually, the number of field periods for a QH one. This results in the following expression for the magnetic field,
\begin{align}
    \label{eq:BoozerCoordsHelical}
    \mathbf{B} = &\nabla\psi \times\nabla\vartheta + \iota_N \nabla\varphi \times\nabla\psi, \\
     = &\beta \nabla\psi + I \nabla\vartheta + (G+NI) \nabla\varphi. \nonumber
\end{align}
\noindent To derive the NAE's main features at first order, we begin by writing the general position vector $\mathbf{r}$ as 
\begin{align}
    \mathbf{r}(r,\theta, \varphi) = \mathbf{r}_0(\varphi) + r X_1(\theta, \varphi ) \mathbf{n}(\varphi) + r Y_1(\theta, \varphi ) \mathbf{b}(\varphi) + \mathcal{O}(r^2),
\end{align}
\noindent with $r = \sqrt{2|\psi|/\bar{B}}$ an effective minor radius, $\Bar{B}$ a reference magnetic field and $\mathbf{r}_0$ the position vector along the axis, with a given set of Frenet-Serret orthonormal basis vectors ($\mathbf{t}$, $\mathbf{n}$, $\mathbf{b}$), a local curvature $\kappa=\kappa(\varphi)$ and torsion $\tau=\tau(\varphi)$. At first order, we additionally take the profile functions $G(r)$ and $I(r)$ to be $G_0$ and $I_2 r^2$, respectively \citep{landremanDirectConstructionOptimized2018}. In quasisymmetry, $\mathbf{r}$ can be written as \citep{landremanConstructingStellaratorsQuasisymmetry2019}
\begin{align}
    \mathbf{r}(r,\vartheta,\varphi) = \mathbf{r_0}(\varphi) + \frac{r \bar{\eta}}{\kappa(\varphi)}\cos\vartheta \ \mathbf{n}(\varphi) + \frac{r s_\psi s_G \kappa(\varphi)}{\bar{\eta}} \left[ \sin\vartheta + \sigma(\varphi) \cos\vartheta \right] \mathbf{b}(\varphi) +O(r^2/\mathcal{R}),
    \label{eq:position_vector_r1}
\end{align}
\noindent where $s_\psi = \mathrm{sign}(\psi)$, $s_G = \mathrm{sign}(G_0)$, and $\bar{\eta}$ is a constant reference field strength that parametrizes $B= |\mathbf{B}|$ as
\begin{align}
    B = B_0 \left( 1 + r \bar{\eta} \cos\vartheta \right) + O((r/\mathcal{R})^2),
    \label{eq:B_first_order}
\end{align}
\noindent where $\Bar{B}=s_\psi B_0$ and $\sigma(\varphi)$ as a periodic function related to the flux surface shape, that satisfies the Riccati-type equation
\begin{align}
    \label{eq:sigma_quasisymmetry}
    \frac{d\sigma}{d\varphi} + (\iota_0 - N) \left[ \frac{\bar{\eta}^4}{\kappa^4} + 1 + \sigma^2 \right]
    -\frac{2 G_0 \bar{\eta}^2}{B_0 \kappa^2} \left[ \frac{I_2}{B_0} - s_\psi \tau \right] = 0.
\end{align}
\noindent Equation (\ref{eq:B_first_order}) only enables the construction of stellarators with elliptical cross sections, so a higher order is needed to express the stronger shaping of existing stellarators. At the second order, nine new functions of $\varphi$ arise in the surface shapes that can now possess triangularity and a Shafranov shift. However, these functions are constrained by 10 new equations of $\varphi$, a mismatch that results in the fact that most axis shapes are not consistent with quasisymmetry at this order. In  \cite{landremanConstructingStellaratorsQuasisymmetry2019}, this is circumvented by allowing quasisymmetry to be broken at second order in the field strength, which is the method used here. Furthermore, a detailed description of the application of this method to the construction of stellarator shapes is provided, built on a third-order method with the shaping details that need to be taken into account to generate a boundary surface that is consistent with the desired field strength and eliminate unwanted mirror modes.
\par As the present work performs comparisons between equilibria generated with the quasisymmetric NAE code \textit{pyQSC} \citep{Landreman_pyQSC} and with \textit{VMEC} \citep{hirshmanSteepestdescentMomentMethod1983}, an MHD equilibrium code, it is important to note that these do not use the same coordinates. Unlike the description made for the radial coordinate in the NAE, \textit{VMEC} uses a normalized radial coordinate $s=\psi/\psi_b$, with $\psi_b$ the toroidal flux at the last closed flux surface. Therefore, to obtain $r$ in the NAE system, we use the relation
\begin{align}
    r = r_{\text{max}} \sqrt{s},  
\end{align}
\noindent where $r_{\text{max}}=\sqrt{2 \psi_b /\bar{B}}$. Additionally, the toroidal angle in \textit{VMEC}, $\phi$, corresponds to the azimuthal cylindrical coordinate defined as $\arctan(y/x)$, while \textit{pyQSC} operates with a cylindrical coordinate on-axis, $\phi_0$, that corresponds to the asymptotic value on axis, i.e. 

\begin{equation}
    \phi_0= \lim_{r \to 0} \tan^{-1} \bigg(\frac{y}{x}\bigg) ,
    \label{eq:phi0}
\end{equation}
which is distinct from the Boozer coordinate described in the NAE derivation above. In our workflow, we use the Boozer coordinate $\varphi$ and transform it into $\phi_0$ through $\phi_0 = \varphi - \nu$, where $\nu$ is an output of the \textit{pyQSC} code, given by \cite{landremanDirectConstructionOptimized2018}. The coordinate $\phi$ is further computed through a root-finding function. Finally, the poloidal coordinate in \textit{VMEC}, $\theta_V$, is in the range $[-\pi,\pi]$, whereas $\theta$ in \textit{pyQSC} is defined from $0$ to $2 \pi$ in the opposite direction, leading to the relation $\theta_V=\pi - \theta$. The initialization of a particle in a \texttt{pyQSC} field is, however, done with the helical angle $\vartheta=\theta - N \varphi$.
\section{Guiding-Center Formalism and Orbit Integration}

\par The confinement properties of a given stellarator configuration are commonly estimated employing either proxies or direct methods, such as numerically simulating particle orbits in a plasma. Although proxies for confinement such as QS, $\Gamma_c$, and others have proven fruitful in device optimization due to their minimal computational expense \citep{bader_drevlak_2019, hennebergPropertiesNewQuasiaxisymmetric2019}, they also bring some drawbacks. For example, QS may be too stringent of a condition when dealing with a multi-objective optimization, possibly undermining our capacity to achieve optimal outcomes for other objectives like MHD stability or feasible coil sets, and proxies like $\Gamma_c$ and $\Gamma_\alpha$ might not capture the full physical picture of drift orbits \citep{albertAcceleratedMethodsDirect2020}. Furthermore, orbit following is necessary if one wants to either understand the mechanisms that lead to losses or take those into account in the loss estimation. Thus, we apply the equations of motion for charged particles in a magnetized plasma to assess such mechanisms.
\par Computing particle trajectories in a magnetic field is commonly not a straightforward task, often requiring numerical treatment. To do so, we leverage the fact that in the magnetized plasmas of interest, the magnetic field is strong enough that we can decouple the high-velocity cyclotronic motion of charged particles from the slowly varying guiding-center coordinates and parallel velocity, therefore simplifying our problem. We begin by writing the position of a particle $\mathbf{r}$ in terms of its guiding-center position $\mathbf{R}$ and its gyration radius vector $\boldsymbol{\rho}$ in the following way \citep{caryHamiltonianTheoryGuidingcenter2009}
\begin{align}
    \mathbf{r}(t) = \mathbf{R}(t) + \boldsymbol{\rho}(t).
\end{align}
\noindent If the gyroradius $\rho_c=mv_\perp/qB$ of a particle of mass $m$ and charge $q$ is small enough compared to a field gradient length $L_B$, we can average out the particle's motion over a gyration time $t_c=1/\Omega_c$, where $\Omega_c=qB/m$ is the gyrofrequency, and still retain most of the information about the particle motion for timescales much greater than $t_c$. To a first approximation, the guiding-center motion can be described with the noncanonical gyro-averaged Lagrangian \citep{littlejohn_1983}:
\begin{align}
    \mathcal{L} =  q \mathbf{A} \cdot \Dot{\mathbf{R}} +  m v_{\parallel}\mathbf{b} \cdot \Dot{\mathbf{R}} + \frac{m \mu}{q} \Omega_c - \frac{m v_{\parallel}^2}{2}  - \mu B ,
    \label{eq:little}
\end{align}
\noindent where $\mu = m v_\perp^2 / 2 B$ is an adiabatic invariant (magnetic moment), $v_\parallel$ and $v_\perp$ are, respectively, the components of the velocity parallel and perpendicular to the magnetic flux surfaces, $\mathbf{b} = \mathbf{B}/B$ the magnetic unit vector and $\mathbf{A}$ is the magnetic potential. As we are not considering the effects of electric fields from local charge densities or varying magnetic fields, the electric potential is absent from the Lagrangian above. The application of Lagrangian mechanics to this system allows us to use the Euler-Lagrange equations to derive equations of motion that not only conserve energy for time-independent systems but also possess Poincaré integral invariants and allow for derivations of conservation laws from N{\"o}ether's theorem in the presence of spatial symmetries \citep{caryHamiltonianTheoryGuidingcenter2009}. The guiding-center Lagrangian in Boozer coordinates has the following form
\begin{align}
    \mathcal{L} =  q (\psi \Dot{\theta} -\psi_p \Dot{\varphi}) +  \frac{m v_{\parallel}}{B} \left(\dot{\psi} \beta + \dot{\theta} I + \dot{\varphi} G \right) + \frac{m \mu}{q} \Omega_c - \frac{m v_{\parallel}^2}{2}  - \mu B ,
    \label{eq:littleBoozer}
\end{align}
\noindent where we used the fact that, as $\mathbf{B}=\boldsymbol{\nabla} \times \mathbf{A} = \nabla \psi \times \nabla \theta + \nabla \varphi \times \nabla \psi_p$ with $d \psi_p / d \psi=\iota$, $\mathbf{A}$ can be written as $\mathbf{A}=\psi \nabla \theta - \psi_p \nabla \varphi$, where $\psi_p$ is the poloidal flux. In the case of quasisymmetry, where the magnitude of $\mathbf{B}$ depends only on $(\psi,\chi)$, with $\chi=M \theta - N \varphi$, \cref{eq:littleBoozer} exhibits a continuous symmetry in a third coordinate $\eta=M' \theta - N' \varphi$, with $M'/N' \neq M/N$. In compliance with the N\"{o}ether's theorem, this symmetry leads to the conservation of a quantity analogous to the canonical angular momentum in the tokamak, which constraints the trajectories of the particles in the plasma, improving their overall confinement up to modern tokamak levels or greater \citep{landremanMagneticFieldsPrecise2022}.
\par Once we have the equations of motion there is no unique method of performing their integration to obtain particle orbits. Although (\ref{eq:little}) is valid independently of the applied system of coordinates, the standard approach to particle tracing is to write this Lagrangian in terms of coordinates that are more suitable to the problem in question, such as Boozer coordinates, as shown in (\ref{eq:littleBoozer}), usually leading to simplified equations of motion. This is the case of SIMPLE \citep{albertSIMPLESymplecticIntegration2020}, a symplectic (energy conserving) orbit tracer used in this work. The symplectic approach to particle following has significant advantages regarding numerical stability and energy conservation, but its implementation is not straightforward. Symplectic integrators rely on canonical coordinates and an explicit expression of the guiding-center Hamiltonian, 
\begin{align}
    H(\mathbf{z}) = \frac{m v_\parallel^2(\mathbf{z})}{2} + \mu B(\mathbf{z}),
\end{align}
\noindent which exists only in non-canonical coordinates $\mathbf{z}=(\psi,\theta,\varphi, p_\varphi)$, where $p_\varphi = \partial_{\dot{\varphi}} \mathcal{L}$, so a coordinate transformation to canonical coordinates $(\mathbf{q},\mathbf{p})=(\theta,\varphi, p_\theta,p_\varphi)$ is needed and can be found in  \citet{albertSymplecticIntegrationNoncanonical2020}, with $p_\theta = \partial_{\dot{\theta}} \mathcal{L}$. The canonical coordinates can be written as explicit and invertible functions $\mathbf{q} = \mathbf{q} (z)$ and $\mathbf{p} = \mathbf{p} (z)$ with inverses $\mathbf{z} (\mathbf{q}, \mathbf{p})$ given only implicitly. Although this would not be sufficient for an explicit integrator, it is enough for semi or fully implicit integrators. 

The equations of motion of the canonical coordinates in SIMPLE are given by the set of equations:
\begin{align}
    \dot{\theta}(t) & =\frac{\partial H}{\partial \psi}\left(\frac{\partial p_{\theta}}{\partial \psi}\right)^{-1}, 
    \label{eq:qdot-2}\\
    \dot{\varphi}(t) & =\frac{\partial H}{\partial p_{\varphi}}-\frac{\partial H}{\partial \psi}\left(\frac{\partial p_{\theta}}{\partial \psi}\right)^{-1}\frac{\partial p_{\theta}}{\partial p_{\varphi}},\label{eq:phidot}\\
    \dot{p}_{\theta}(t) & =-\frac{\partial H}{\partial\theta}+\frac{\partial H}{\partial \psi}\left(\frac{\partial p_{\theta}}{\partial \psi}\right)^{-1}\frac{\partial p_{\theta}}{\partial\theta},\label{eq:pdot-2}\\
    \dot{p}_{\varphi}(t) & =-\frac{\partial H}{\partial\varphi}+\frac{\partial H}{\partial \psi}\left(\frac{\partial p_{\theta}}{\partial \psi}\right)^{-1}\frac{\partial p_{\theta}}{\partial\varphi},
\end{align}
\noindent which are then solved with symplectic schemes, such as implicit Euler schemes, Verlet, and implicit midpoint, estimating the time evolution of the radial coordinate $\psi$ in the process, as further described in \cite{albertAcceleratedMethodsDirect2020}.

\par The inconveniences related to canonical transformations and their inversions can be avoided by employing the Euler-Lagrange equations of motion that result from the Lagrangian in \cref{eq:little}, as these are invariant regarding any specific choice of coordinates. This is the approach followed by one of the equations-of-motion models implemented in the library \textit{gyronimo}, with the corresponding dynamical system being written in terms of the general curvilinear coordinates $\mathbf{X}$ as

\begin{align}
    \Omega^{*} \ \frac{d\mathbf{X}}{d\tau} = \tilde{\Omega} \ 
  \tilde{v}_\parallel \mathbf{b} +
    \tilde{v}_\parallel^2 \ \tilde{\nabla} \times \mathbf{b} -
    \mathbf{b}\times\biggl( \tilde{\mathbf{E}}
       - \tilde{v}_\parallel\partial_\tau\mathbf{b} -
      \frac{\tilde{\mu}\tilde{\nabla}\tilde{B}}{2} \biggr),
      \label{eq:X_GC}
\end{align}
\vspace{-1.5em}
\begin{align}
     \Omega^{*} \ \frac{d\tilde{v}_\parallel}{d\tau} = \biggl( \tilde{\Omega} \  \mathbf{b} + \tilde{v}_\parallel
      \tilde{\nabla}\times\mathbf{b} \biggr) \cdot \biggl( \tilde{\mathbf{E}} -
          \tilde{v}_\parallel\partial_\tau\mathbf{b} - \frac{\tilde{\mu}\tilde{\nabla}\tilde{B}}{2} \biggr),
    \label{eq:v_par_GC}
\end{align}
\noindent where $ \Omega^{*} = \tilde{\Omega} + \tilde{v}_\parallel \Bigl( \mathbf{b} \cdot \tilde{\nabla} \times \mathbf{b} \Bigr)$, the position $\mathbf{X}$ of the guiding-center is normalized to a reference length $L_\text{ref}$, the time $\tau$ to $T_\text{ref}$ and the parallel velocity to $V_\text{ref}=L_\text{ref}/T_\text{ref}$. In these expressions  $\tilde{B} = B/B_\text{ref}$ is a normalized magnetic field magnitude, $\tilde{\mathbf{E}} = \tilde{E}_\text{ref} (\mathbf{E}/E_\text{ref})$ a normalized electric field and $\tilde{\Omega} = \Omega_\text{ref} \tilde{B}$ a normalized gyrofrequency. Moreover, $\tilde{\nabla} = L_\text{ref} \nabla$, $\tilde{E}_\text{ref} = \tilde{\Omega}_\text{ref} (E_\text{ref} V_\text{ref}^{-1} B_\text{ref}^{-1})$ and the magnetic moment $\mu$ is normalized to the ratio $U_\text{ref}/B_\text{ref}$ and $U_\text{ref}$ is the kinetic energy corresponding to $V_\text{ref}$.  

This enables the use of different kinds of magnetic equilibria in any given geometry. Using these equations of motion, the integration of the particles' dynamics can be performed independently of the specific geometry and coordinate system of the problem, taking only the tensor metric $\mathbf{g}$ and either the co- or contravariant components of $\mathbf{B}$ as inputs. All magnetic configurations analyzed in this work are equilibrium vacuum configurations with no electric fields resulting in $\partial_\tau \mathbf{b}=0$, $\mathbf{b} \cdot \tilde{\nabla} \times \mathbf{b} = 0$ and $\tilde{\mathbf{E}}=0$, effectively reducing the equations of motion to:
\begin{equation}
    \frac{d\mathbf{X}}{d\tau} = 
     \ \tilde{v}_\parallel  \mathbf{b} +
      \frac{\tilde{\mu}}{2 \tilde{\Omega}} \ \mathbf{b}\times \tilde{\nabla}\tilde{B},
\end{equation}
and
\begin{equation}
    \frac{d\tilde{v}_\parallel}{d\tau}= - \frac{\tilde{\mu}}{2} \ \mathbf{b} \cdot \tilde{\nabla}\tilde{B}.
\end{equation}
\par
\begin{figure}
    \centering
    \includegraphics[width=0.4\textwidth]{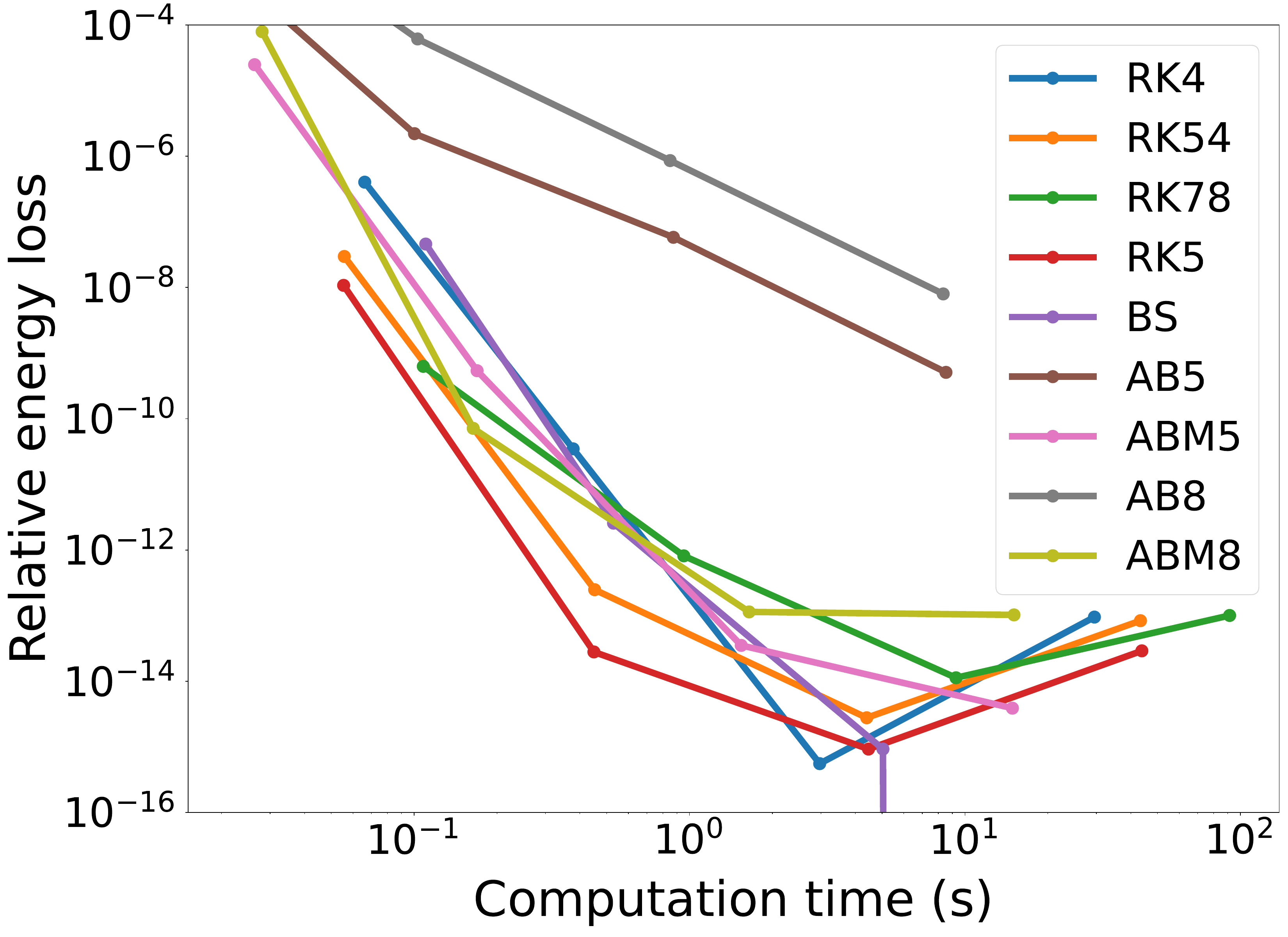}
    \includegraphics[width=0.4\textwidth]{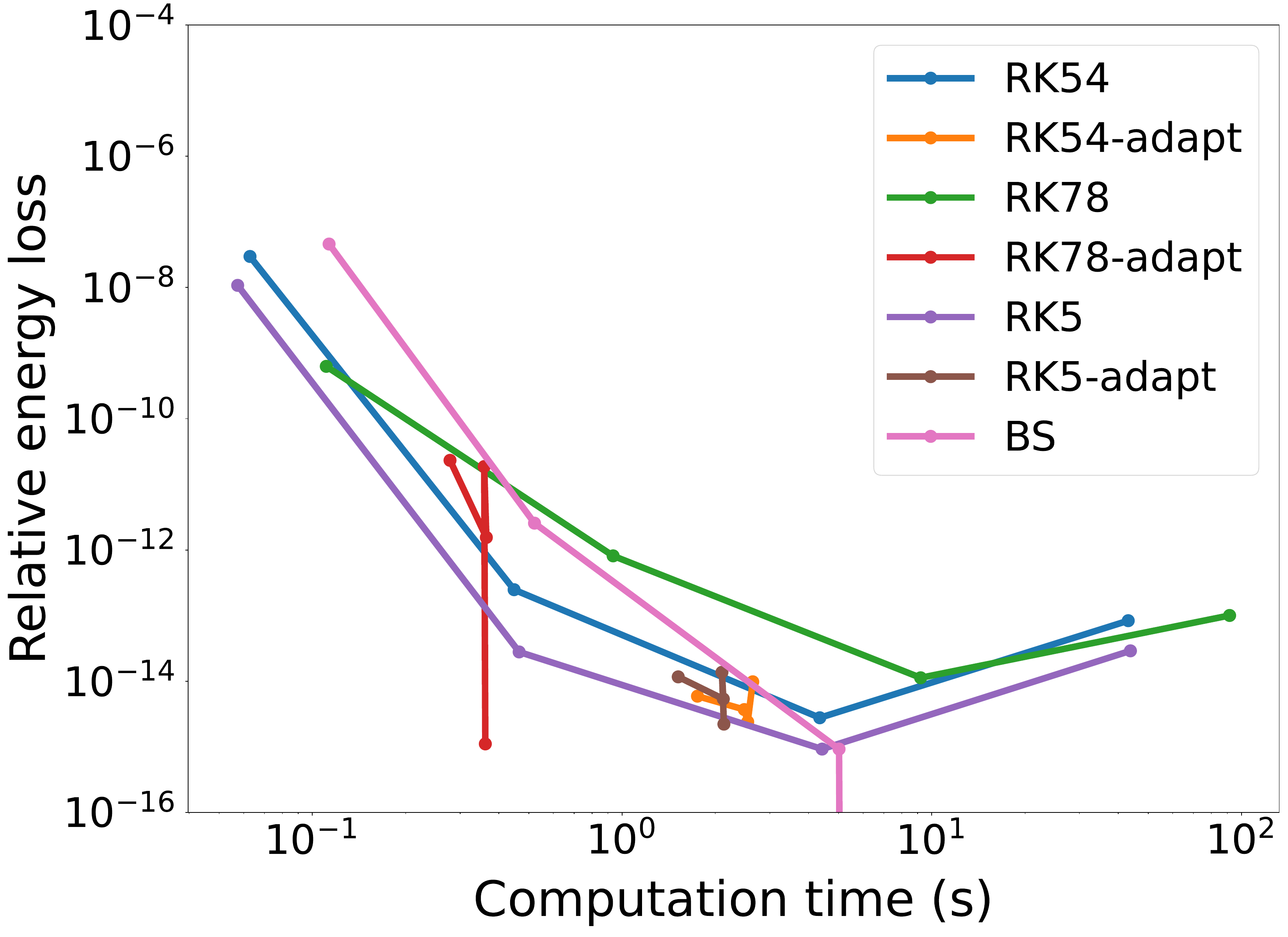}
    \caption{Evaluation of the performance of different integration algorithms. Left: Constant step Runge-Kutta (RK), Adams-Bashforth (AB), and Bulirsch–Stoer (BS) integrators. Right: Best performing constant step integrators and their adaptive step counterparts available in the \textit{boost::odeint} library.}
    \label{fig:integrations}
\end{figure}

 The equations of motion \cref{eq:X_GC,eq:v_par_GC} can be integrated with a wide assortment of algorithms made available by the ODE library \textit{boost::odeint} \citep{AhnertMulansky2011}, either with constant or adaptive time step, which can be easily selected from within the \textit{gyronimo} framework and allow the flexibility to choose the algorithm best adapted to the problem being solved. Henceforth, we will work with the Dormand-Prince (RK5) method from the Rung-Kutta family of ODE solvers. This decision is based in \cref{fig:integrations}, where it is shown on the panel of the left the trade-off between the energy loss $\Delta E= (E-E_0)/E_0$, which is a proxy for the computational error since the energy should be conserved, and the computation time required by different constant time-step algorithms to integrate up to the same final instant. On the right panel, one compares the best-performing constant-step algorithms with their adaptive-step counterparts when available. 

Although the Bulirsch-Stoer (BS) method and the forth-order Runge-Kutta (RK4) constant step methods exhibit better performances, that entails increased computational costs, which is a significant bottleneck when evaluating loss fractions for a wide range of magnetic configurations in optimization workflows. Even though the adaptive step Runge-Kutta-Fehlberg (RK78 adapt) appears to accomplish this, the parameters that control the energy loss do not show a linear dependence and so were left to later statistical and systematic analysis. It is also interesting to note that there is a limit on how small we can make the step size and still decrease the relative energy loss, a limit that may be due to the numerical precision used in the code.

\section{Single Particle Motion}

\par We begin by performing integration of the guiding-center trajectories of alpha particles at an energy of $3.52$ MeV. All of the equilibria were scaled to have an effective minor radius $a_A=1.7044$ m, equal to the one of the ARIES-CS reactor \citep{najmabadiARIESCSCompactStellarator2008}, and major radius larger than ARIES-CS, $R_A=7.7495$ m, with a magnetic field of $5.3267$ T. In here, ARIES-CS is used as a baseline scale as it was one of the latest viability studies for a power plant-grade compact stellarator. However, as it is a compact device, its aspect ratio $A$, the ratio between its major radius $R_A$ and its minor radius $a_A$, may stretch the limit of applicability of the NAE to be effective all the way up to the boundary, so we take higher values of the major radius in order to have a better behavior of this approximation, albeit at the cost of a less compact reactor. The effective minor radius in this work is given as an average of this quantity throughout the last closed surface, which can be written as $\pi a_{A}^2 = 1/2 \pi \int_0^{2\pi} S(\phi) d\phi$, where $S(\phi)$ is the area of the surface cross section at each point of the cylindrical angle $\phi$ \citep{jorgeIontemperaturegradientStabilityMagnetic2021}. We note that the minor radius value is substantially larger than the alpha particles gyroradius as expected of this kind of fusion reactor.

\par Finally, we initialize the particles velocities with two parameters: $s_{v_\parallel}$, the sign of the initial parallel velocity, which can take the values +1 and -1, and $\lambda$, the ratio between the initial perpendicular kinetic energy $E_{\perp_i}= m v_\perp^2 / 2$, and the total kinetic energy $E=m v^2 / 2$. The normalized adiabatic invariant is given by $\tilde{\mu}=\lambda (E/E_\text{ref}) (B_\text{ref}/B(r_i))$, where $r_i$ represents the initial position of the particle.

\par
\begin{figure}
    \centering
    \hfill
    \includegraphics[width=0.32\textwidth]{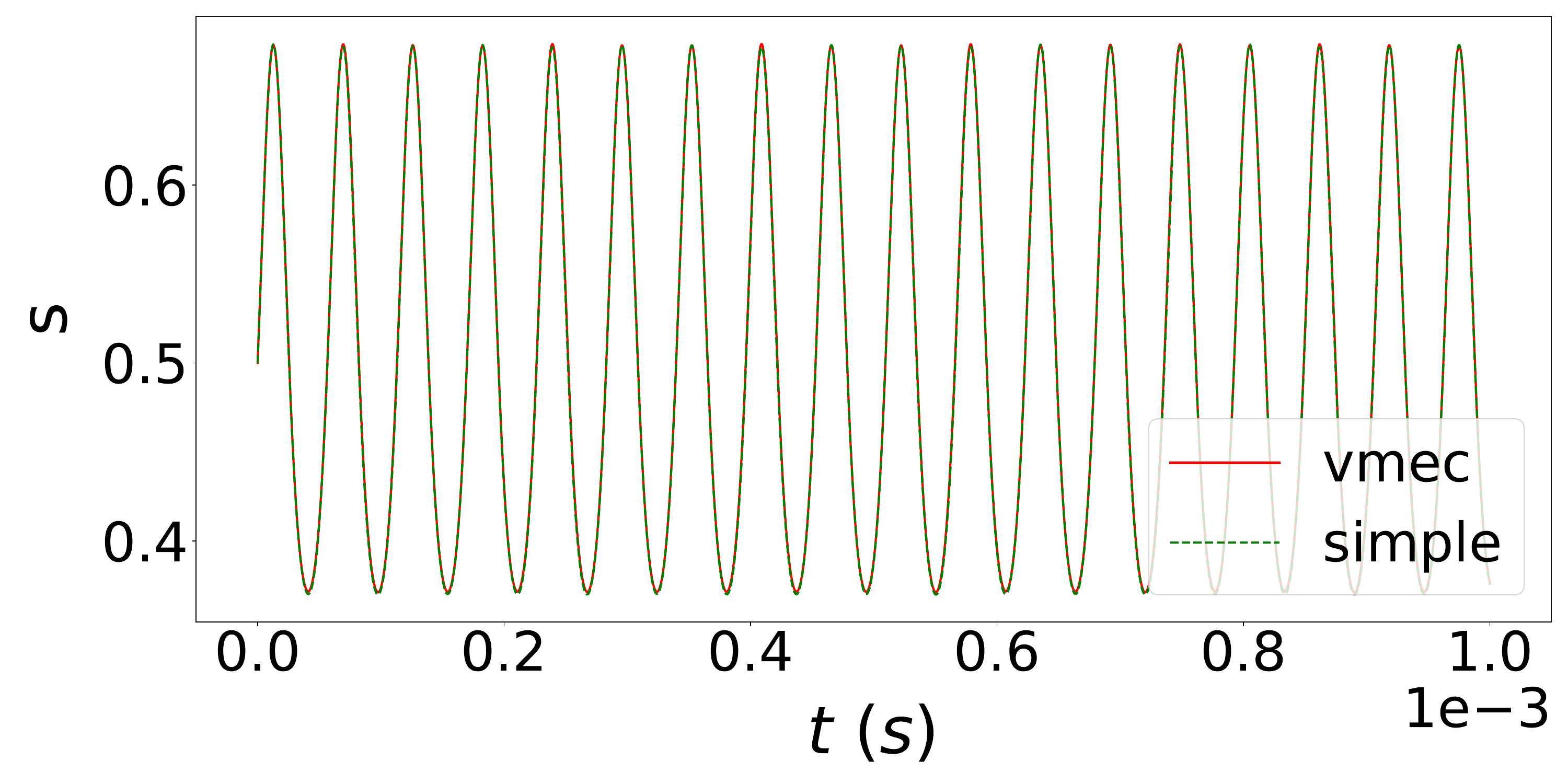}
    \hfill
    \includegraphics[width=0.32\textwidth]{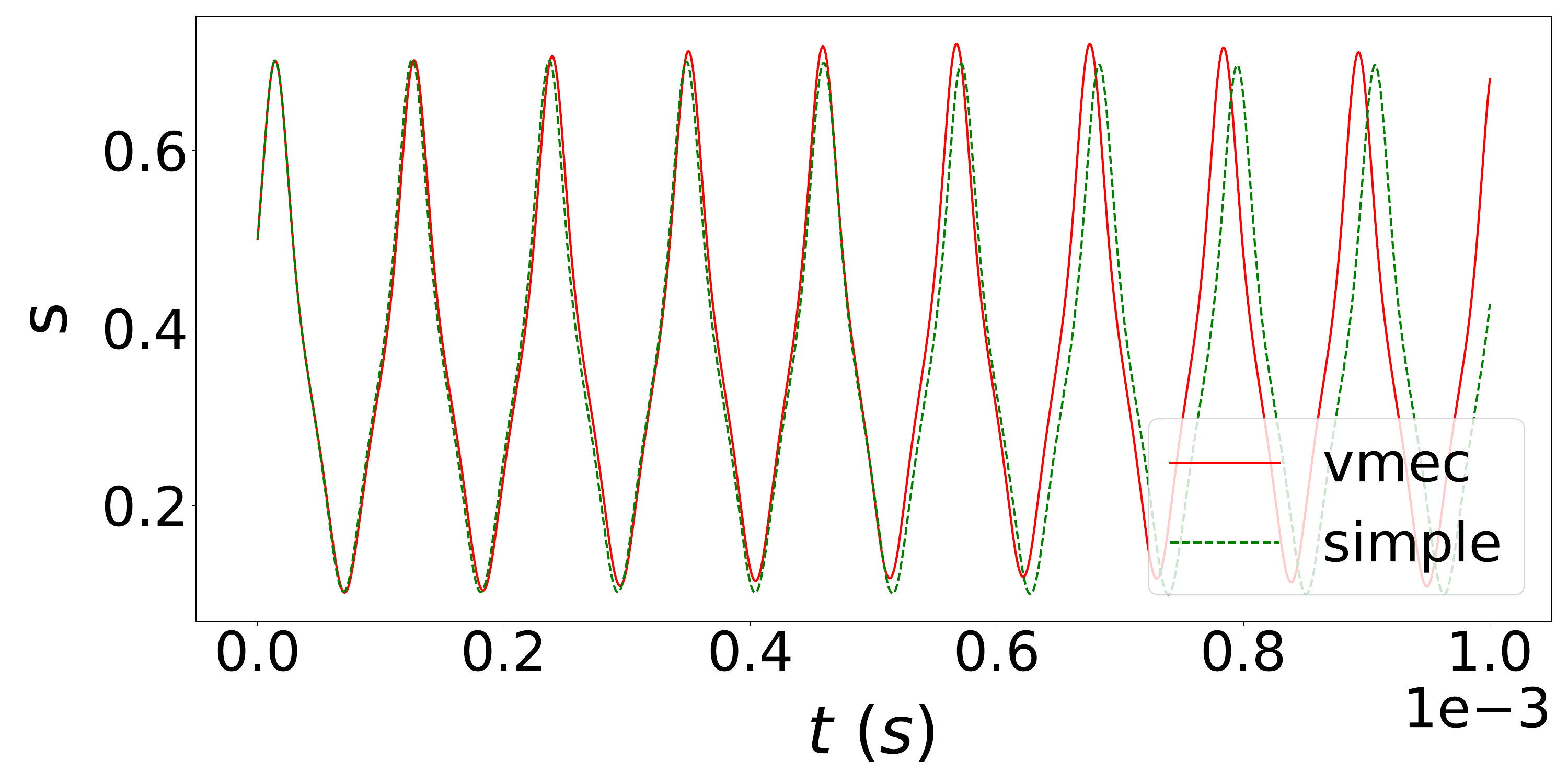}
    \hfill
    \includegraphics[width=0.32\textwidth]{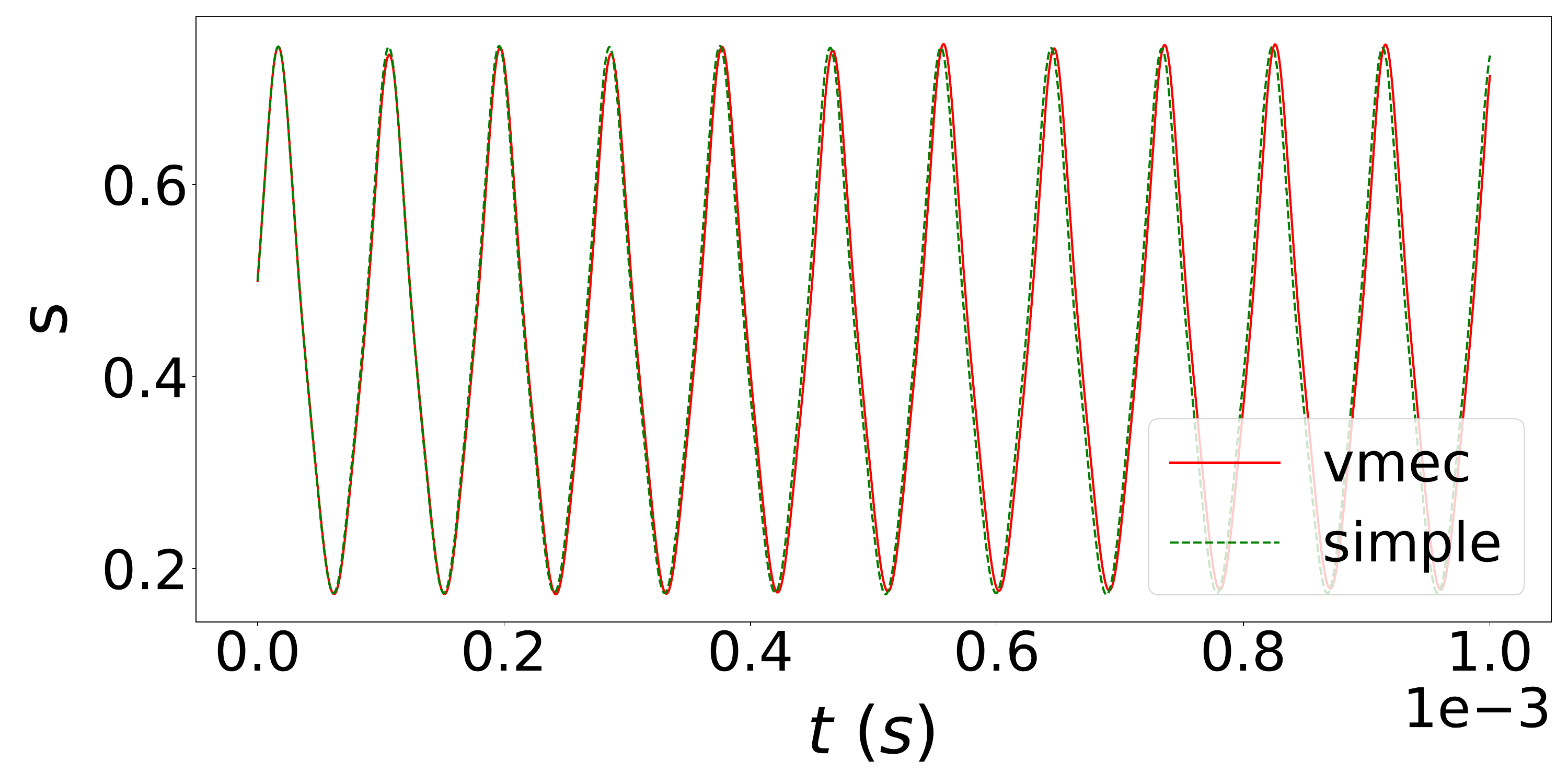}
    
    \vspace{0.1em}

    \includegraphics[width=0.25\textwidth]{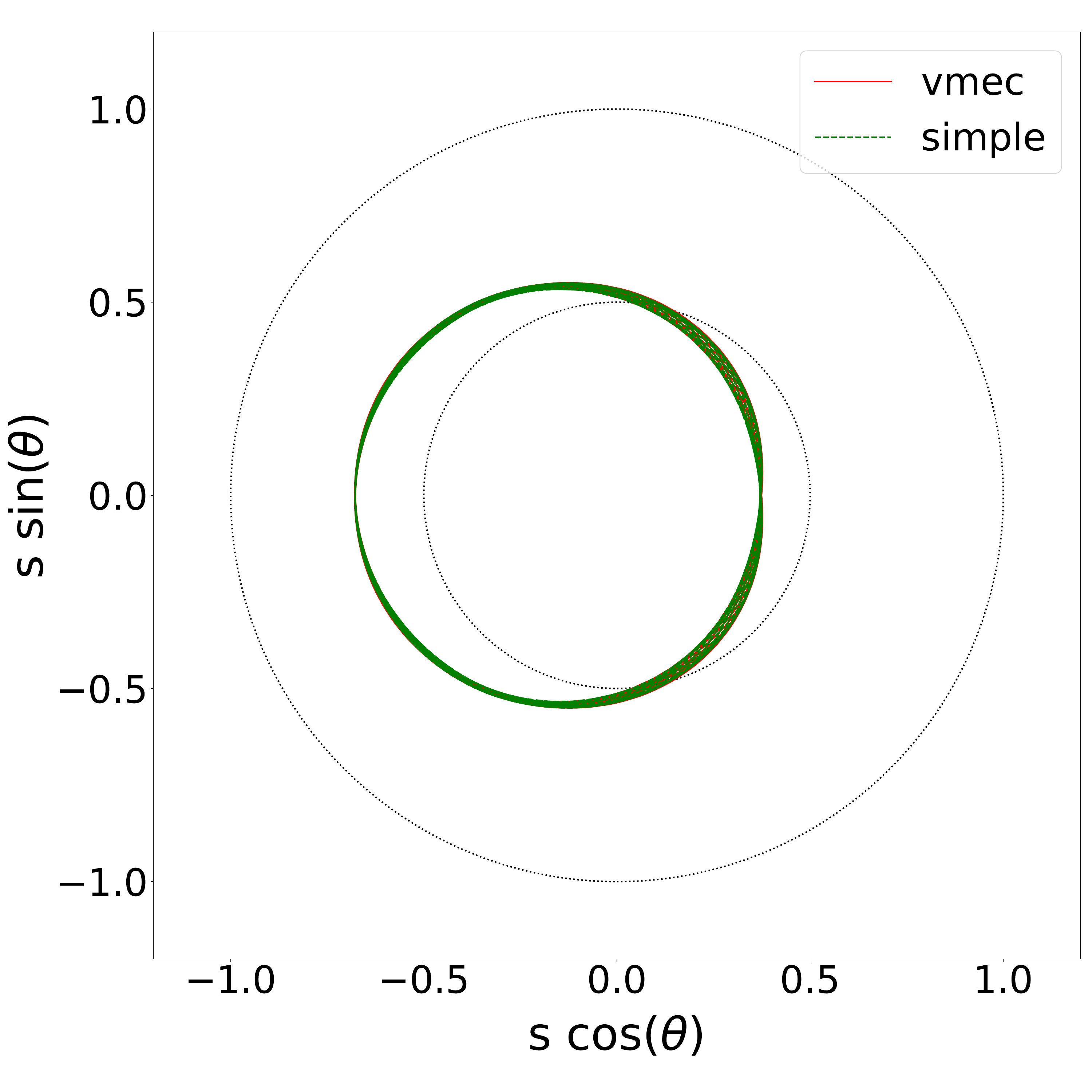}
    \hspace{2.5em}
    \includegraphics[width=0.25\textwidth]{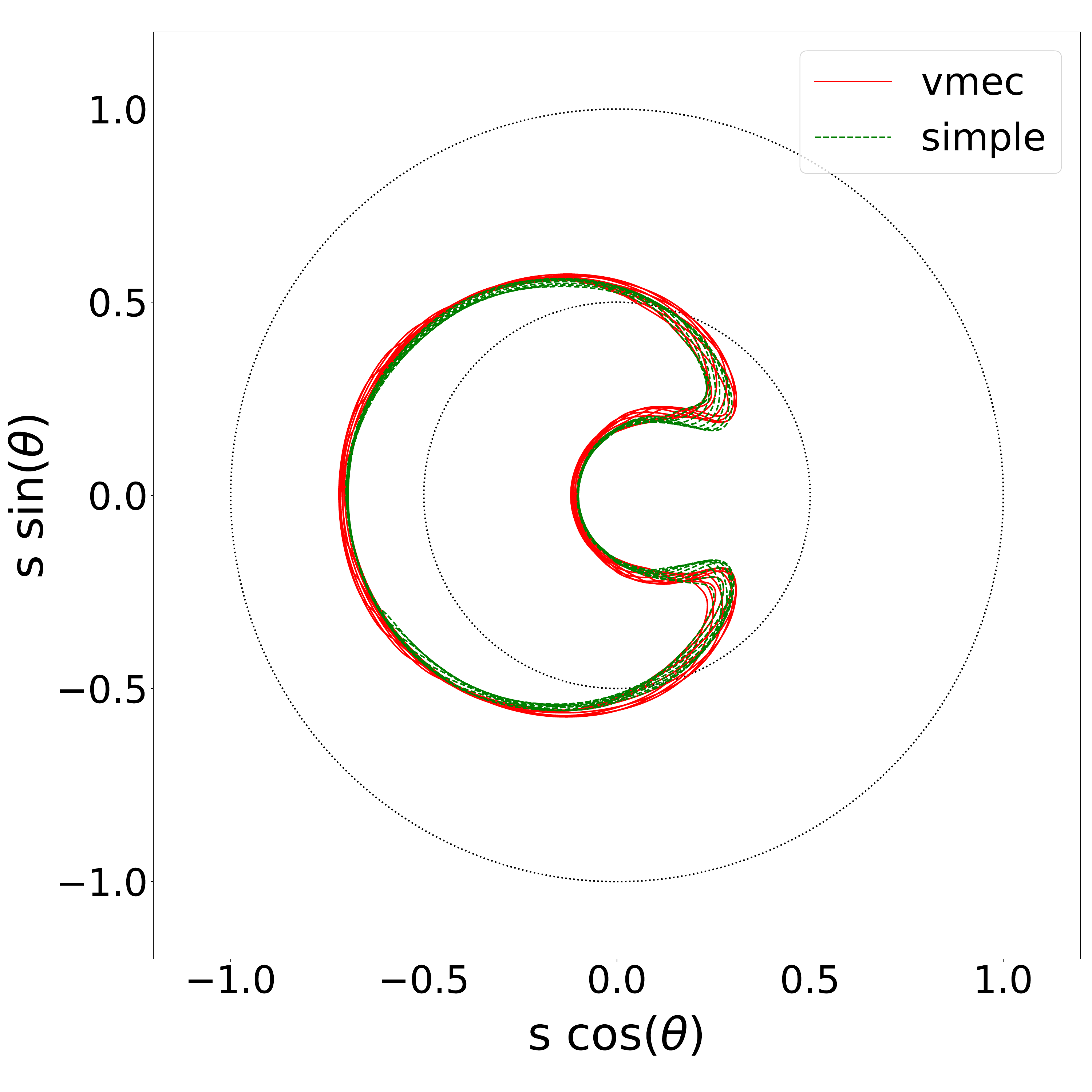}
    \hspace{2.5em}
    \includegraphics[width=0.25\textwidth]{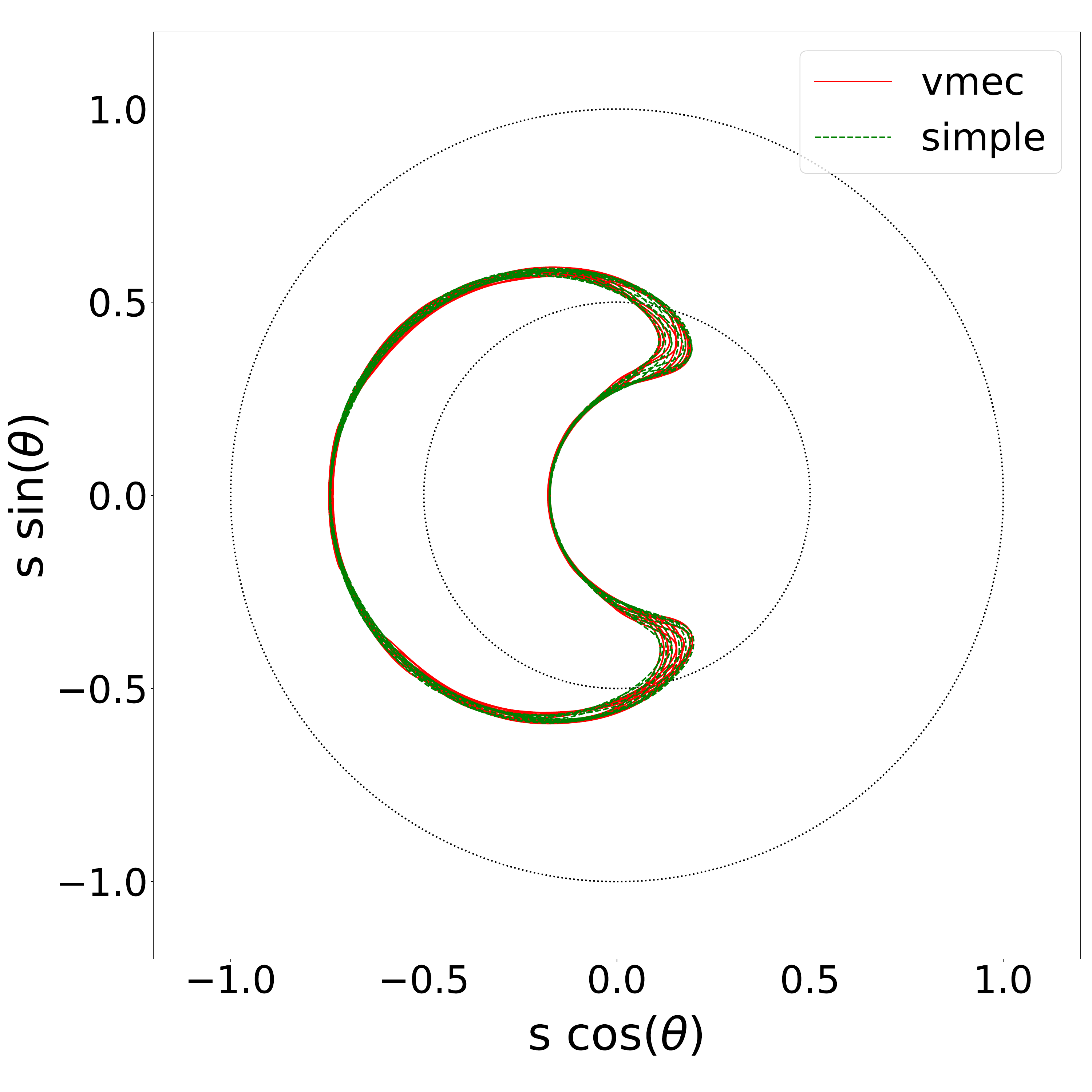}
    
    \caption{Comparison between \textit{gyronimo} (vmec) and \textit{SIMPLE} (simple) tracers for original \textit{precise QA}, scaled to have ARIES-CS minor radius $a_A$ and $B_0$, the field at the plasma axis. With an initial position $(s,\theta, \phi)=(0.25, 0.1, 0.1)$ and $\lambda=0.95$ (Left), the particle describes a passing orbit, and for $\lambda=0.97$ (Center) and  $\lambda=0.99$ (Right), trapped orbits are observed. Above is the temporal evolution of the radial coordinate $s$ and below is the poloidal view of the orbit in \textit{VMEC} coordinates, with an inner dashed circle for the initial flux surface and an outer dashed circle for the last closed flux surface. Only approaching the passing-trapped separatrix ($\lambda=0.97$), do the tracers slightly diverge.}
    \label{fig:simple_comp}
\end{figure}

 In order to benchmark \textit{gyronimo} with \textit{SIMPLE}, the same alpha particles were followed with both tracers for $1$ ms in the precise QA VMEC equilibrium in \citet{landremanMagneticFieldsPrecise2022} scaled to the conditions described above, except for the major radius, which is $R = 6 \ a_A$ to preserve the original aspect ratio of $A = 6$. Figure \ref{fig:simple_comp} exhibits the orbits of particle with  $s_{v_\parallel}=1$ and initial position $(s,\theta, \phi)=(0.25, 0.1, 0.1)$. The angular coordinates $\theta$ and $\phi$ were chosen to be finite values to ensure these are not a special case where the orbits coincide, which could happen for $\theta=\phi=0$. Although this benchmark included the variation of all initial values we only exhibit here a change in the value of $\lambda$ which was observed to be one of the most important factors in the alignment between integrators. In the presented results the values of $\lambda$ are $0.95$, $0.97$, and $0.99$. 

\par The radial oscillations predicted by the two different approaches agree for the cases $\lambda=0.95$ and $\lambda=0.99$, which correspond, respectively, to a passing and a trapped particle. Such agreement, however, starts to decrease over extended temporal intervals for the case $\lambda=0.97$. This corresponds to a trapped particle that starts to approach the passing-trapped transition where orbits tend to be more sensitive to slight changes in the integration conditions. For those values of $\lambda$, the amplitude of oscillations in $s$ remains the same but its temporal variation contains a delay between the codes, while the average radial position exhibits a small deviation. While the temporal deviation may be due to the symplectic nature of SIMPLE, the source of the difference in the average radial position, among other factors, may include the fact that the motion of a particle in the transition from the trapped to the passing regime depends strongly on higher order terms of the GC motion. The segment of the banana orbits around reflection points (banana tips) correspond to low values of $v_\parallel$. We note that such higher order terms may have a non-negligible effect in points along the orbit where the values of $v_\parallel$ stay close to zero.

\par
\begin{figure}
    \centering
    \hspace{-1em}
    \includegraphics[width=0.32\textwidth]{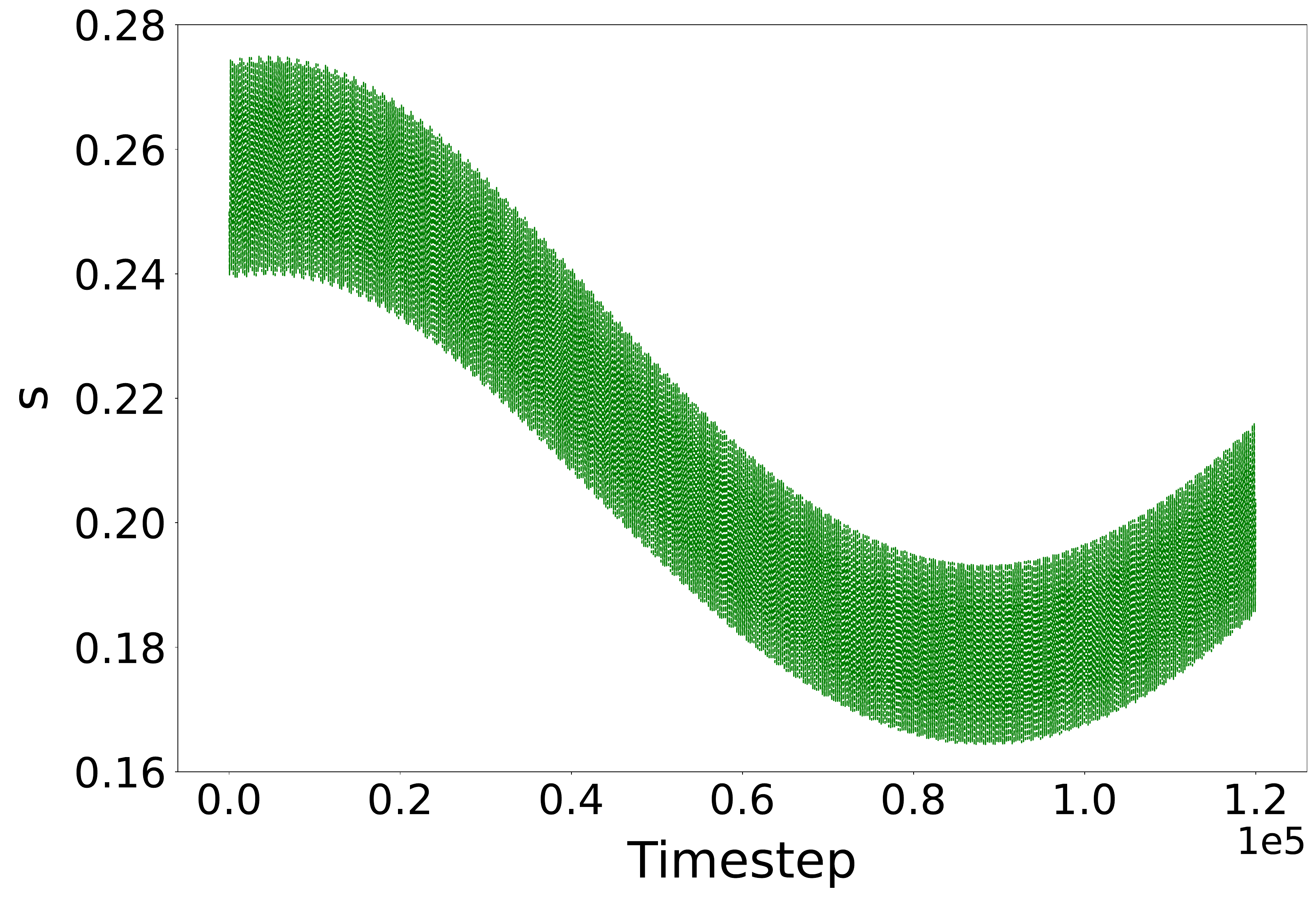}
    \hfill
    \includegraphics[width=0.32\textwidth]{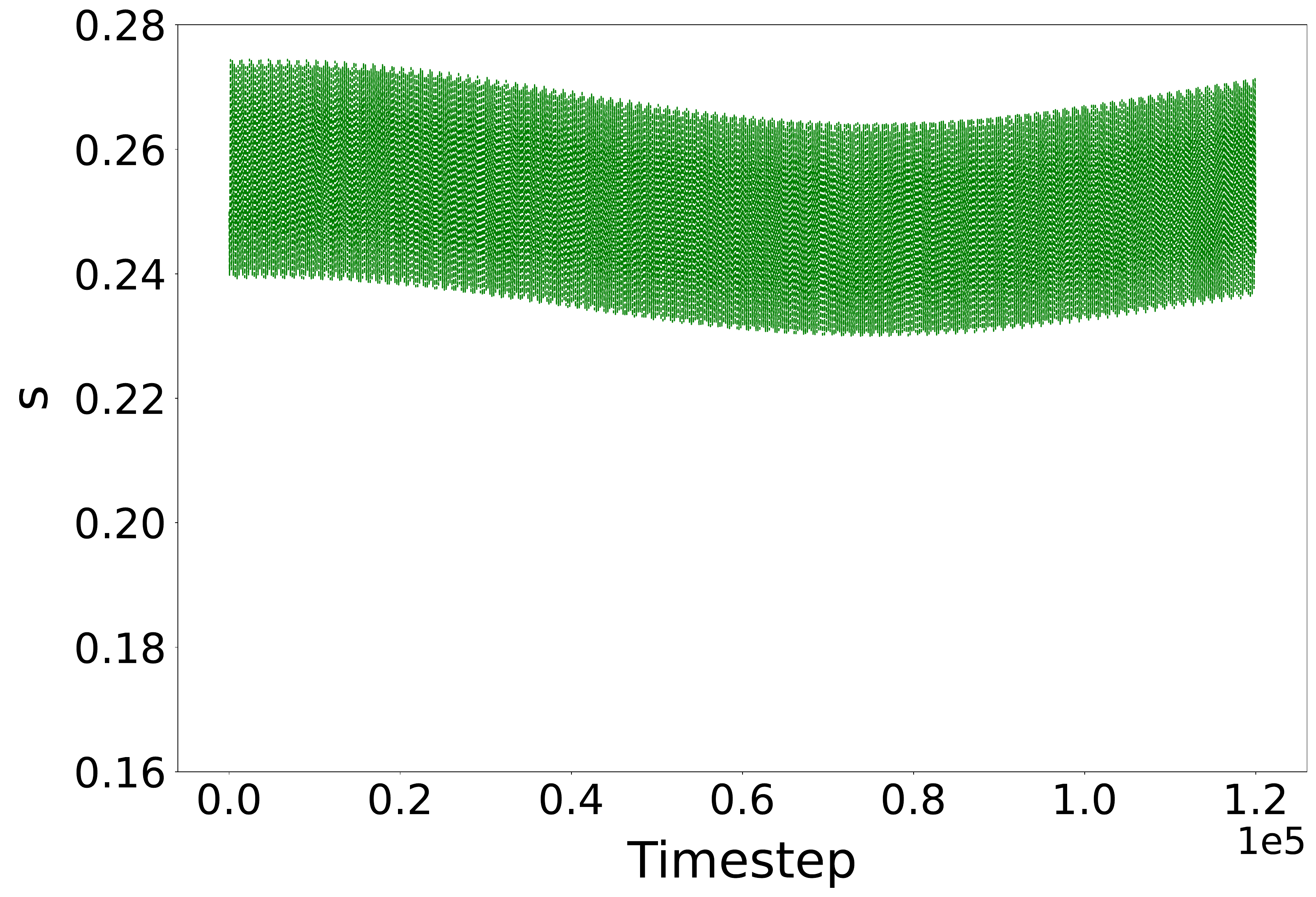}
    \hfill
    \includegraphics[width=0.32\textwidth]{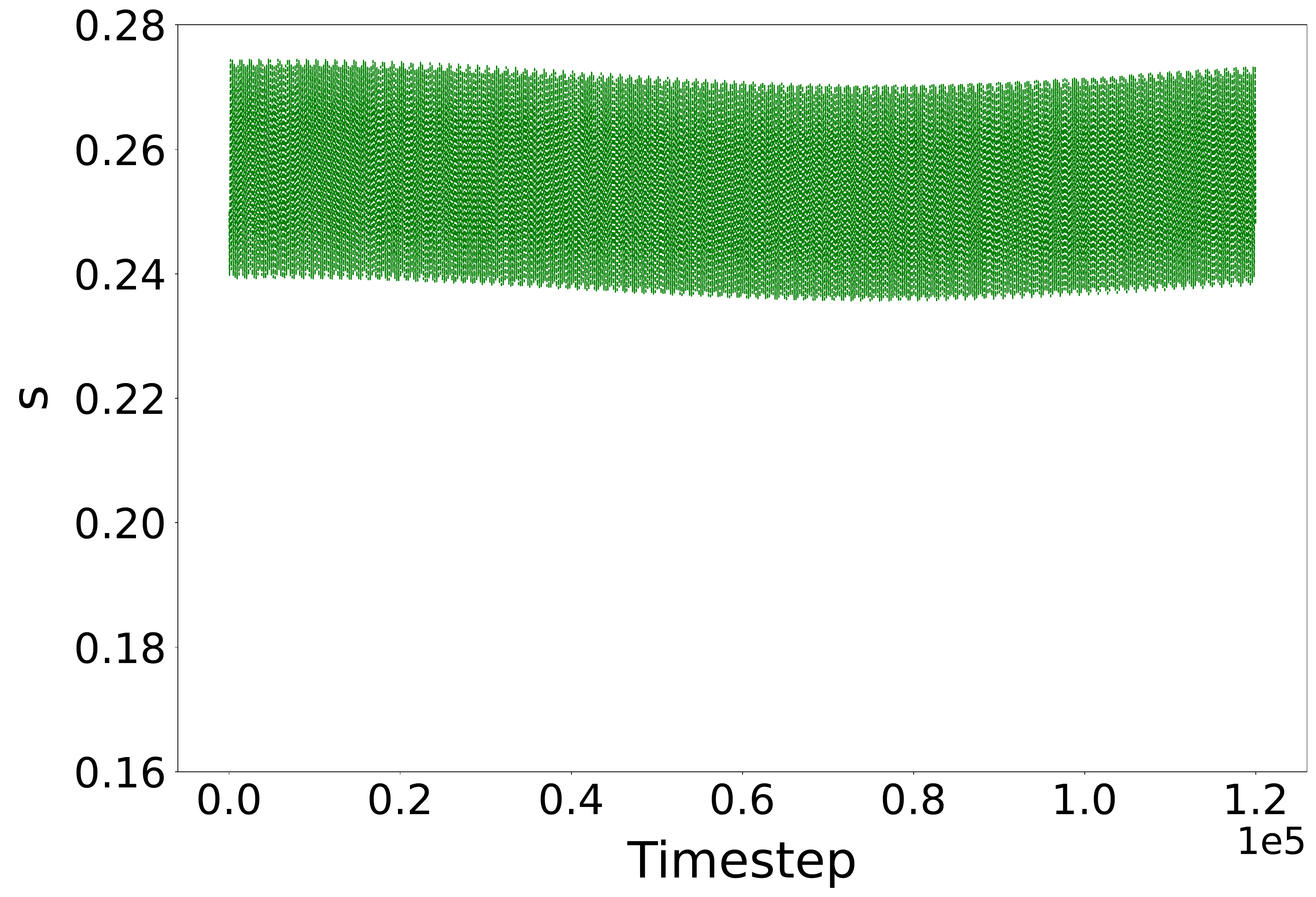}
    \caption{Orbits obtained with SIMPLE with $(s,\theta, \phi)=(0.25, 2.89, 1.84)$ and $v_\parallel / v = 0.44$ as initial conditions, by varying the angular grid factor multharm. Left: multharm $=3$. Center: multharm $=4$. Right: multharm $=5$.}
    \label{fig:simple_changes}
\end{figure}

 We conclude that we may work with \textit{gyronimo} as it compares well in most cases with \textit{SIMPLE}. Throughout the rest of the analysis we will not be showing the results of the SIMPLE particle tracing as such a good match for some orbits is only achievable using an increased amount of memory due to the number of grid points necessary for the interpolation used by SIMPLE. The effect of varying an interpolation angular grid factor that controls the number of poloidal and toroidal grid points within SIMPLE can be seen in \cref{fig:simple_changes} for a particle initialized with $s_i=(0.25, 2.89, 1.84)$ and $v_\parallel$ $/ v = 0.44$, where drastic changes on its passing orbit. This is partially explained by the fact that the main goal of SIMPLE is to achieve long-term conservation of invariants like energy rather than minimizing the spatial deviation from a reference orbit \citep{albertSymplecticIntegrationNoncanonical2020}. Despite that, all single particle results were compared with the corresponding values in SIMPLE with a low grid factor to ensure that there are no major divergences between tracers, including the ones shown in this work.

\par We will divide our comparison for QA and QH configurations due to the nuanced analysis needed between them. The reference case for a QA stellarator used was the precise QA from \citet{landremanMagneticFieldsPrecise2022} and the one for QH stellarators was the four field period QH vacuum configuration with magnetic well from section 5.4 of \citet{landremanMappingSpaceQuasisymmetric2022}. Although we will use the terms QH and QA for all of the magnetic configurations that are close to precise QS, some have finite deviations from it. This is the case because we take the outer surface of NAE configurations calculated by \textit{pyQSC} and create an input file for \textit{VMEC} with the relevant configuration parameters, following the procedure in \citep{landremanDirectConstructionOptimized2019}. \textit{VMEC} computes an ideal MHD equilibrium from this input file which may not retain some of the initial features of the NAE configuration. This is especially true for equilibria with smaller aspect ratios, where the outer bounds are not well described by an expansion from the axis.    

\par
\begin{figure}
    \centering
    \hfill
    \includegraphics[width=0.25\textwidth]{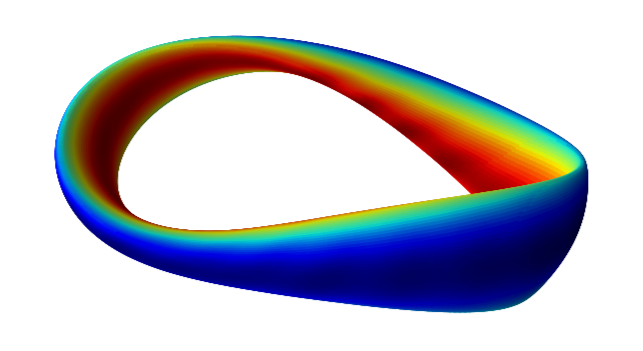}
    \hfill
    \includegraphics[width=0.25\textwidth]{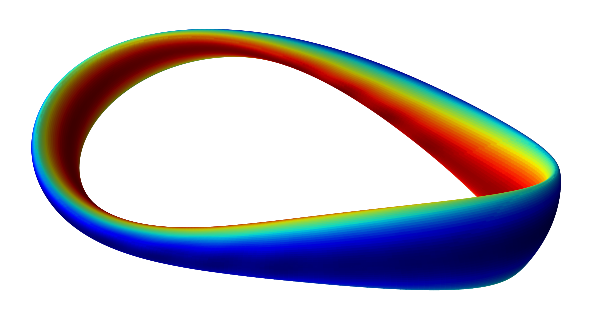}
    \hfill
    \includegraphics[width=0.25\textwidth]{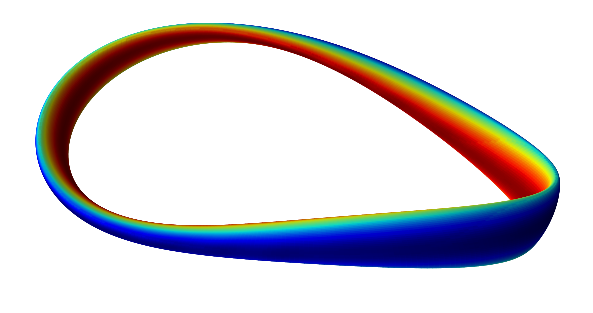}
    \hfill
    \vspace{0.1em}
    \includegraphics[width=0.3\textwidth]{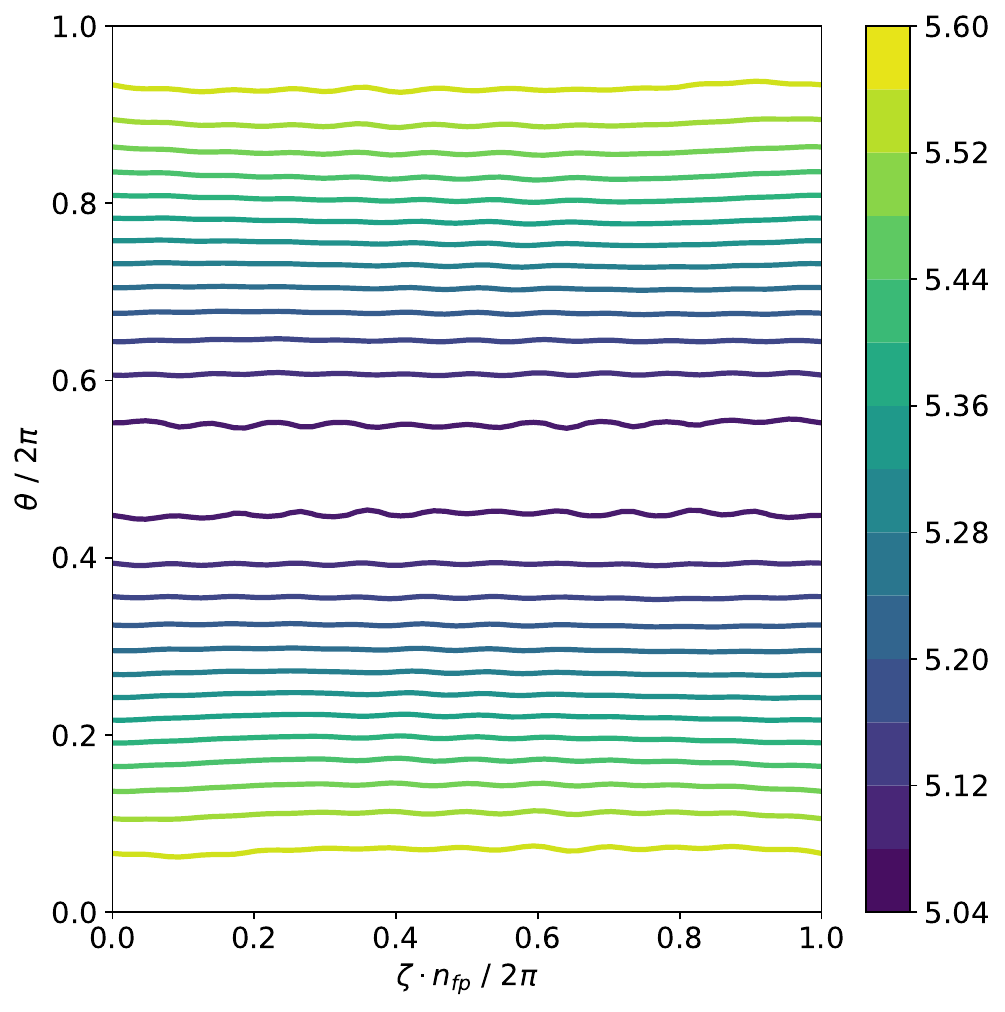}
    \includegraphics[width=0.3\textwidth]{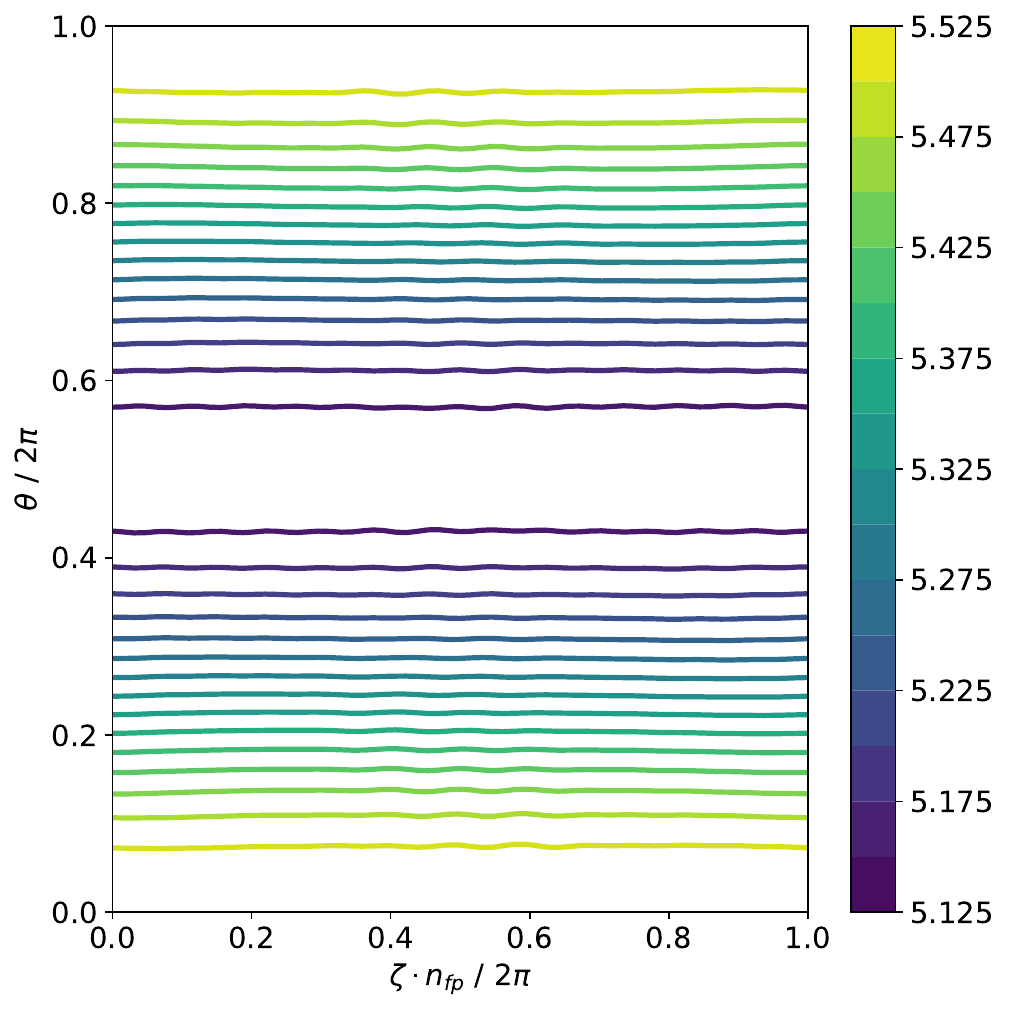}
    \includegraphics[width=0.3\textwidth]{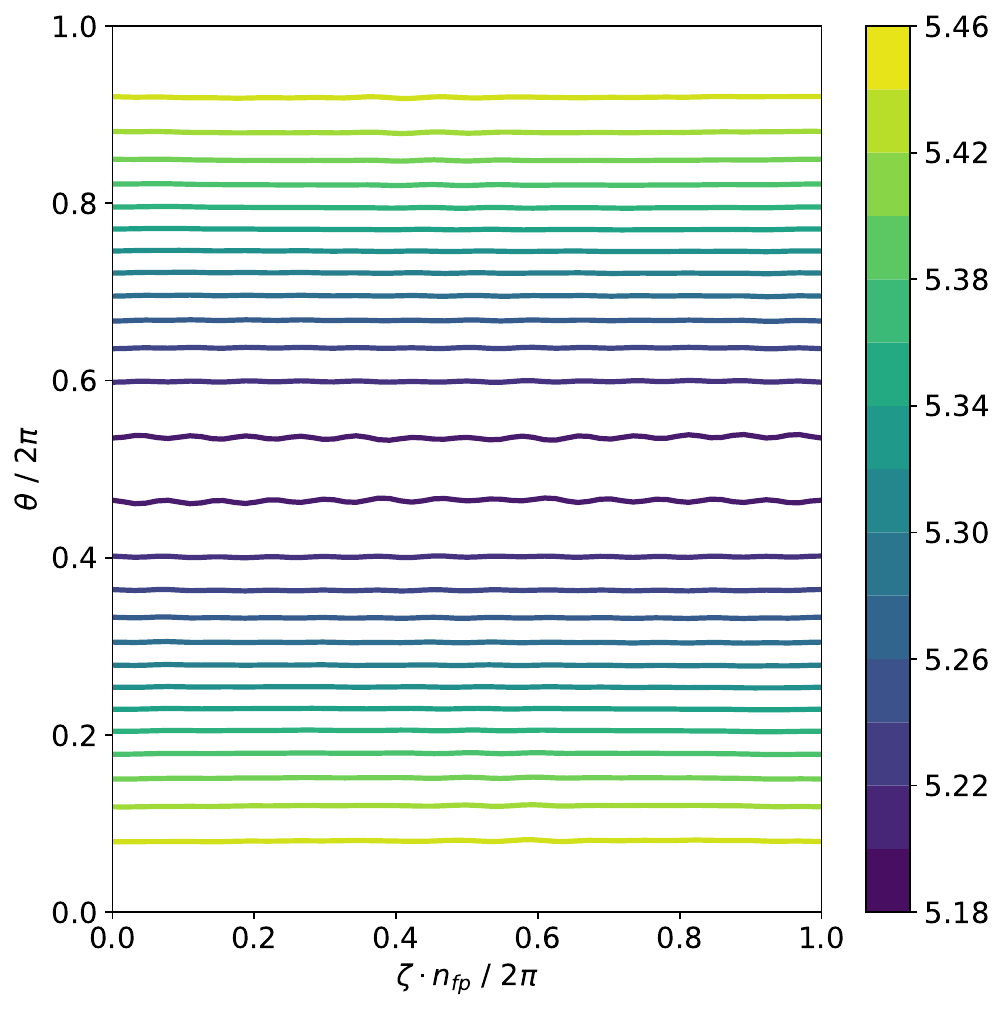}
    \caption{QA \textit{VMEC} equilibria generated from the precise QA from \textit{pyQSC} with different major radius scalings. Above are the 3D versions of the equilibria and below are the contour plots of the magnetic fields on the angular \textit{VMEC} coordinates. Left: Aspect ratio $A=6.8$. Center: $A=9.1$. Right: $A=13.6$.}
    \label{fig:QA_eq}
\end{figure}

\begin{figure}
    \centering
    \hfill
    \includegraphics[width=0.25\textwidth]{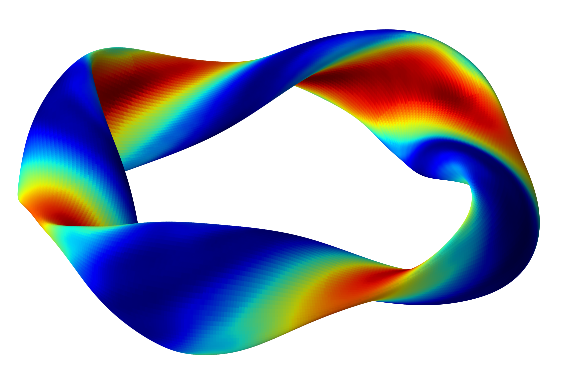}
    \hfill
    \includegraphics[width=0.25\textwidth]{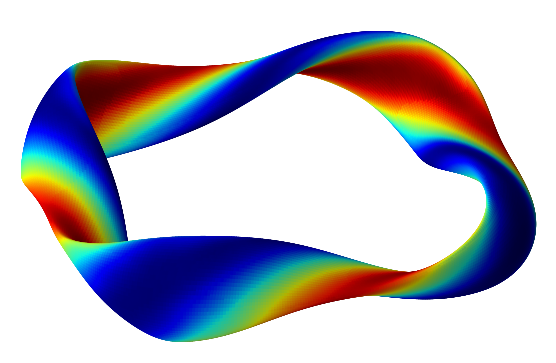}
    \hfill
    \includegraphics[width=0.25\textwidth]{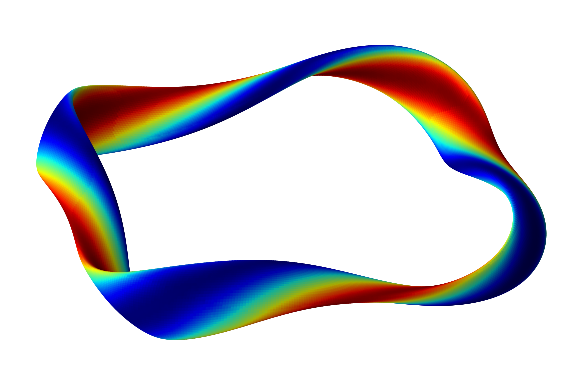}
    \hfill
    \vspace{1em}
    \includegraphics[width=0.3\textwidth]{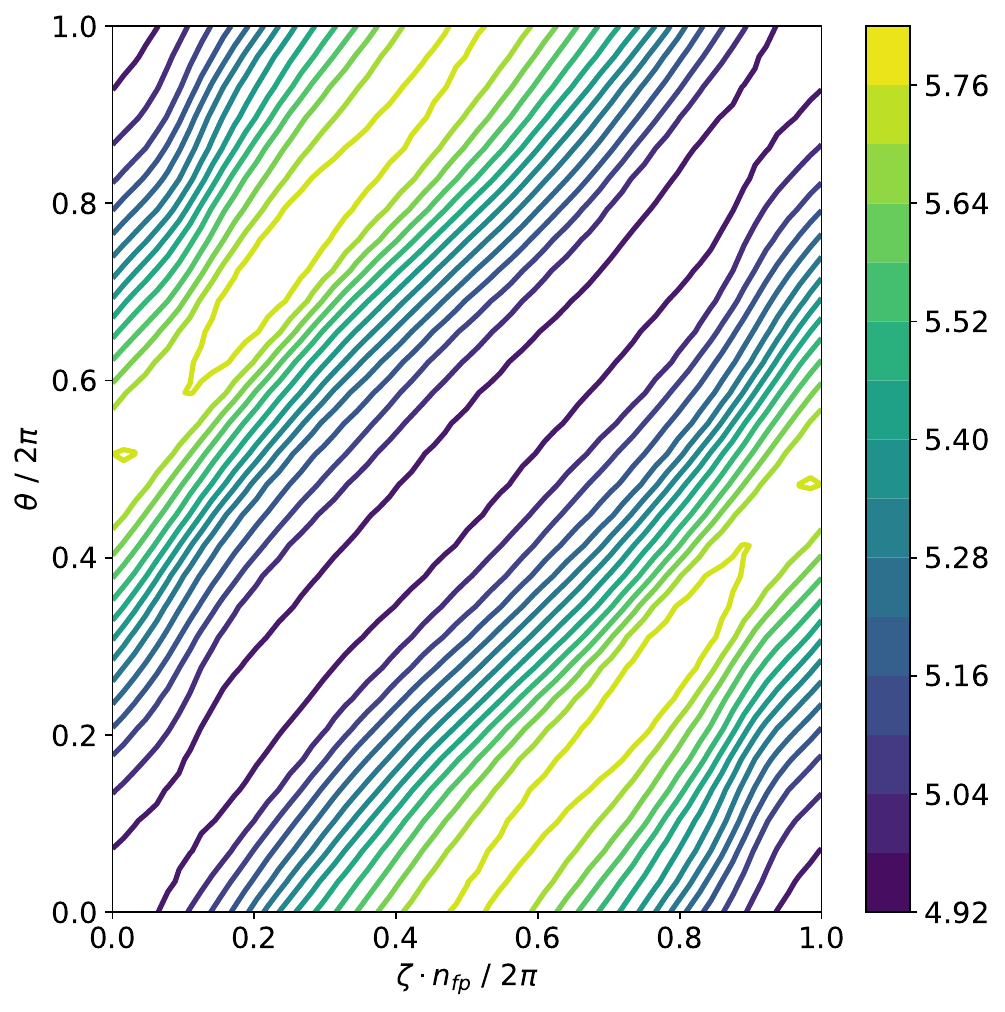}
    \includegraphics[width=0.3\textwidth]{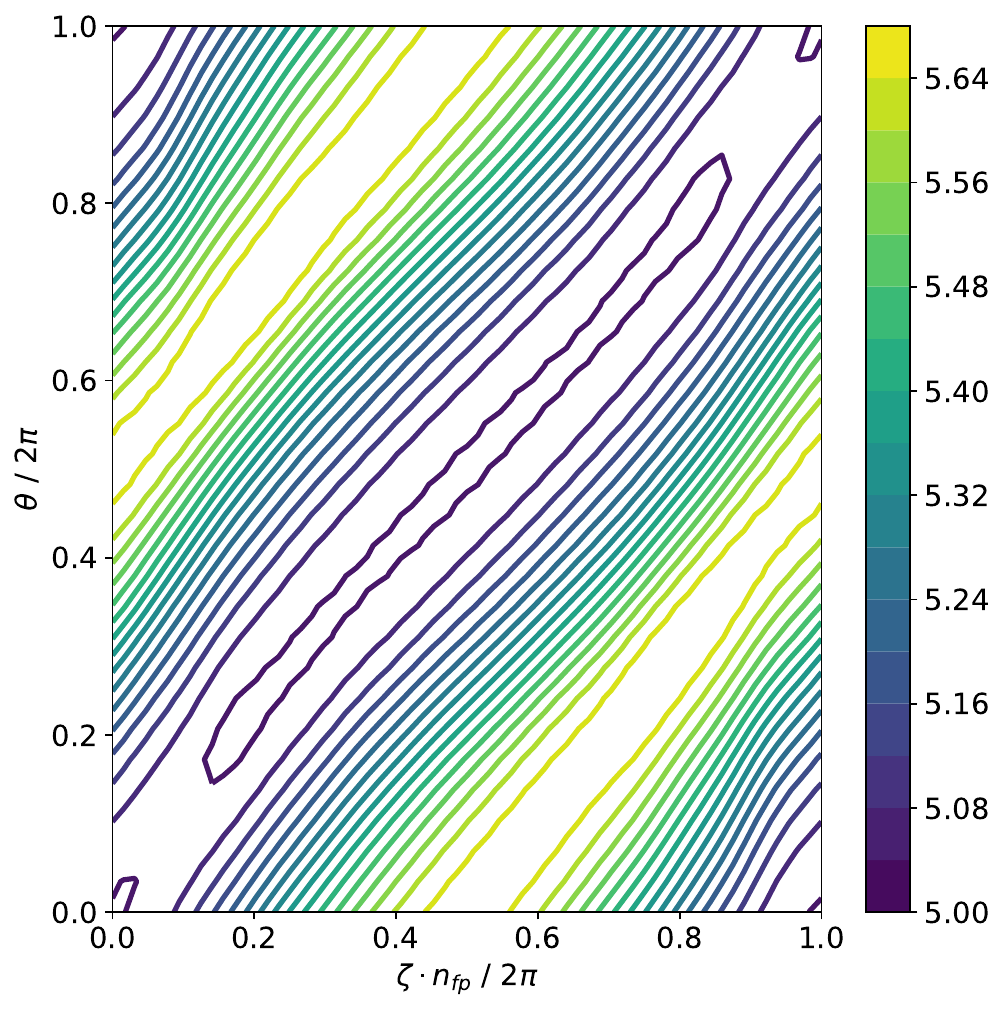}
    \includegraphics[width=0.3\textwidth]{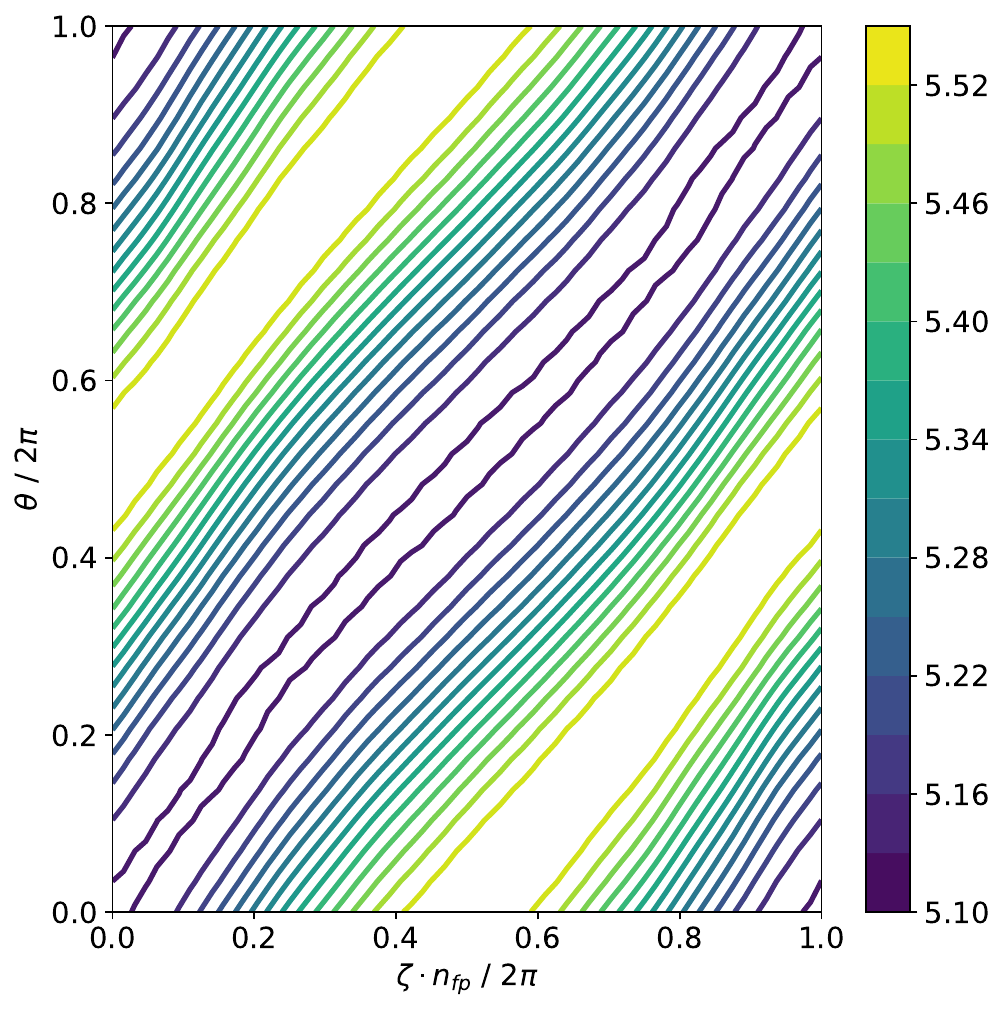}
    \caption{QH \textit{VMEC} equilibria generated from the "2022 QH nfp4 well" from \textit{pyQSC} with different major radius scalings. Above are the 3D versions of the equilibria and below are the contour plots of the magnetic fields on the angular \textit{VMEC} coordinates. Left: Aspect ratio $A=6.8$. Center: $A=9.1$. Right: $A=13.6$.}
    \label{fig:QH_eq}
\end{figure}

 In order to have a good overall picture of how the NAE compares with with different aspect ratio equilibria, the analysis will encompass equilibria with major radius $R=3, 2$ and $1.5$ times the major radius of ARIES-CS, $R_A$. We can see from \cref{fig:QA_eq,fig:QH_eq} how quasisymmetry degrades when generating a \textit{VMEC} equilibrium from the near-axis one for lower aspect ratios. While this effect is more evident for the QH stellarator in \cref{fig:QH_eq}, such degradation is visible also in the QA one in the contour lines representing higher values of $B$ in \cref{fig:QA_eq}. The contour lines representing lower magnitudes of $B$ appear not to follow this trend, but only because they are closer to $\theta / 2 \pi = 0.5$ for the $A=13.6$ equilibrium than for the $A=9.1$ one.

\subsection{Quasi-Axisymmetry}

\par We now proceed to the comparison between orbits of particles for the \textit{VMEC} equilibria and the \textit{pyQSC} ones that originated them. All orbits were traced for $1$ ms, to be compared in the same time frame, although varying the aspect ratio affects the number of toroidal turns and the number of radial oscillations, similarly for changes in the value of $\lambda$. 

\par
\begin{figure}
    \centering
    \hfill
    \includegraphics[width=0.32\textwidth]{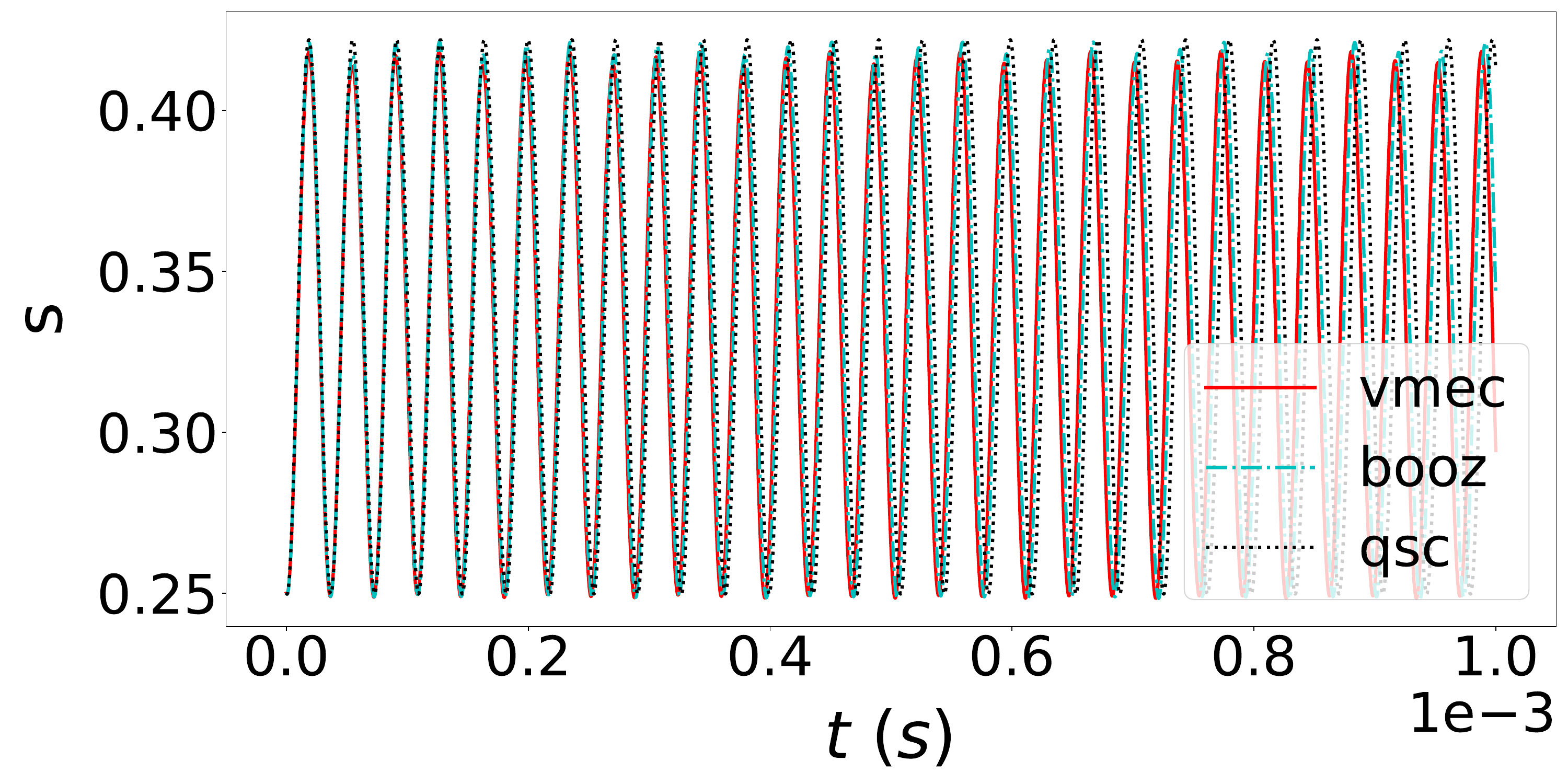}
    \hfill
    \includegraphics[width=0.32\textwidth]{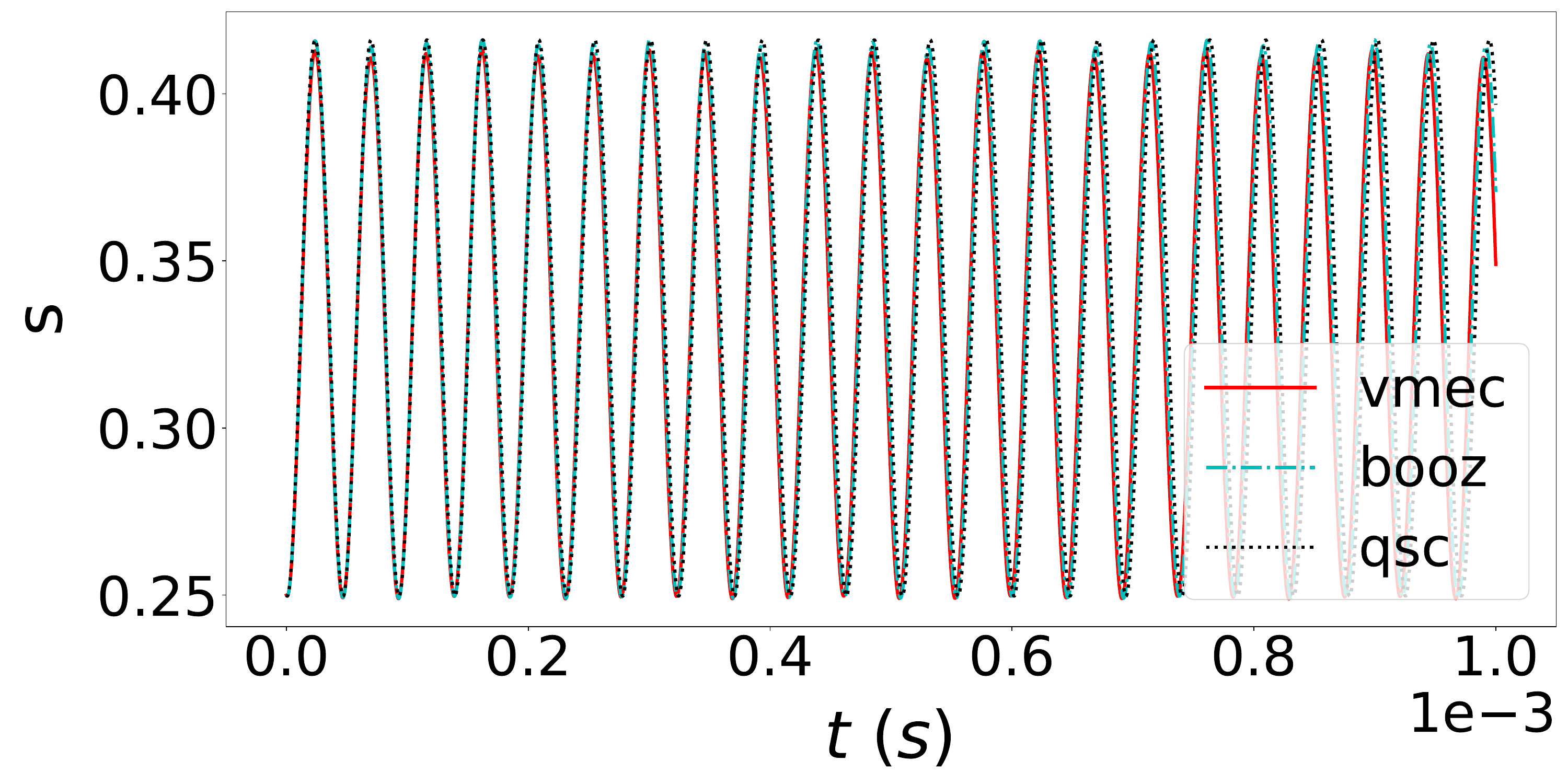}
    \hfill
    \includegraphics[width=0.32\textwidth]{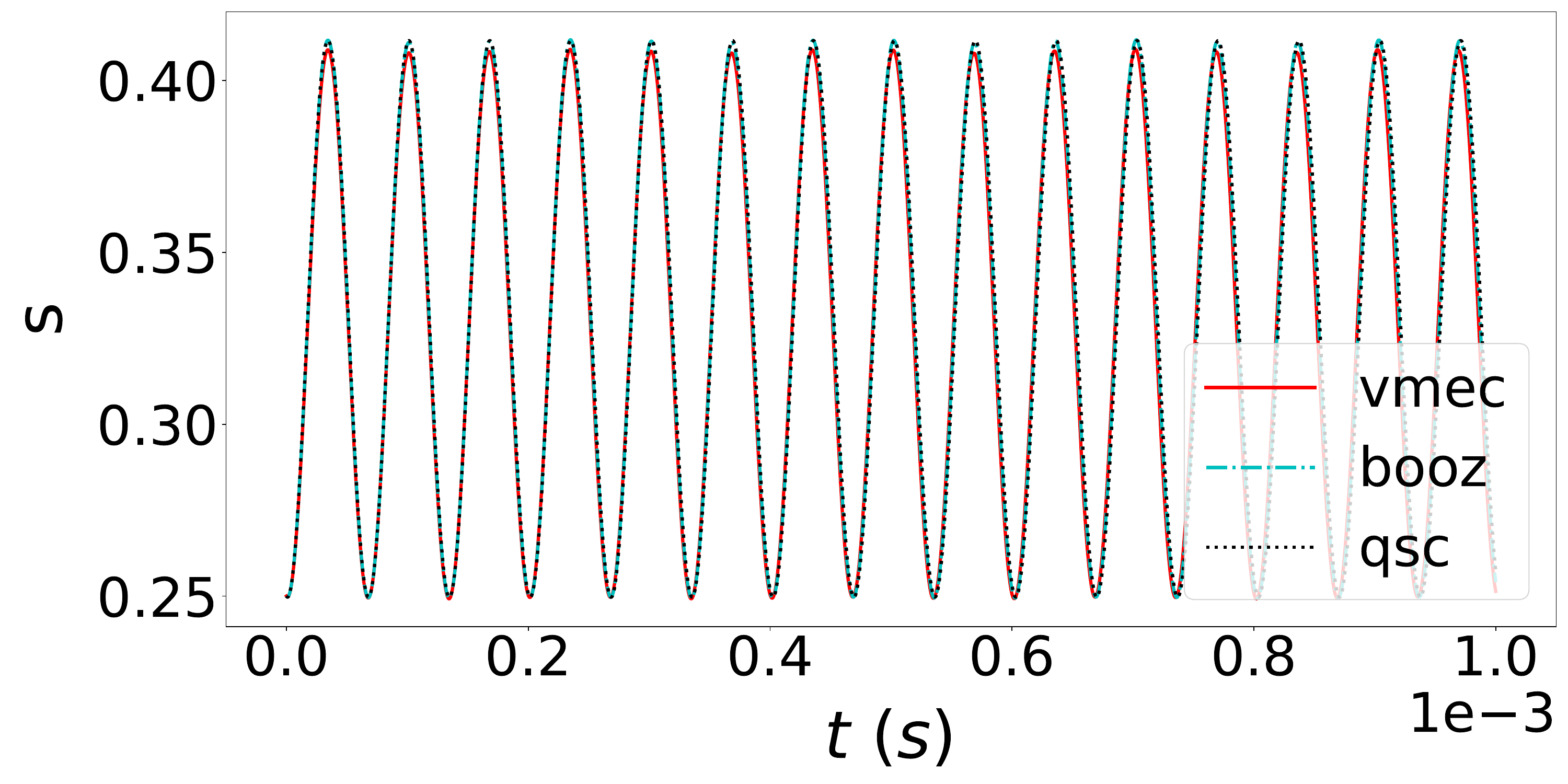}
    \hfill
    \vspace{1em}
    
    \hfill
    \includegraphics[width=0.25\textwidth]{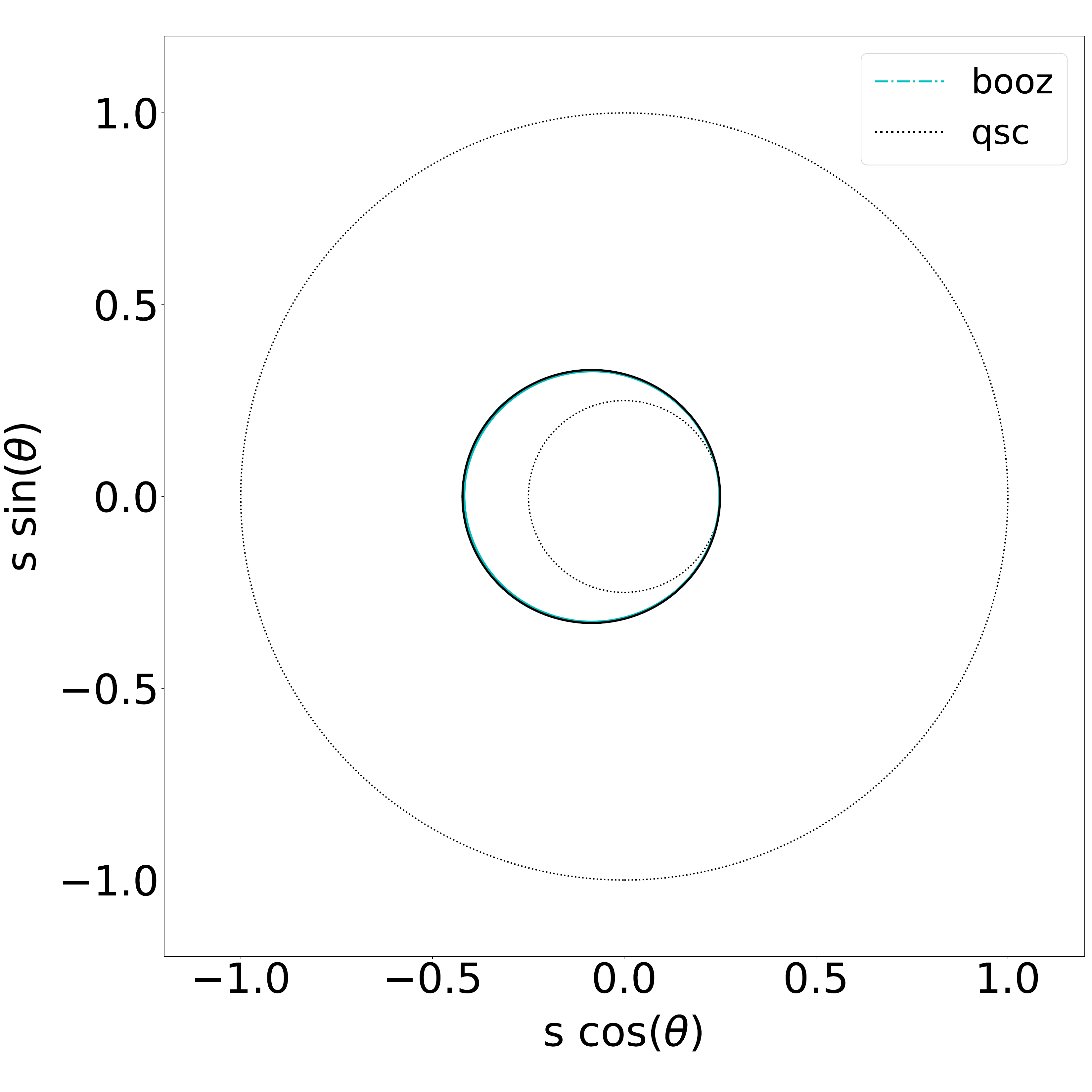}
    \hfill
    \includegraphics[width=0.25\textwidth]{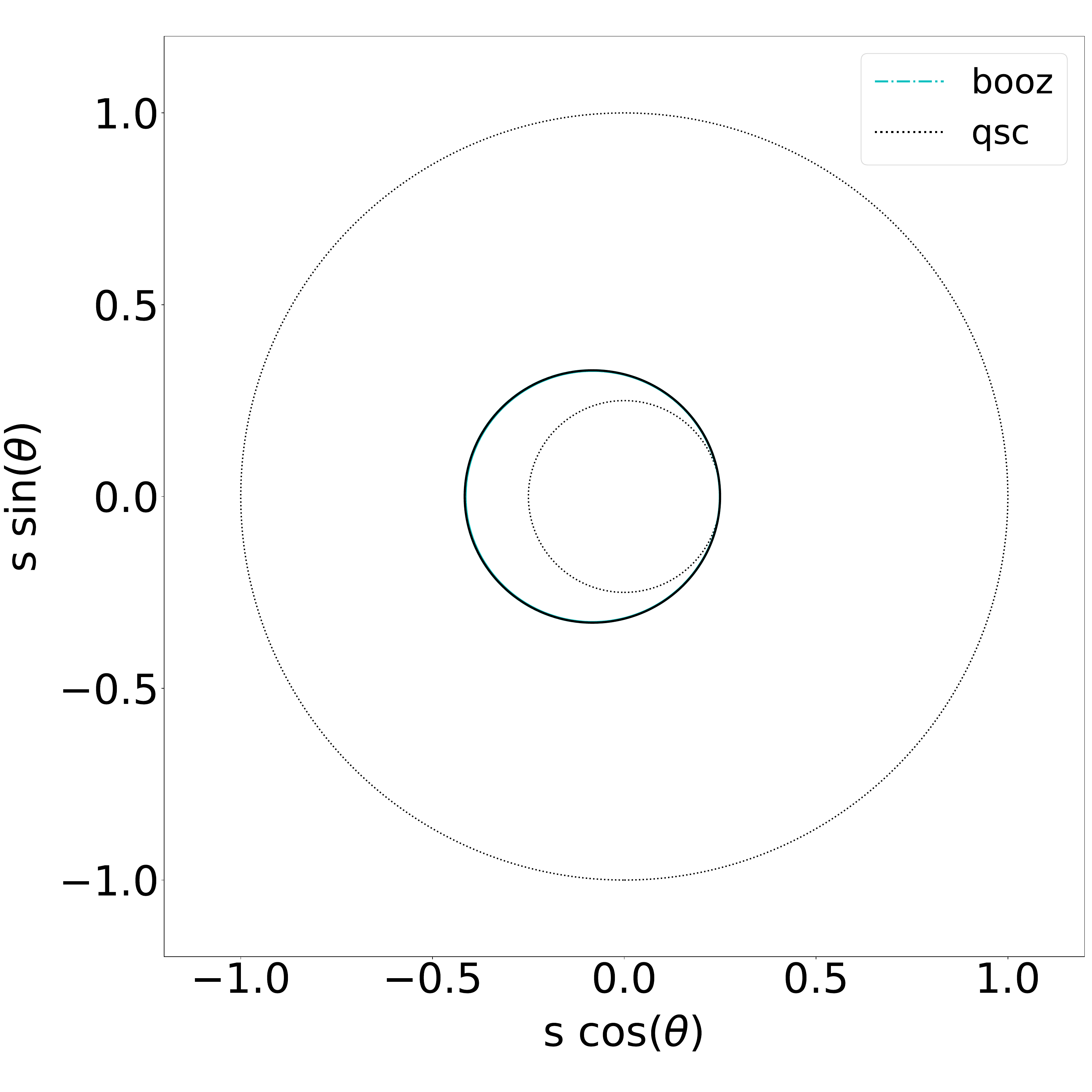}
    \hfill
    \includegraphics[width=0.25\textwidth]{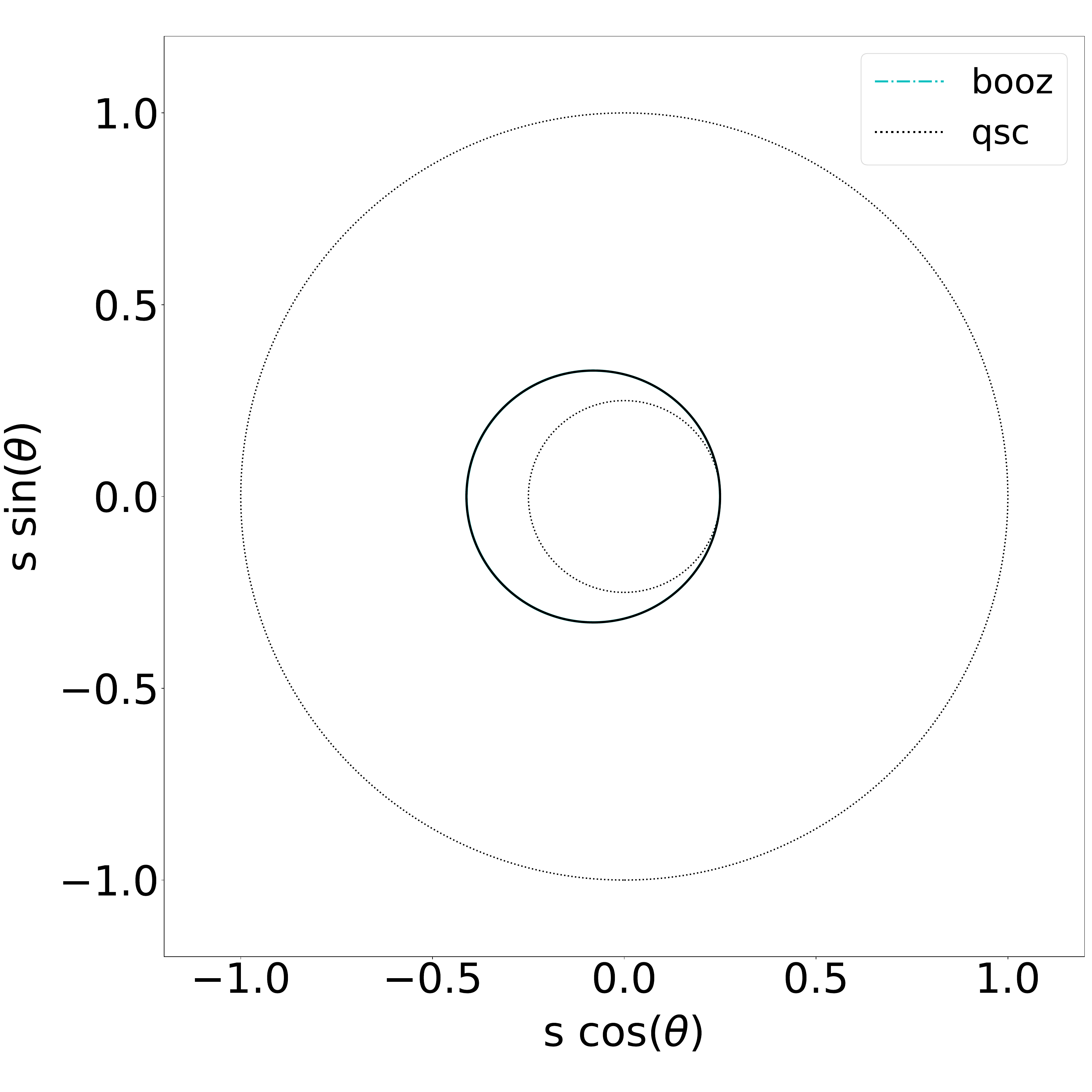}
    \hfill
    \vspace{1em}
    
    \hspace{-1em}
    \includegraphics[width=0.25\textwidth]{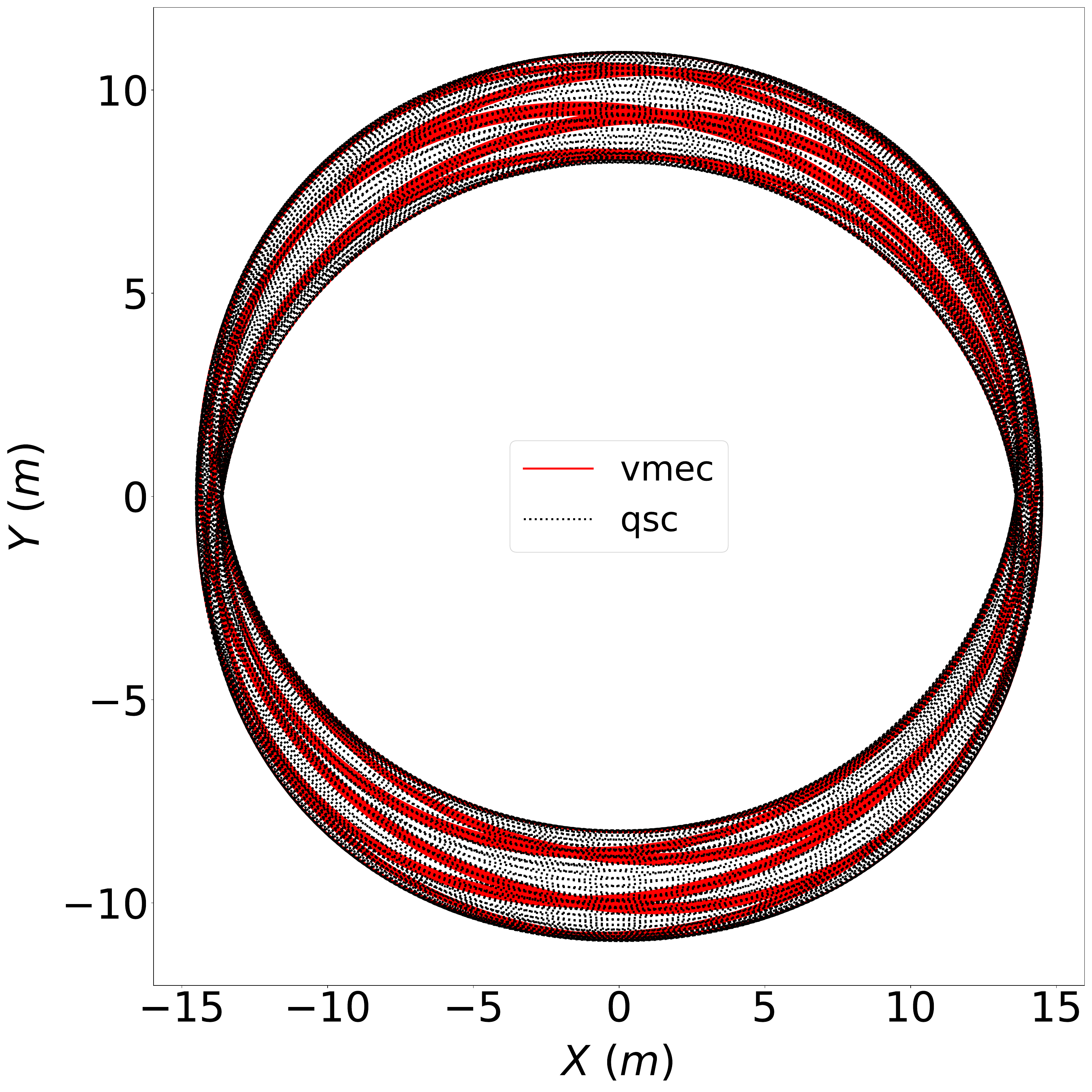}
    \hspace{1.5em}
    \includegraphics[width=0.25\textwidth]{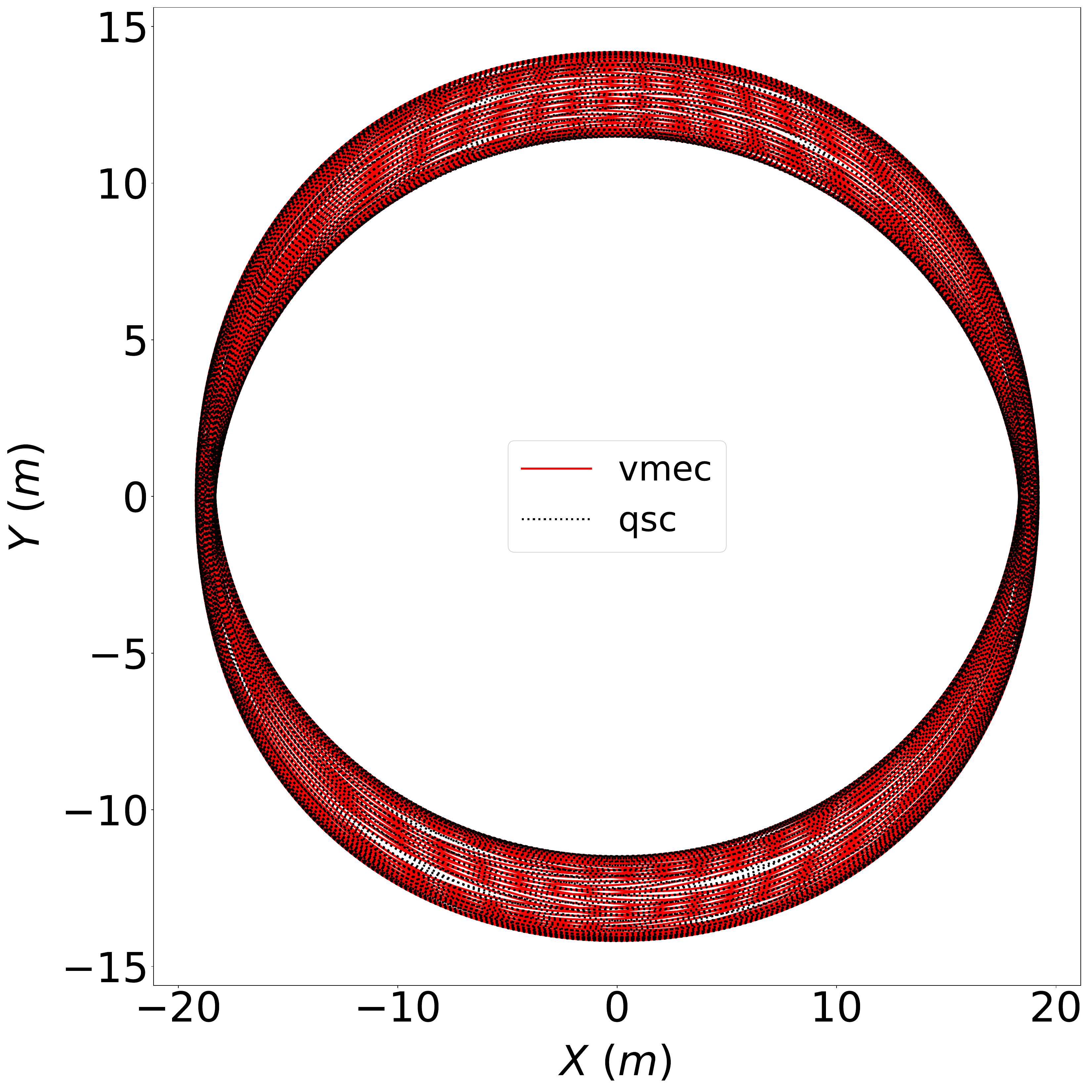}
    \hspace{1.5em}
    \includegraphics[width=0.25\textwidth]{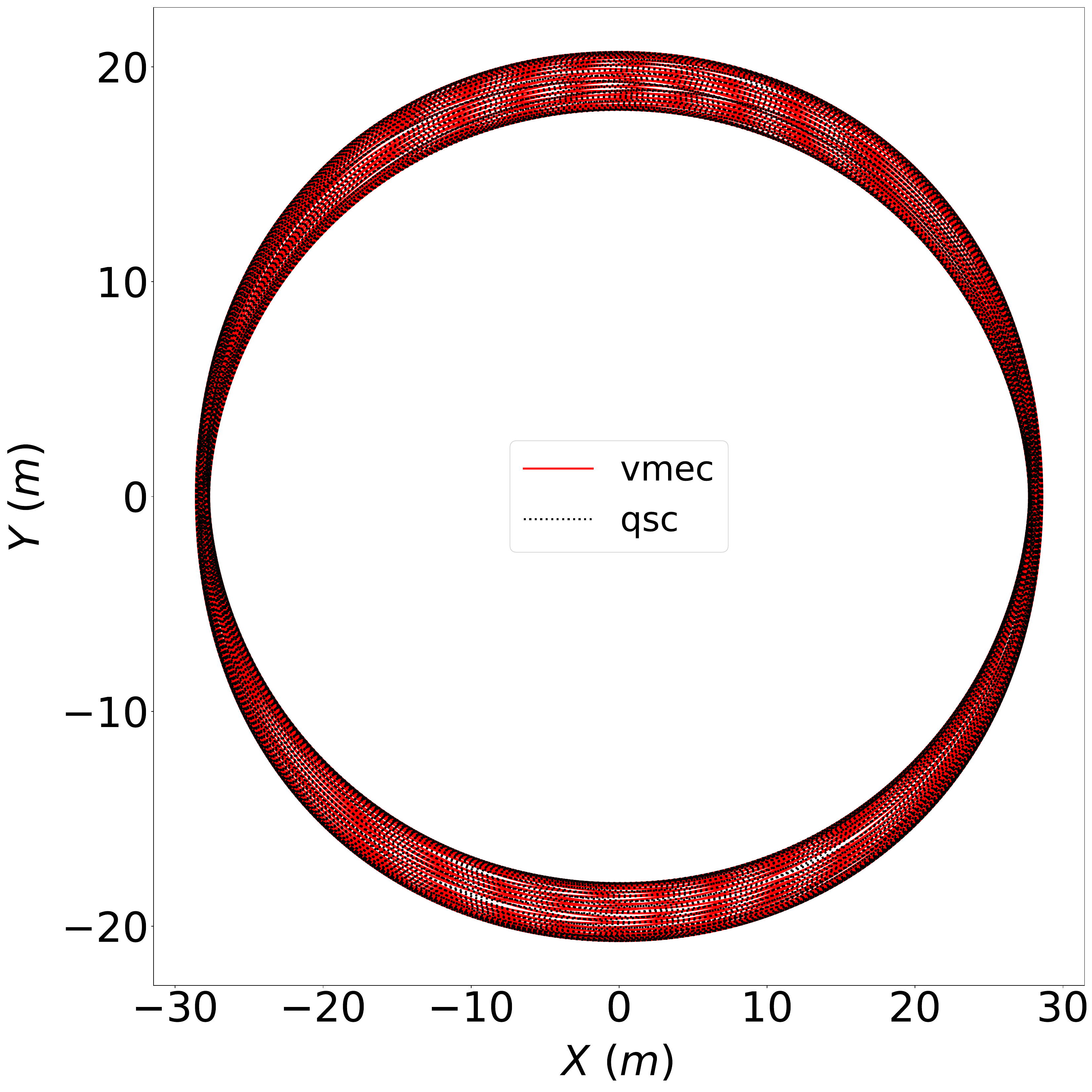}
    \caption{Comparison between alpha particle orbits in \textit{pyQSC} (qsc), \textit{VMEC} (vmec) and \textit{BOOZ\_XFORM} (booz) QA equilibria with initial position $(s, \theta, \varphi) = (0.25, 0.1, 0.1)$ in Boozer coordinates and $\lambda=0.8$. Above: Evolution of the radial position in time. Middle: Normalized poloidal view of the orbit. Below: Top view of the orbit in Cartesian coordinates. Left: $A=6.8$. Center: $A=9.1$. Right: $A=13.6$.}
    \label{fig:QA-passing}
\end{figure}

\begin{figure}
    \centering
    
    \includegraphics[width=0.32\textwidth]{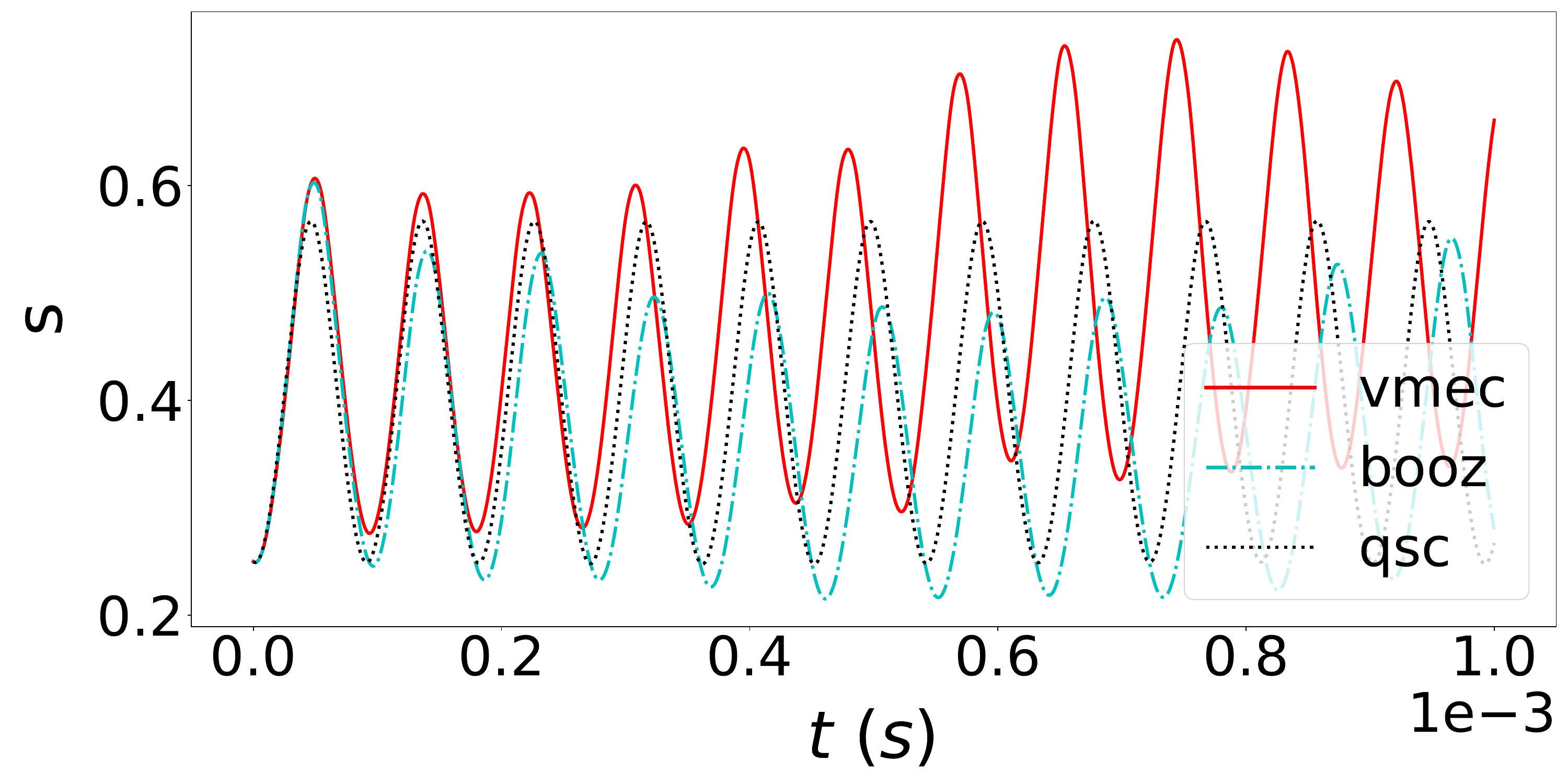}
    \hfill
    \includegraphics[width=0.32\textwidth]{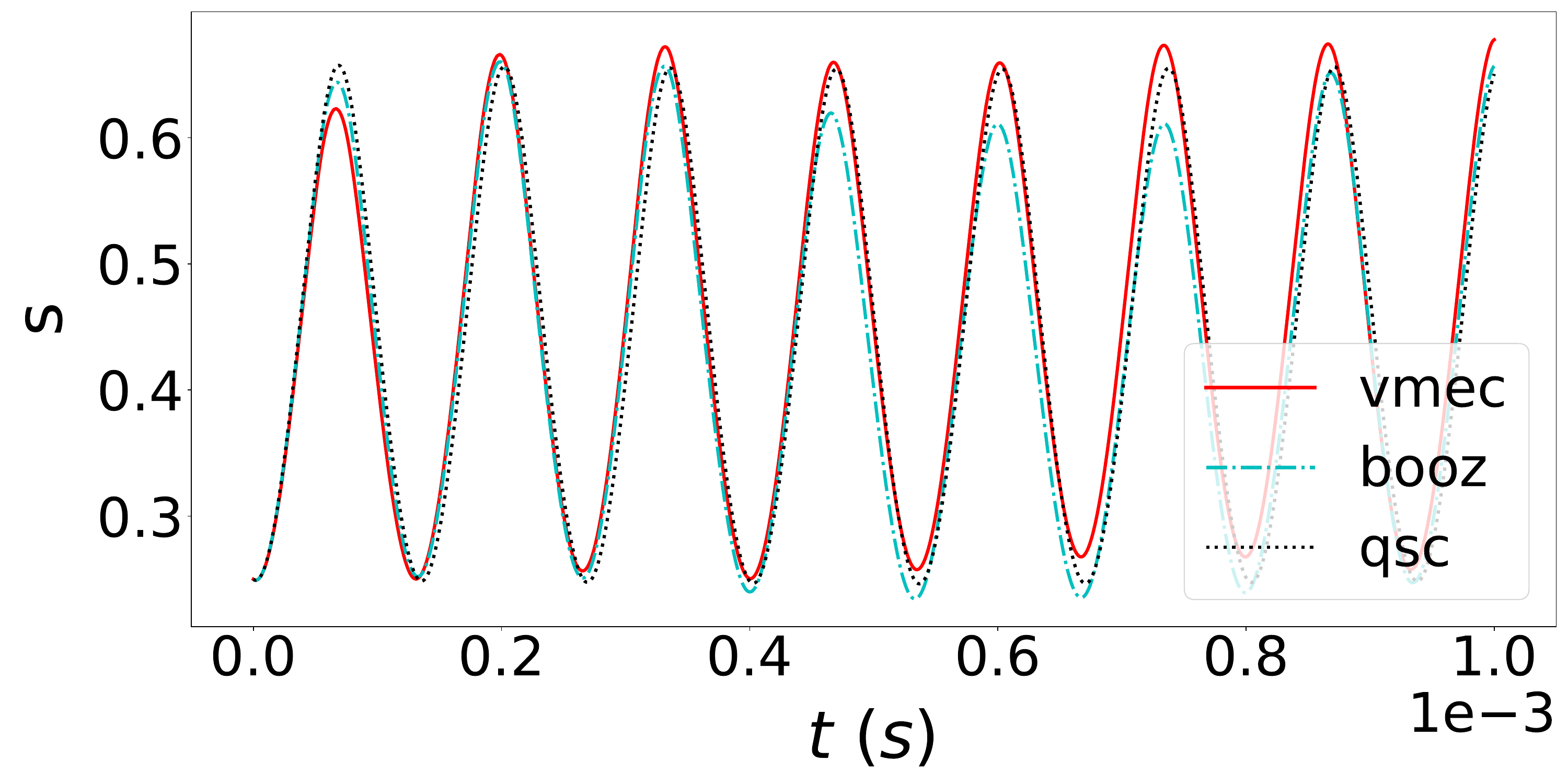}
    \hfill
    \includegraphics[width=0.32\textwidth]{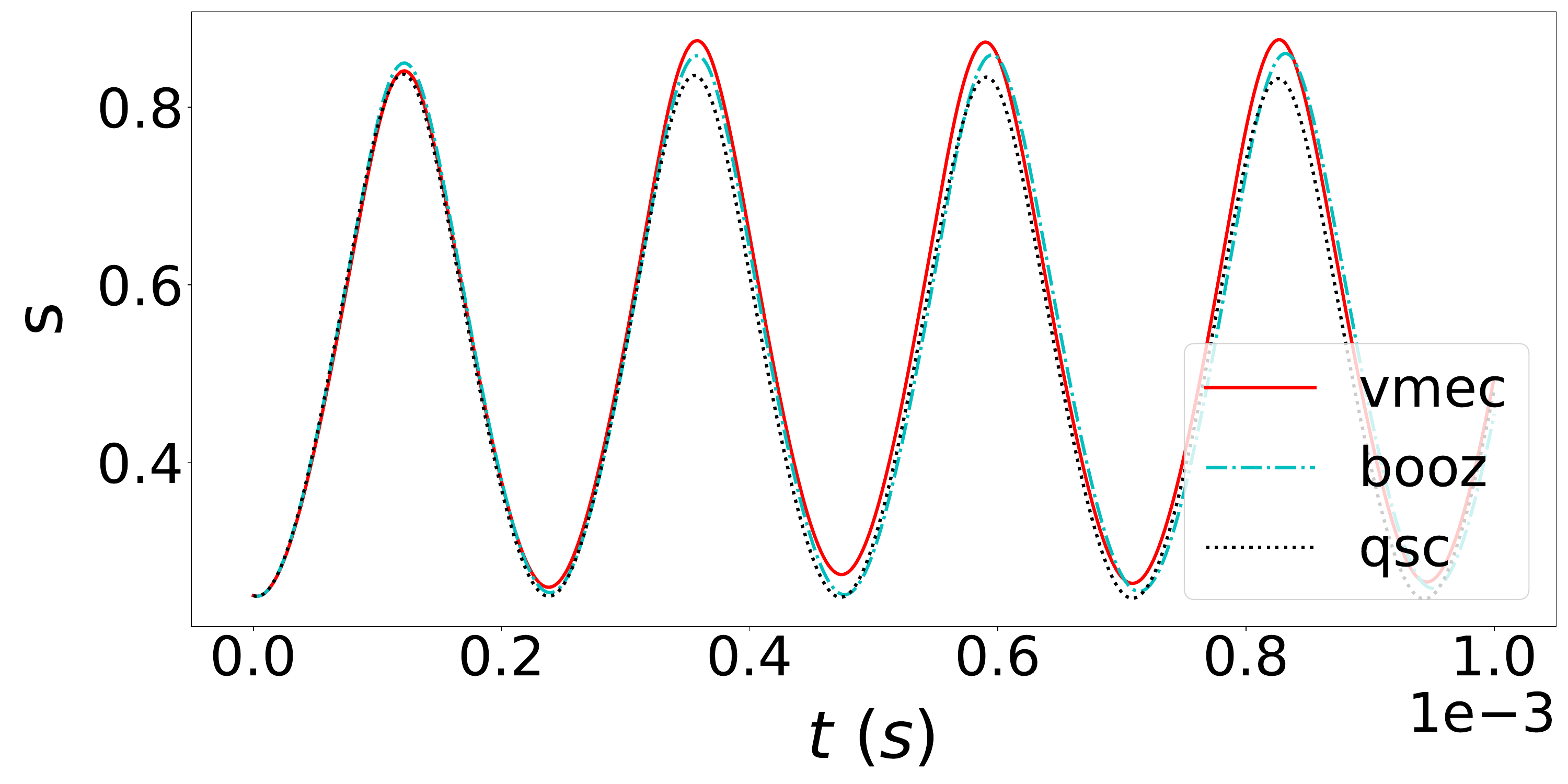}

    \vspace{0.1em}
    
    \hfill
    \includegraphics[width=0.25\textwidth]{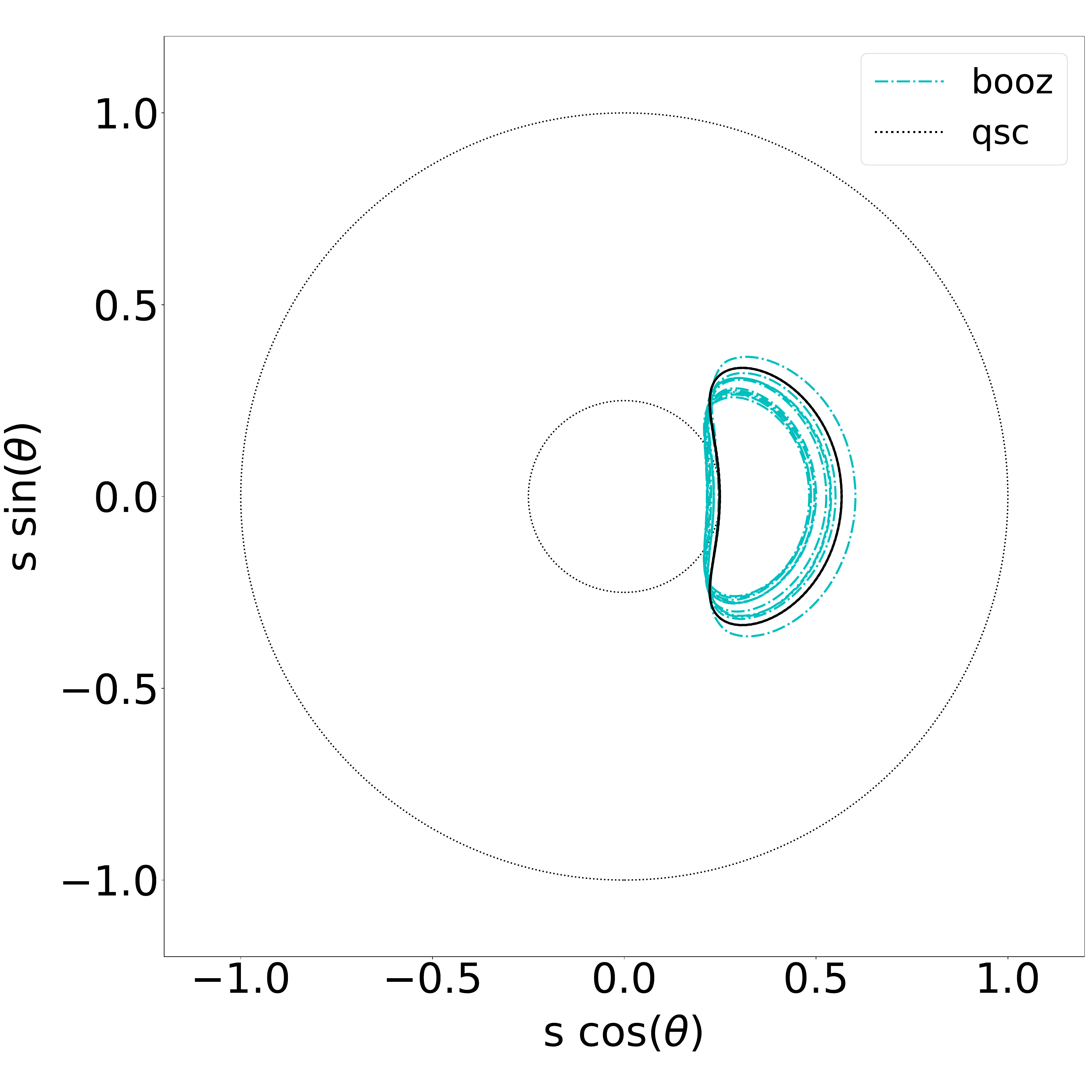}
    \hfill
    \includegraphics[width=0.25\textwidth]{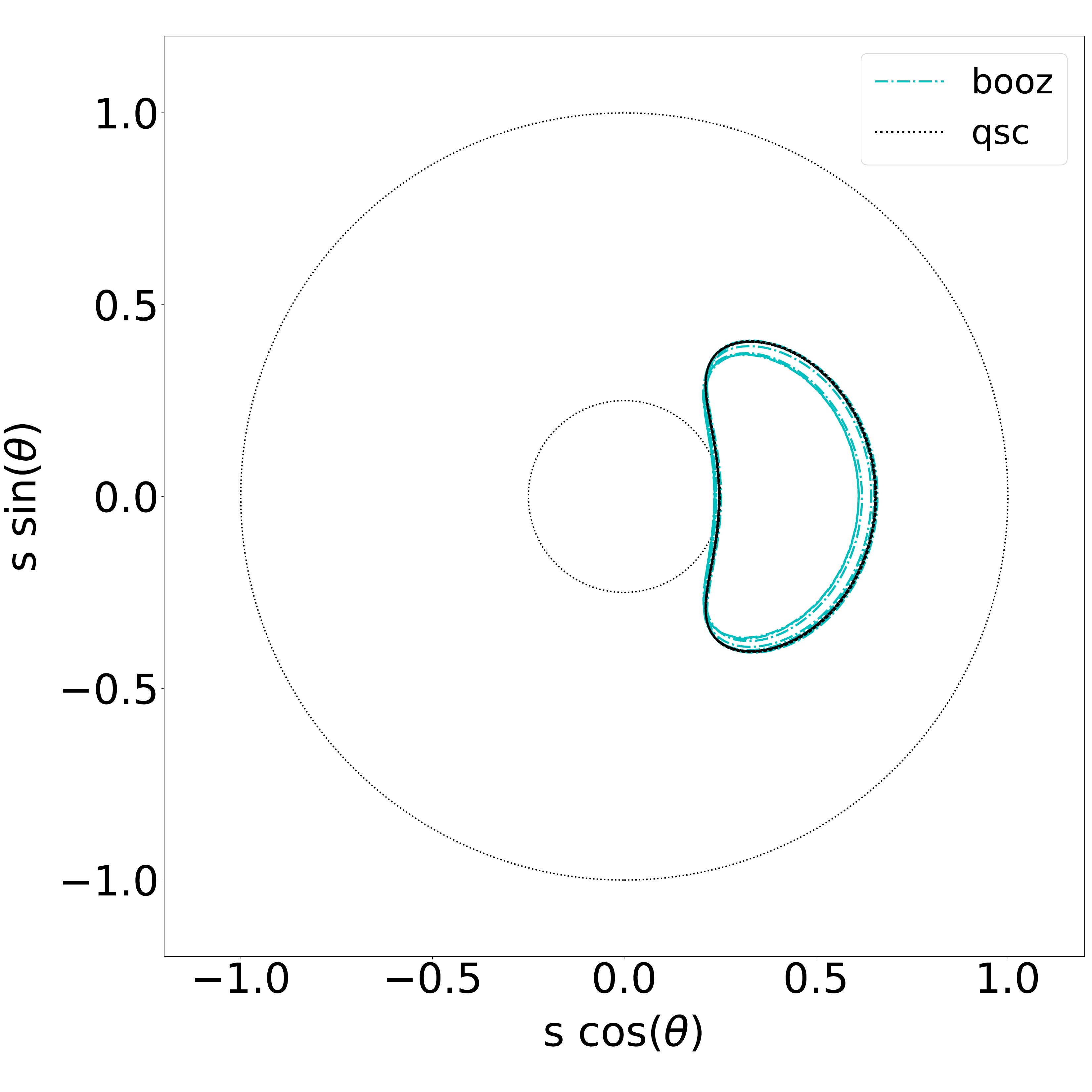}
    \hfill
    \includegraphics[width=0.25\textwidth]{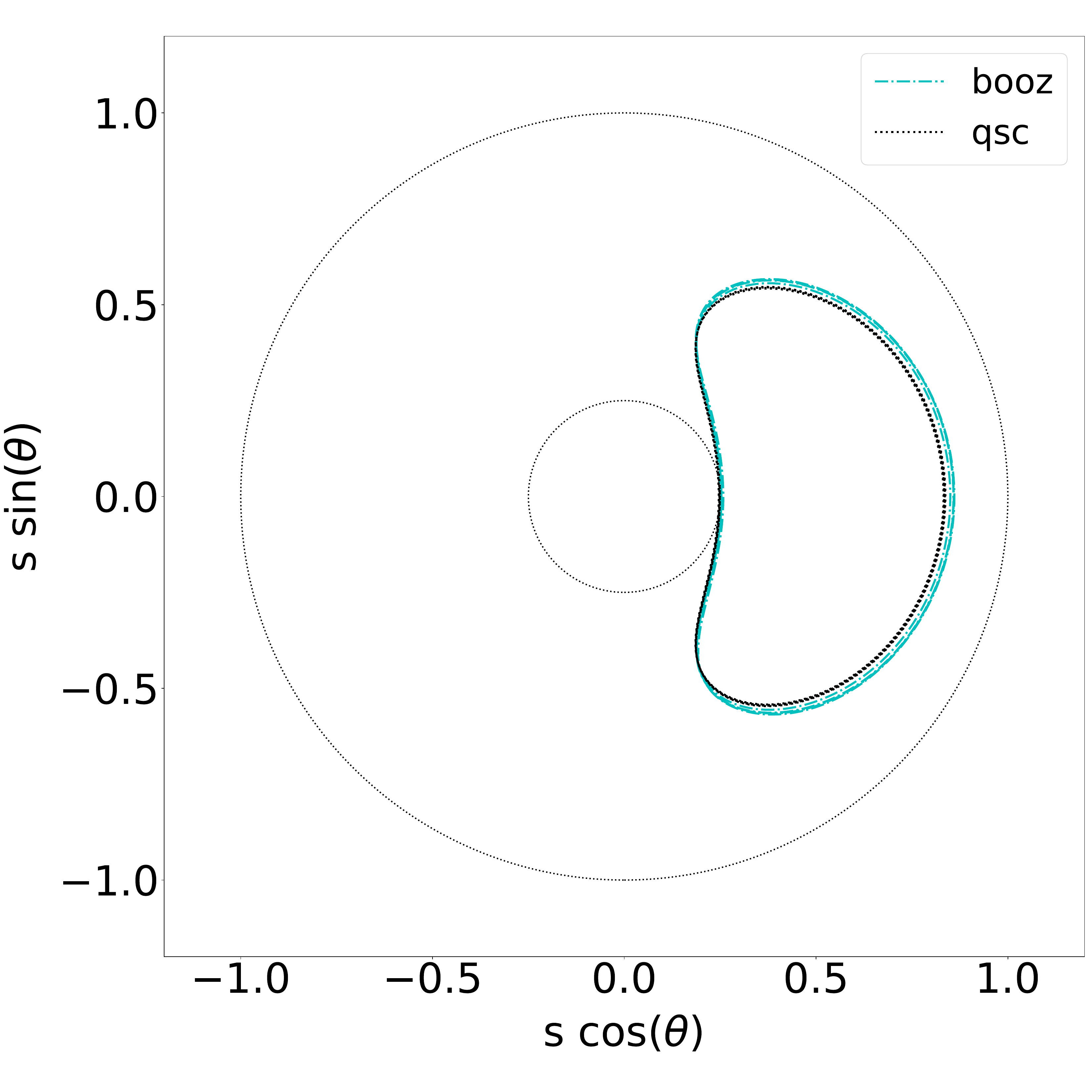}
    \hfill
    \vspace{1em}
    
    \hspace{-1em}
    \includegraphics[width=0.25\textwidth]{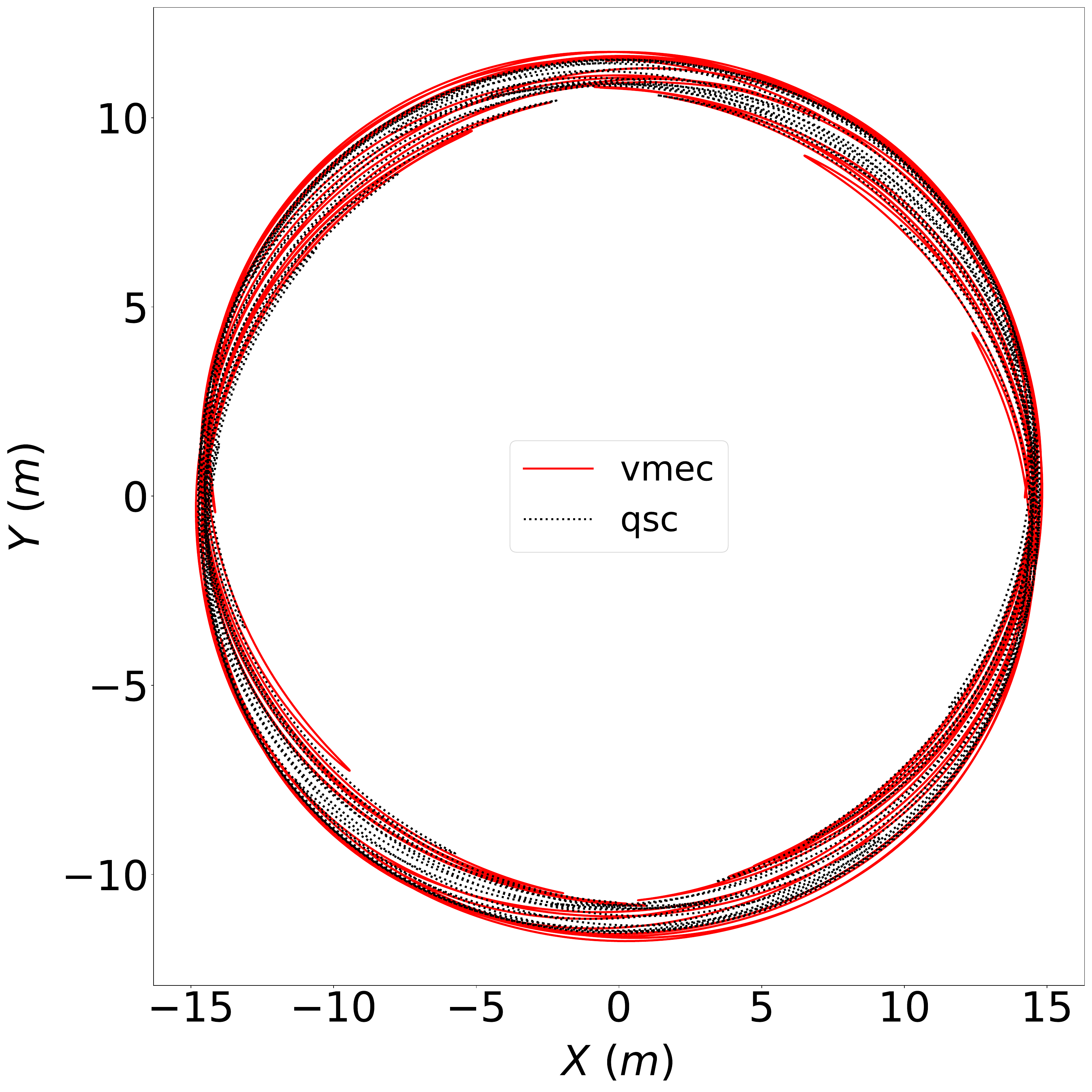}
    \hspace{1.5em}
    \includegraphics[width=0.25\textwidth]{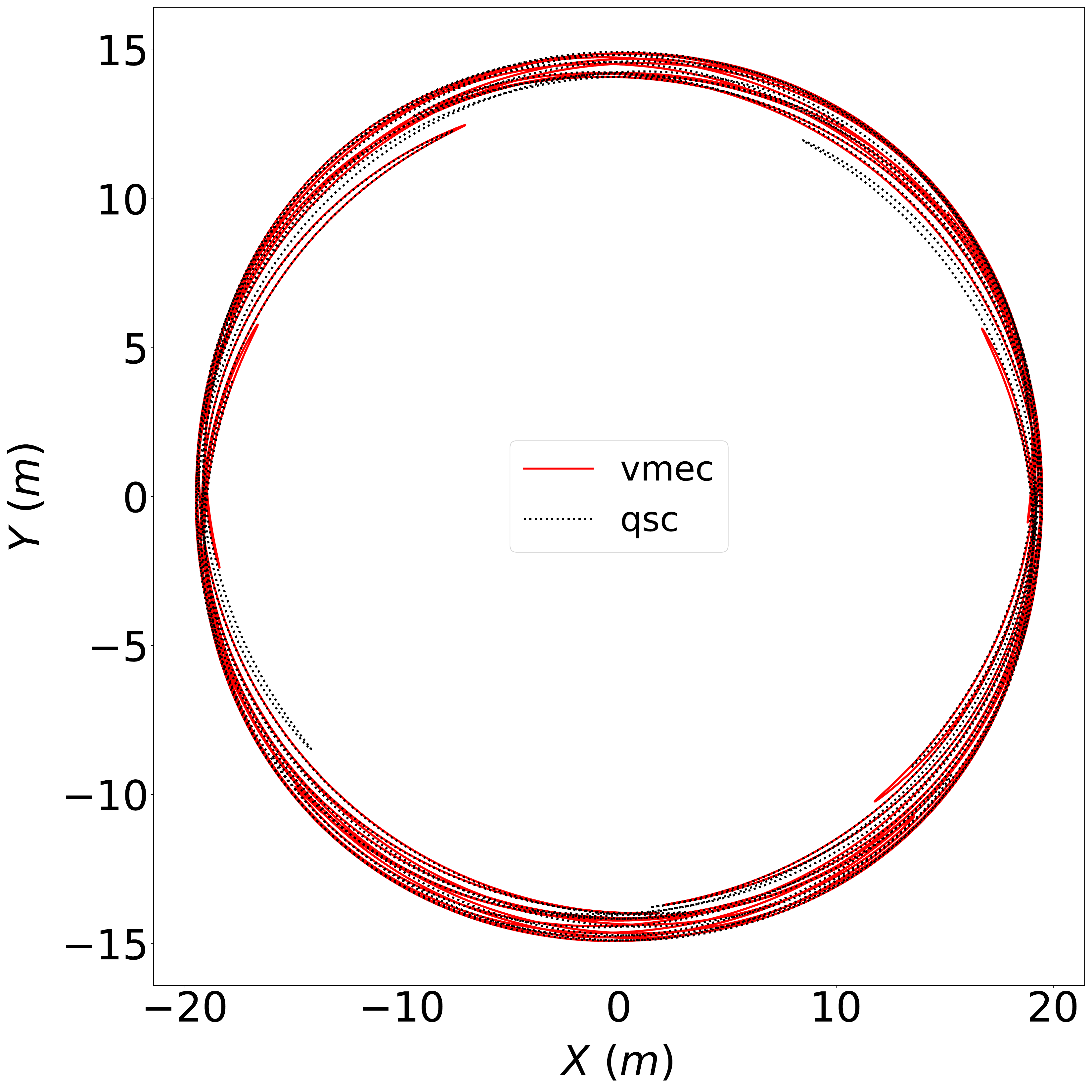}
    \hspace{1.5em}
    \includegraphics[width=0.25\textwidth]{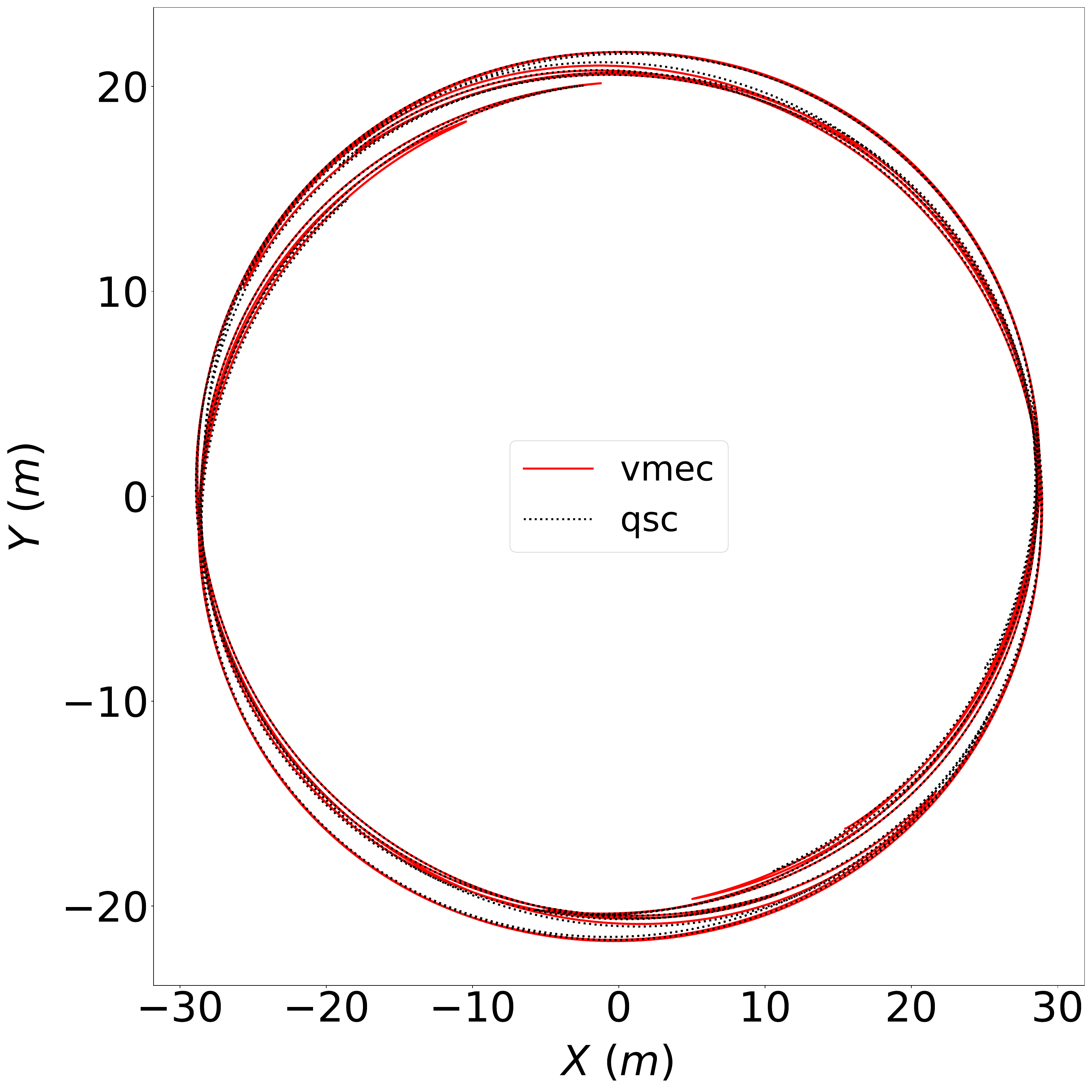}
    \caption{Comparison between alpha particle orbits in \textit{pyQSC} (qsc), \textit{VMEC} (vmec) and \textit{BOOZ\_XFORM} (booz) QA equilibria with initial position $(s, \theta, \varphi) = (0.25, 0.1, 0.1)$ in Boozer coordinates and $\lambda=0.99$. Above: Evolution of the radial position in time. Middle: Normalized poloidal view of the orbit. Below: Top view of the orbit in Cartesian coordinates.  Left: $A=6.8$. Center: $A=9.1$. Right: $A=13.6$.}
    \label{fig:QA-trapped}
\end{figure}

 We show in \cref{fig:QA-passing,fig:QA-trapped} an example of a passing particle with $\lambda=0.9$ and a trapped particle with $\lambda=0.99$, correspondingly. To observe the banana orbits characteristic of trapped particles in tokamaks, we transformed the \textit{VMEC} equilibria to the \textit{BOOZ\_XFORM} \citep{Sanchez2000} format and calculated the particle trajectory for this field, where the angular coordinates are Boozer coordinates similar to \textit{pyQSC}. The plots in Boozer coordinates on the second line of each figure show the comparison between the orbits in these two fields, which exhibit circular curves for passing orbits and bean or banana-shaped curves for trapped orbits. We also show a view from above of the orbits in the \textit{pyQSC} field (qsc) and in the \textit{VMEC} equilibrium (vmec). In this case, the same inputs for the particles result in the same type of orbits, but this does not have to be the case as in different aspect ratios we have different values of initial $B$ and therefore, for some values of $\lambda$, some orbits are trapped and some are not for different aspect ratios.

\par For higher $\lambda$, orbits have some deviations from one another for all aspect ratios, but the most striking differences happen when we have trapped particles, as can be seen in \cref{fig:QA-trapped}, where the average radial positions deviate significantly for the lower aspect ratio which implies larger differences in all the points of view of the orbit. Although no loss of particles is shown in the figures, it is easy to imagine that, for wider banana orbits, we would lose a particle for all equilibria, at least for the higher aspect ratios, making the prompt loss fraction similar between these configurations, despite small differences in the magnetic field. 

\par The fact that the orbits in the \textit{BOOZ\_XFORM} (booz) and \textit{VMEC} (vmec) fields are quite different from each other in \cref{fig:QA-trapped} shows that slight changes on the equilibrium, which can arise from interpolation errors or resolution differences, can cause large differences in the radial excursion of a given particle with the same initial conditions when we are further away from QS or omnigenity. Although increasing the resolution could improve the convergence between fields, the required increase in toroidal and poloidal modes to achieve convergence would substantially affect the computational costs of the procedure. Furthermore, \textit{VMEC} is known to have decreased accuracy near the magnetic axis, so some level of disagreement is expected there. However, even though the orbits are not aligned with each other, the impact of this effect in the confinement of these particles is not obvious as the misalignment in the form of time delays or small average deviations may not have a big impact on the fraction of lost particles. Such small deviations will be assessed in the next section. 


\subsection{Quasi-Helical Symmetry}

\par
\begin{figure}
    \centering
    \hfill
    \includegraphics[width=0.32\textwidth]{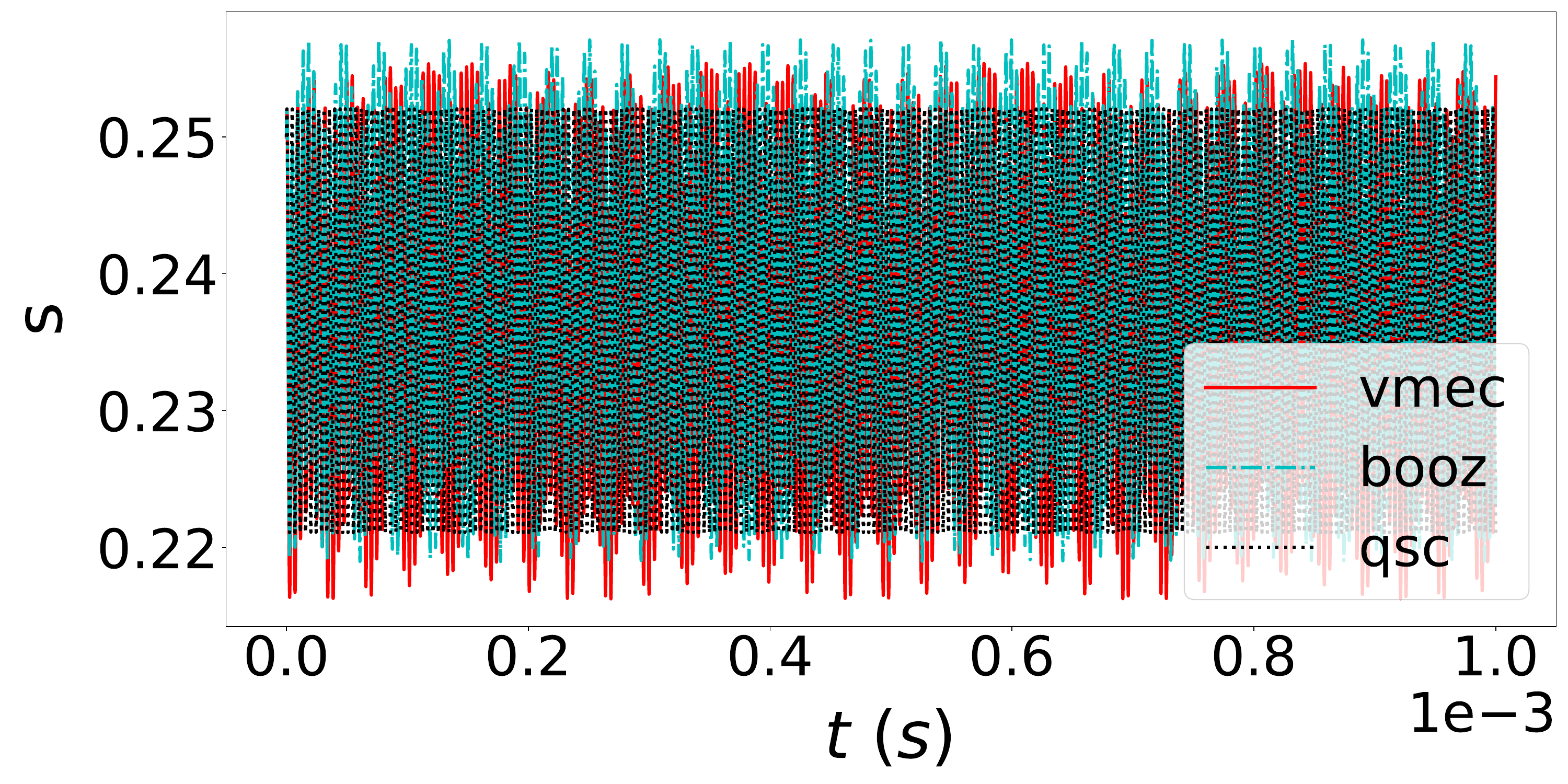}
    \hfill
    \includegraphics[width=0.32\textwidth]{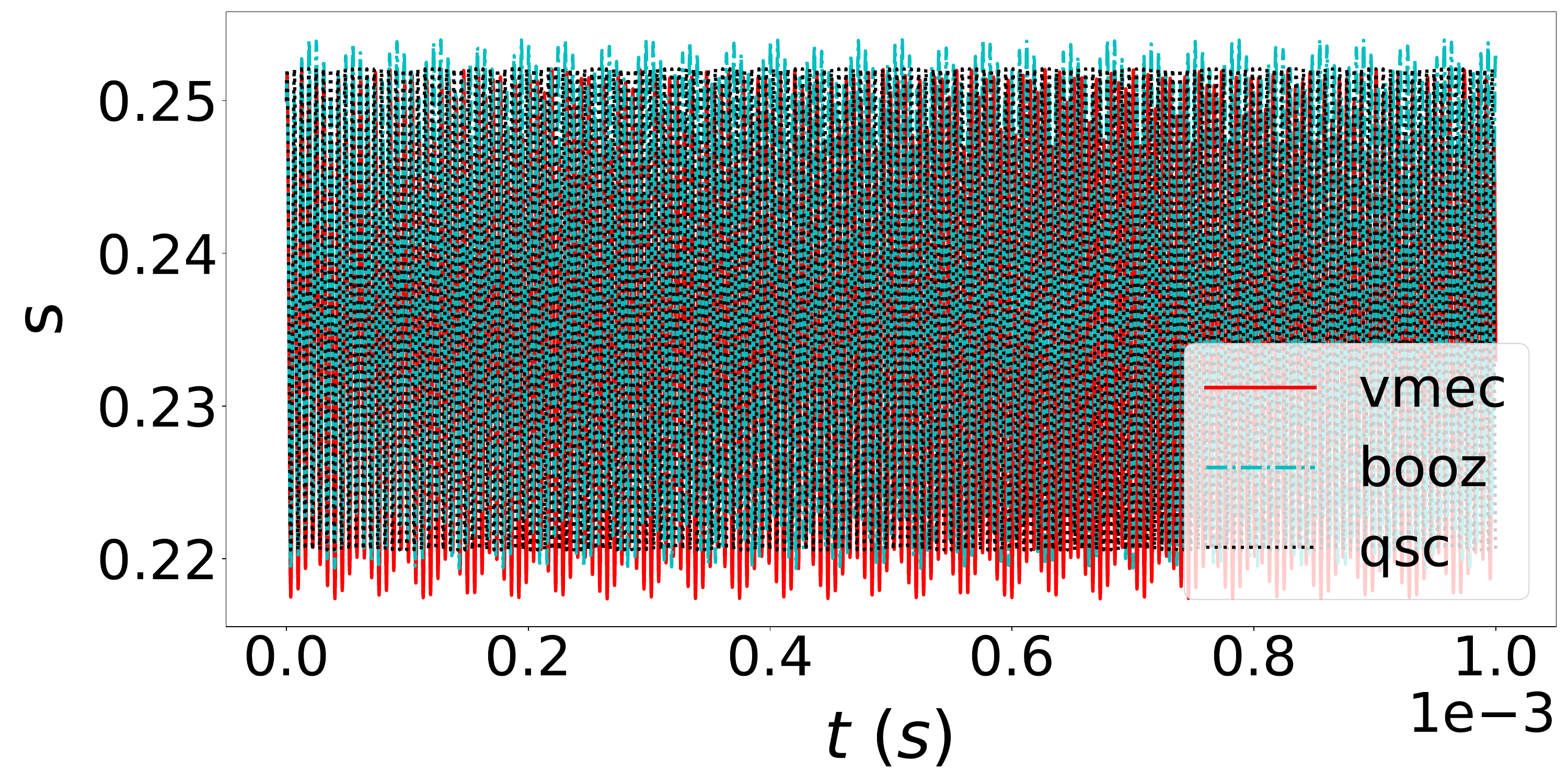}
    \hfill
    \includegraphics[width=0.32\textwidth]{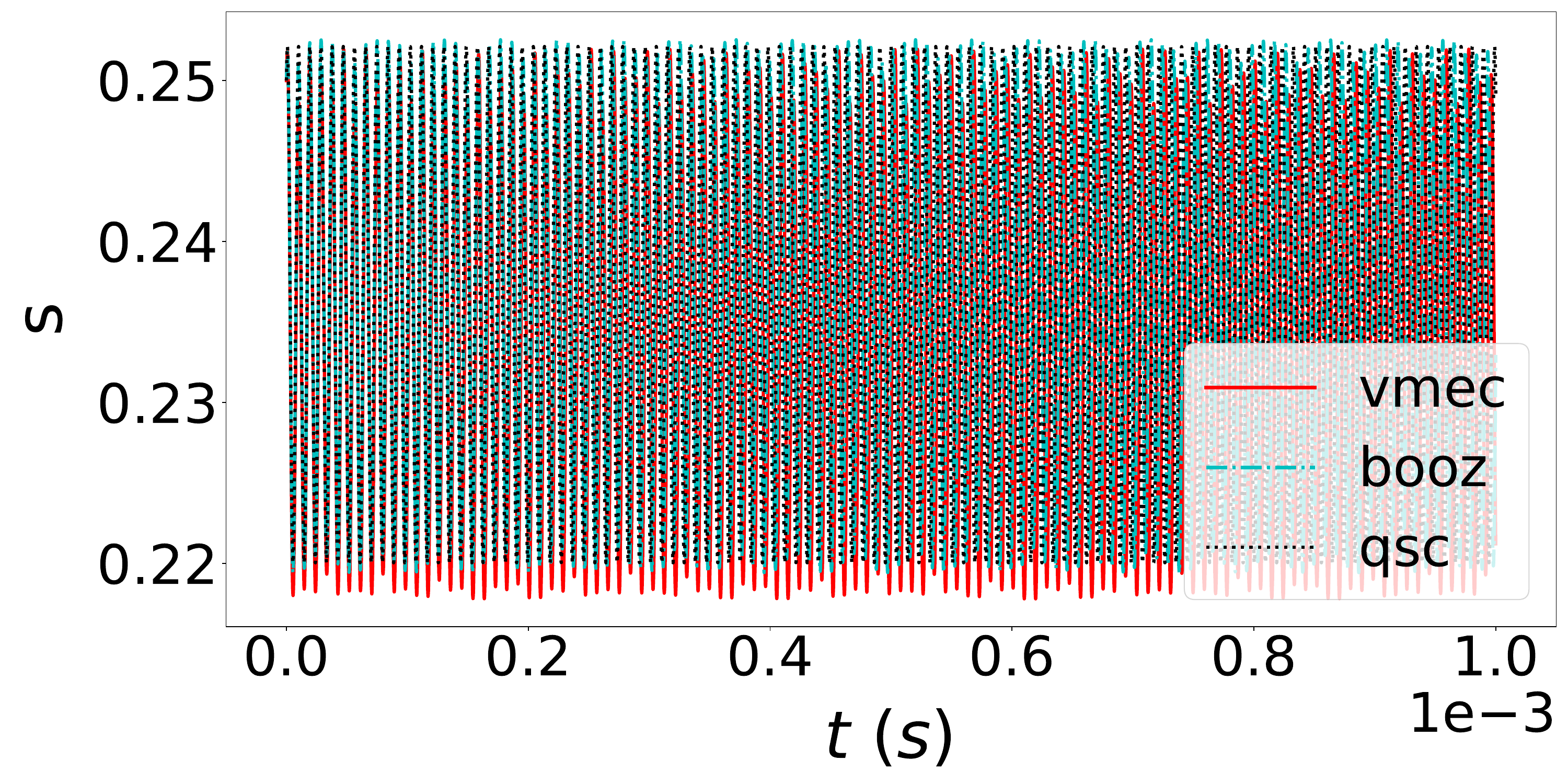}
    \hfill
    \vspace{0.1em}
    
    \hfill
    \includegraphics[width=0.25\textwidth]{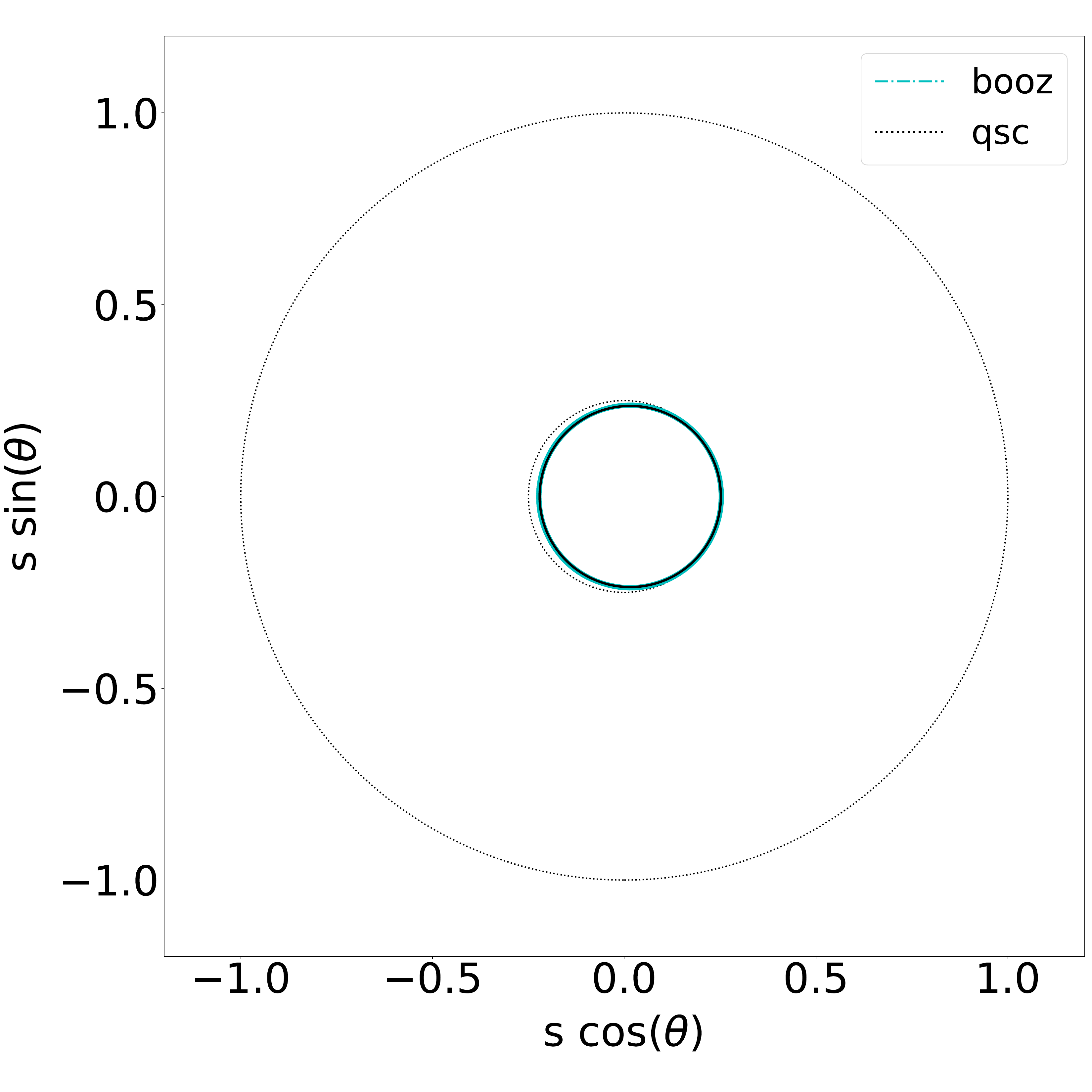}
    \hfill
    \includegraphics[width=0.25\textwidth]{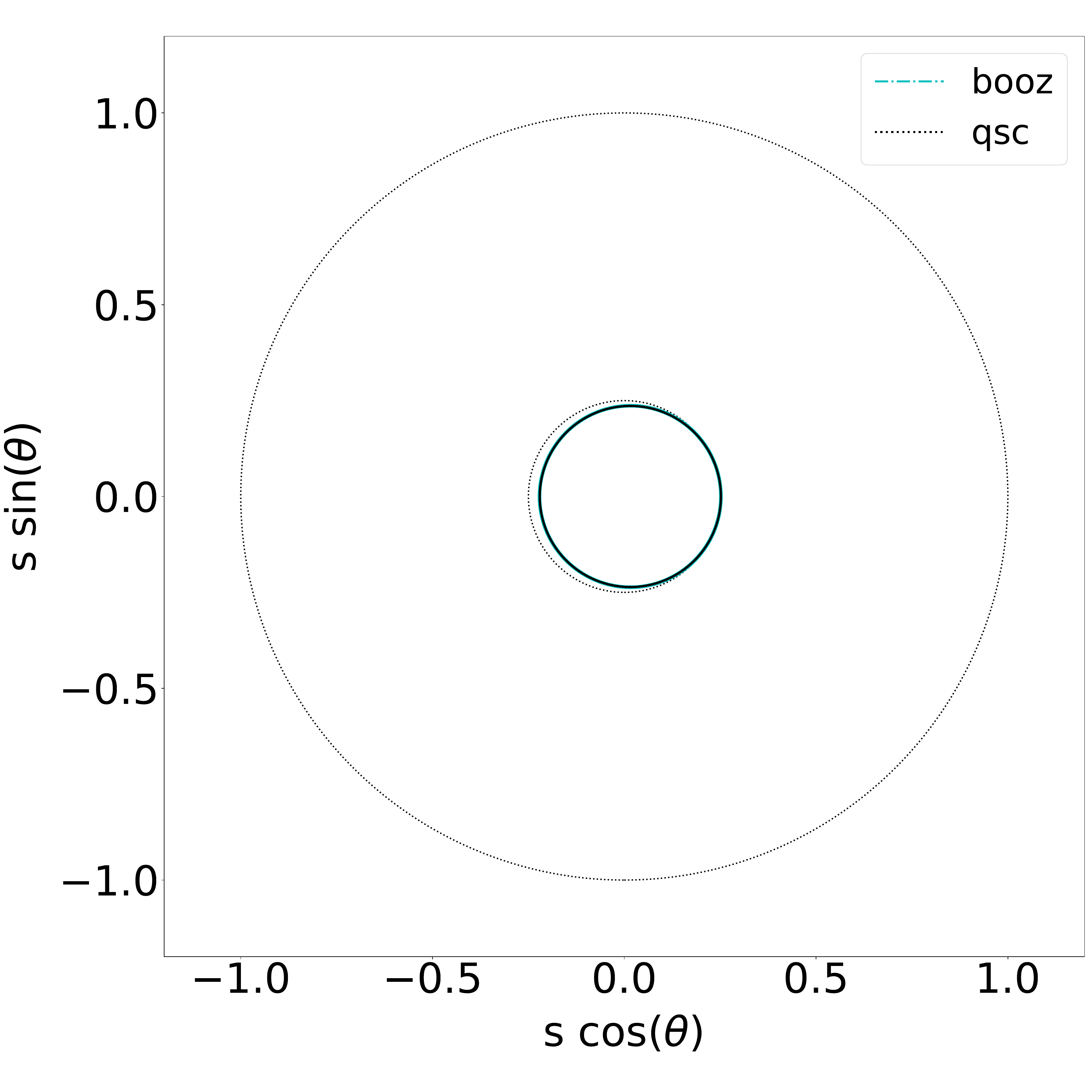}
    \hfill
    \includegraphics[width=0.25\textwidth]{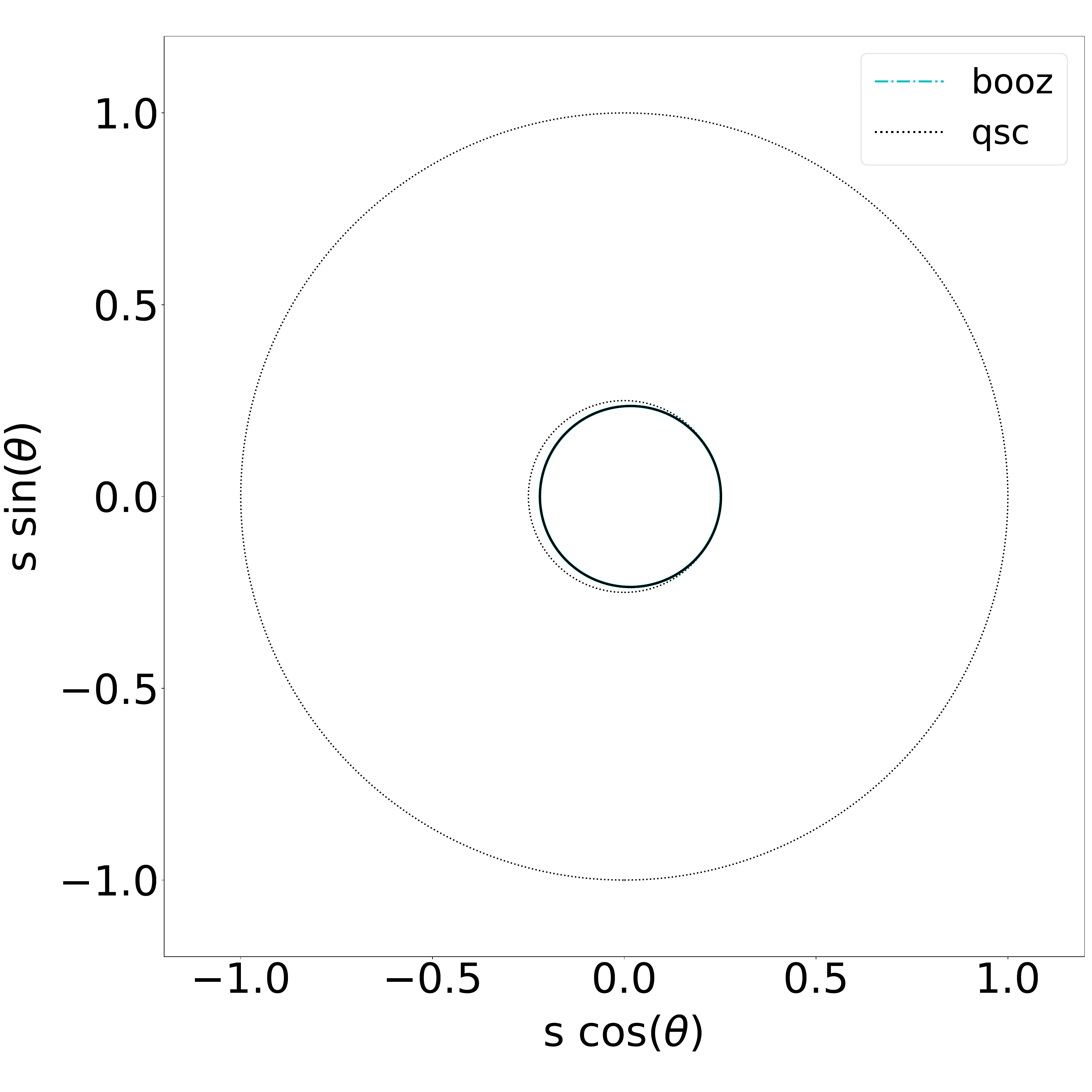}
    \hfill
    \vspace{1em}
    
    \hspace{-1em}
    \includegraphics[width=0.25\textwidth]{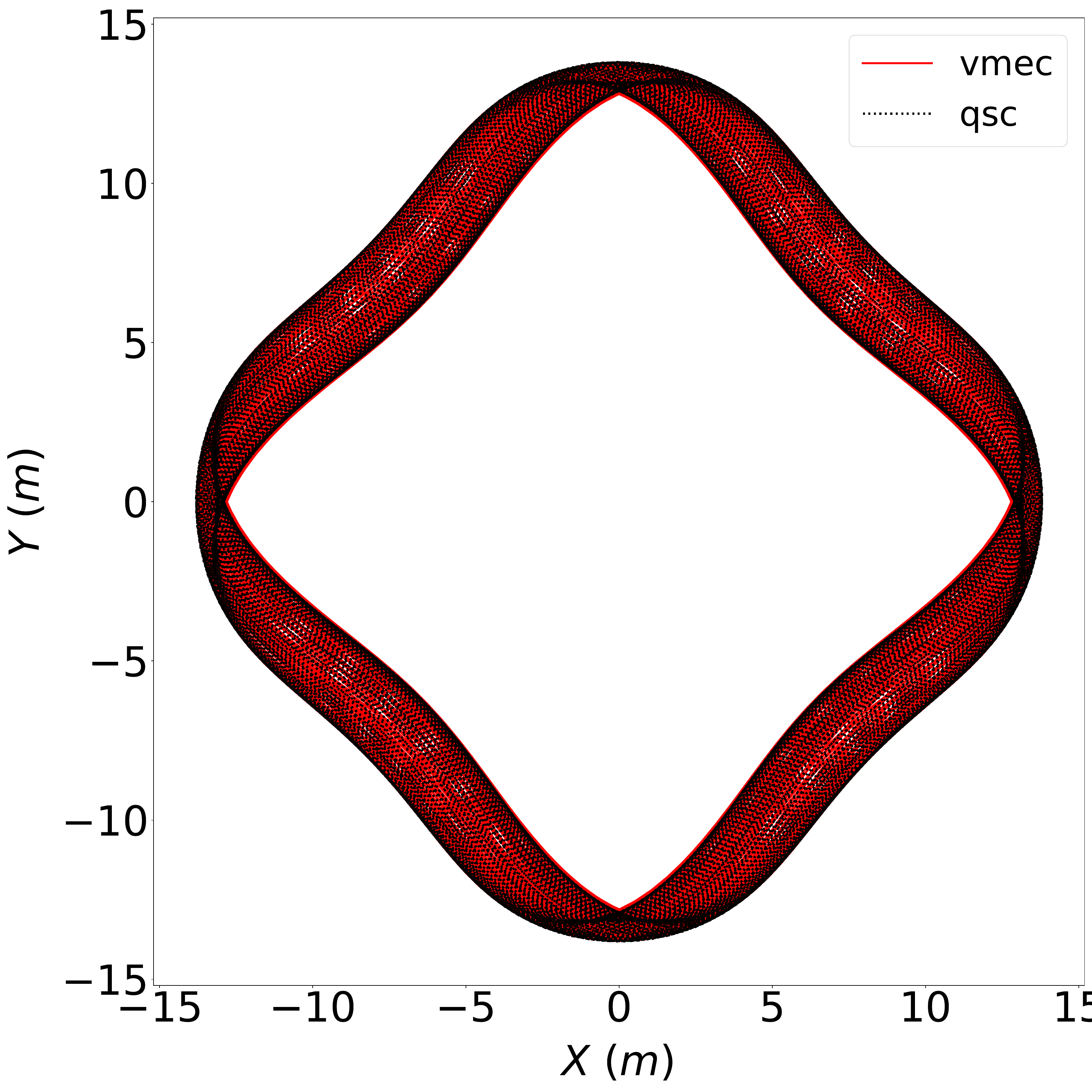}
    \hspace{1.5em}
    \includegraphics[width=0.25\textwidth]{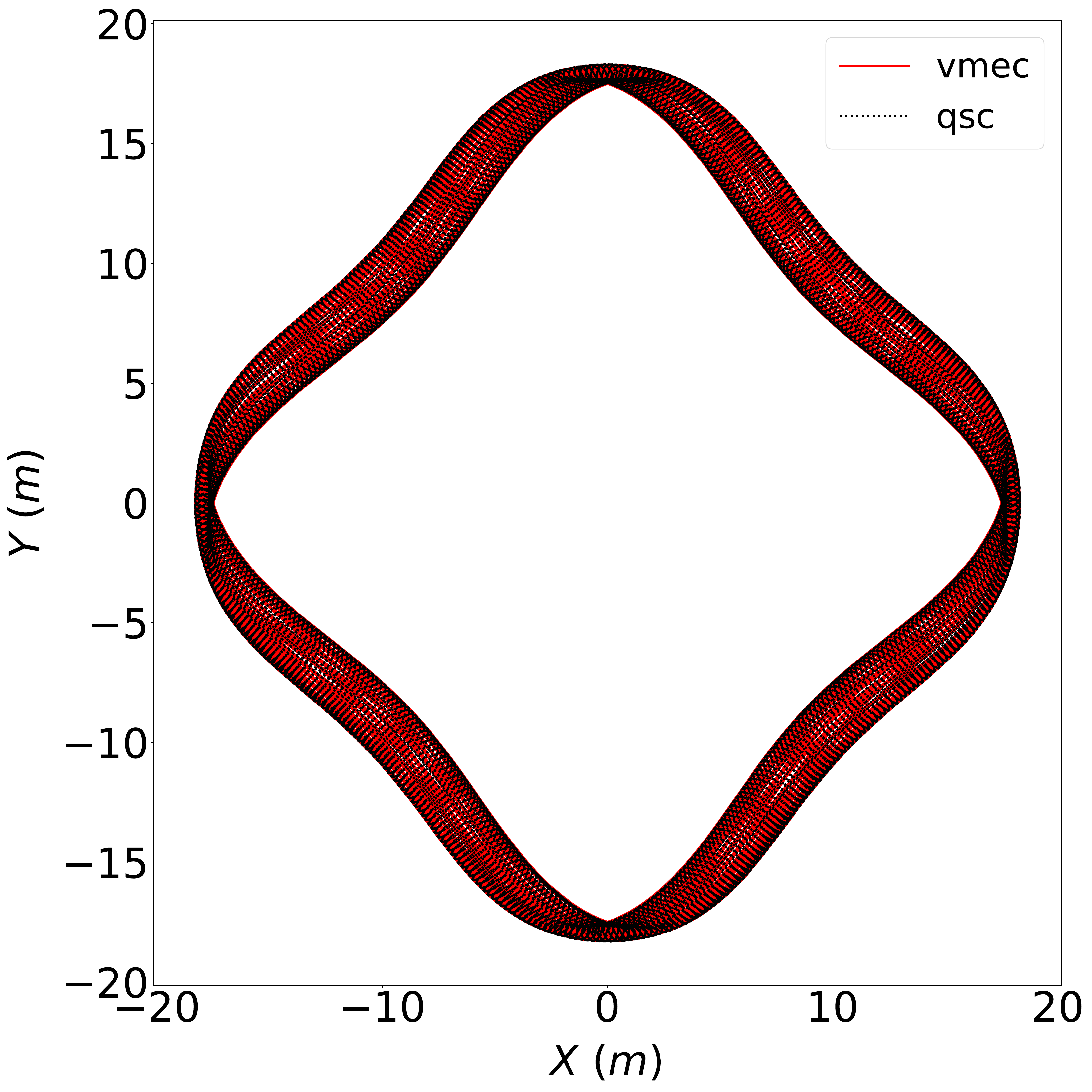}
    \hspace{1.5em}
    \includegraphics[width=0.25\textwidth]{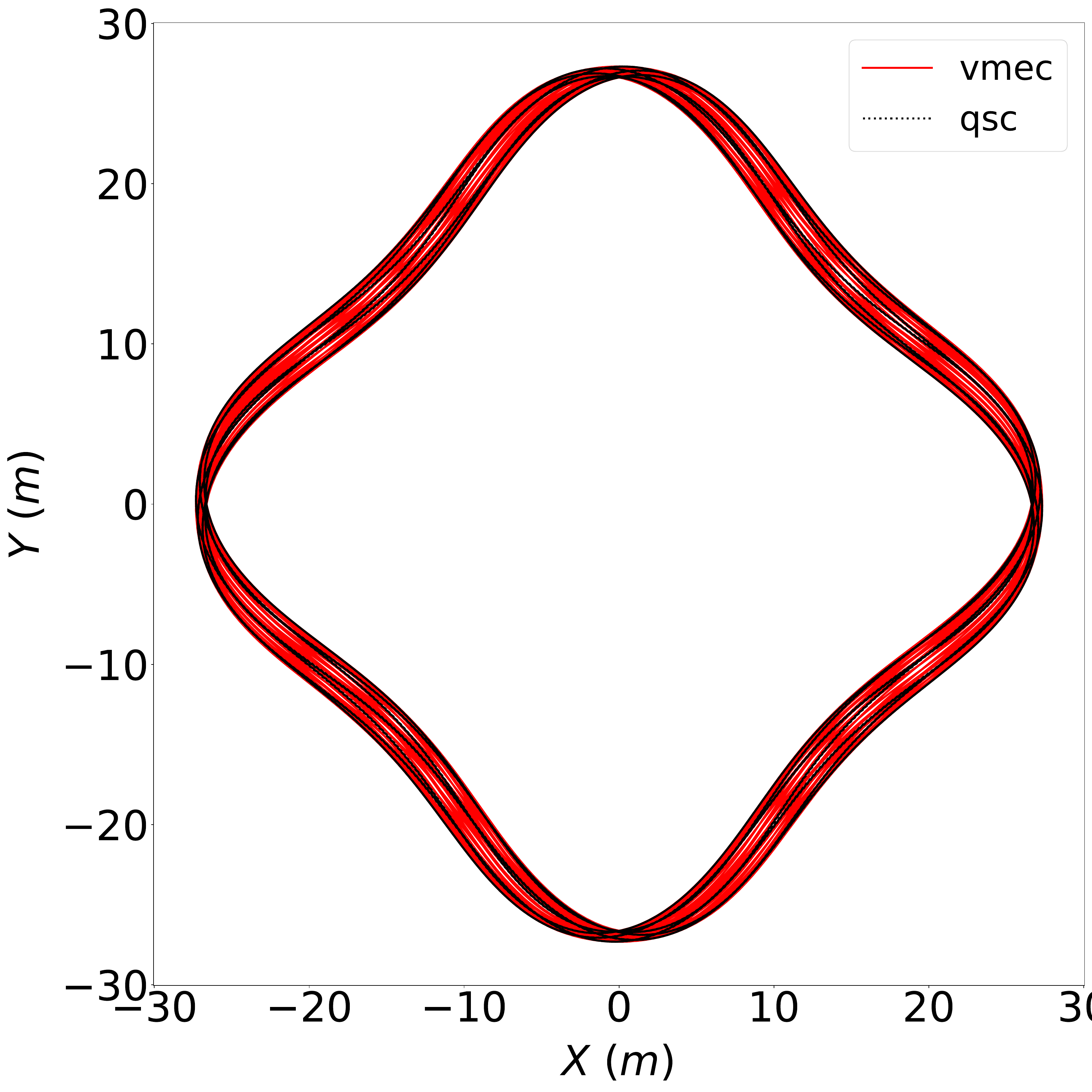}
    \caption{Comparison between alpha particle orbits in \textit{pyQSC} (qsc), \textit{VMEC} (vmec) and \textit{BOOZ\_XFORM} (booz) QH equilibria with initial position $(s, \theta, \varphi) = (0.25, 0.1, 0.1)$ in Boozer coordinates and $\lambda=0.8$. Above: Evolution of the radial position in time. Middle: Normalized poloidal view of the orbit. Below: Top view of the orbit in Cartesian coordinates.  Left: $A=6.8$. Center: $A=9.1$. Right: $A=13.6$.}
    \label{fig:QH-passing}
\end{figure}

\begin{figure}
    \centering
    \hfill
    \includegraphics[width=0.32\textwidth]{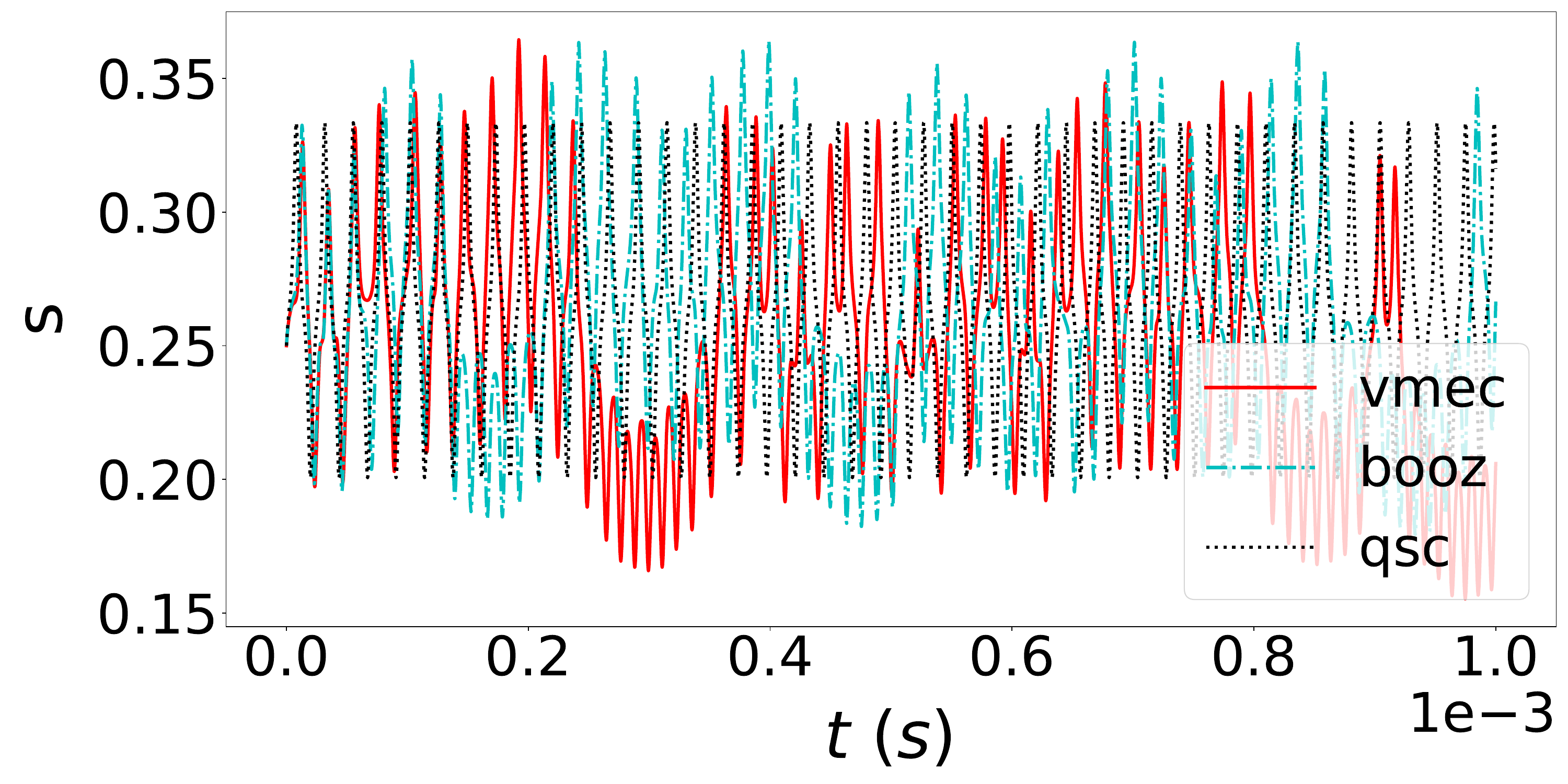}
    \hfill
    \includegraphics[width=0.32\textwidth]{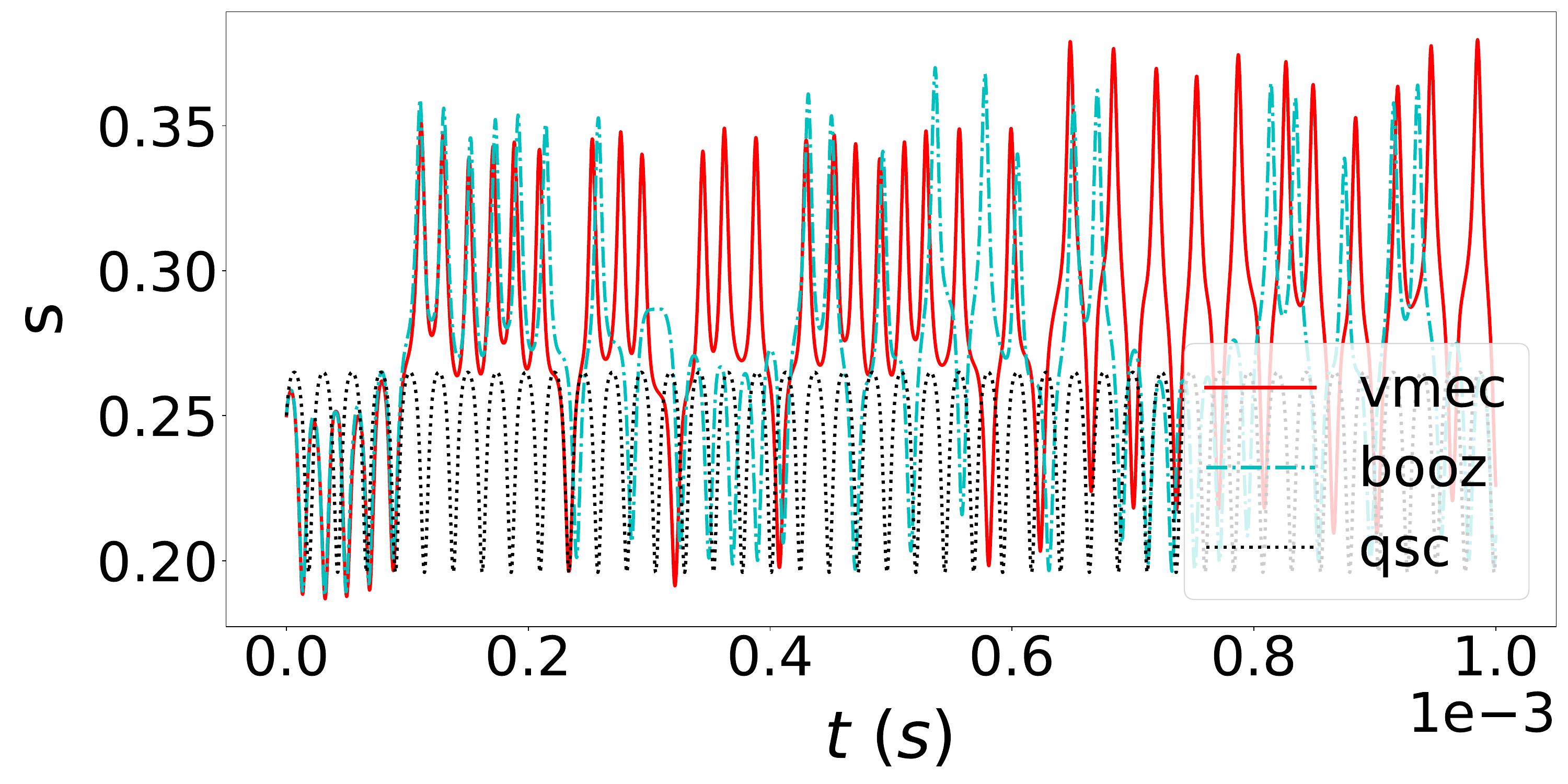}
    \hfill
    \includegraphics[width=0.32\textwidth]{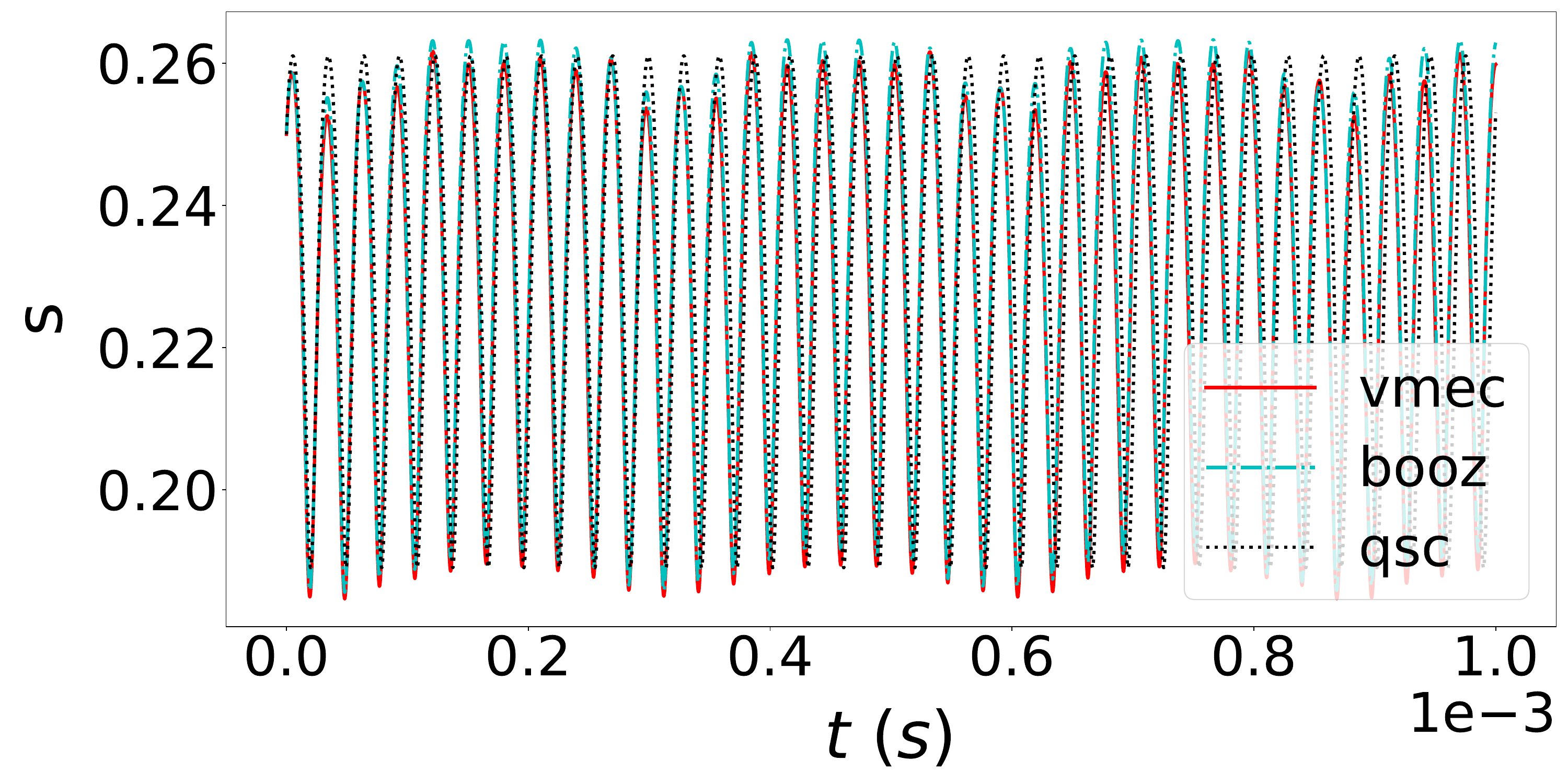}
    \hfill
    \vspace{0.1em}
    
    \hfill
    \includegraphics[width=0.25\textwidth]{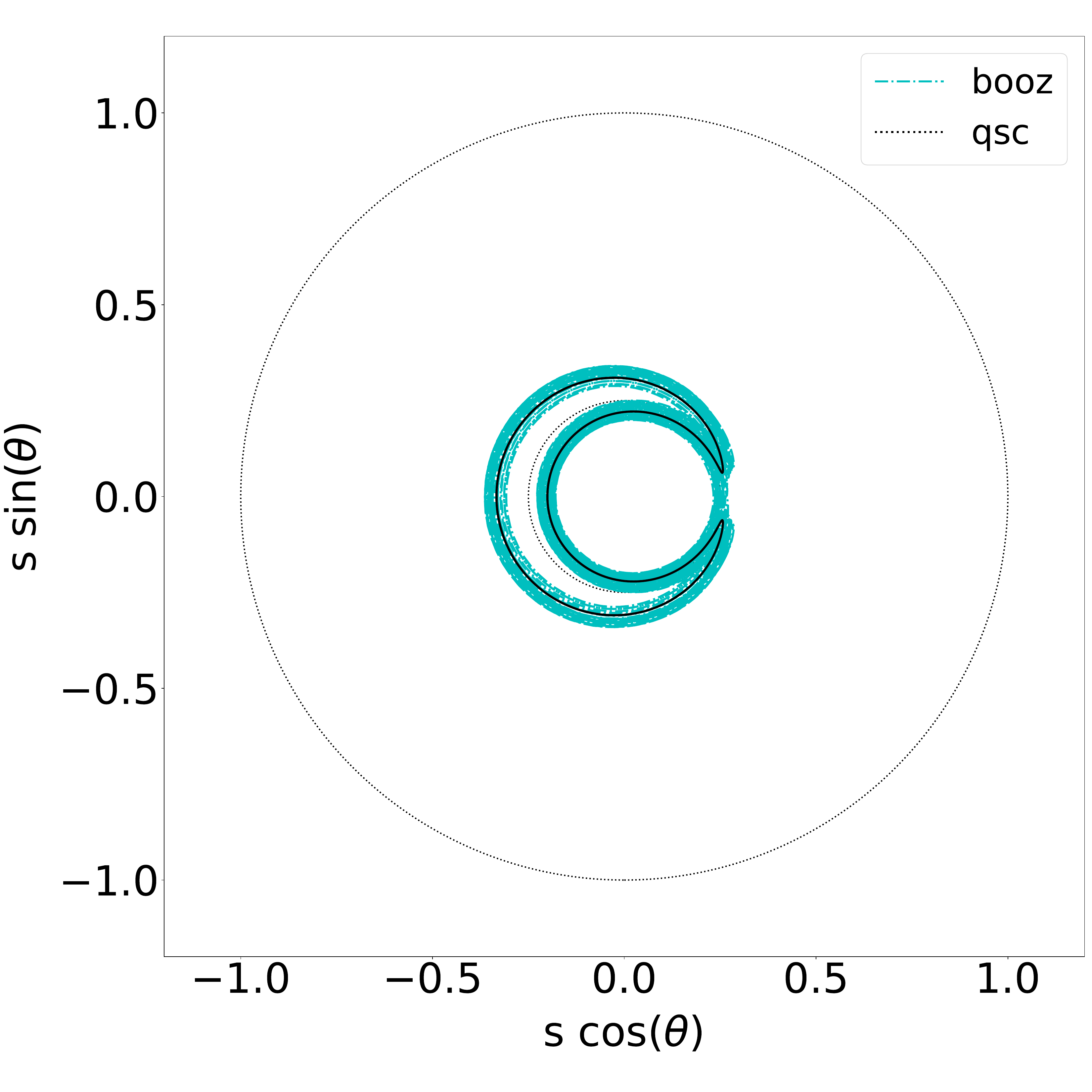}
    \hfill
    \includegraphics[width=0.25\textwidth]{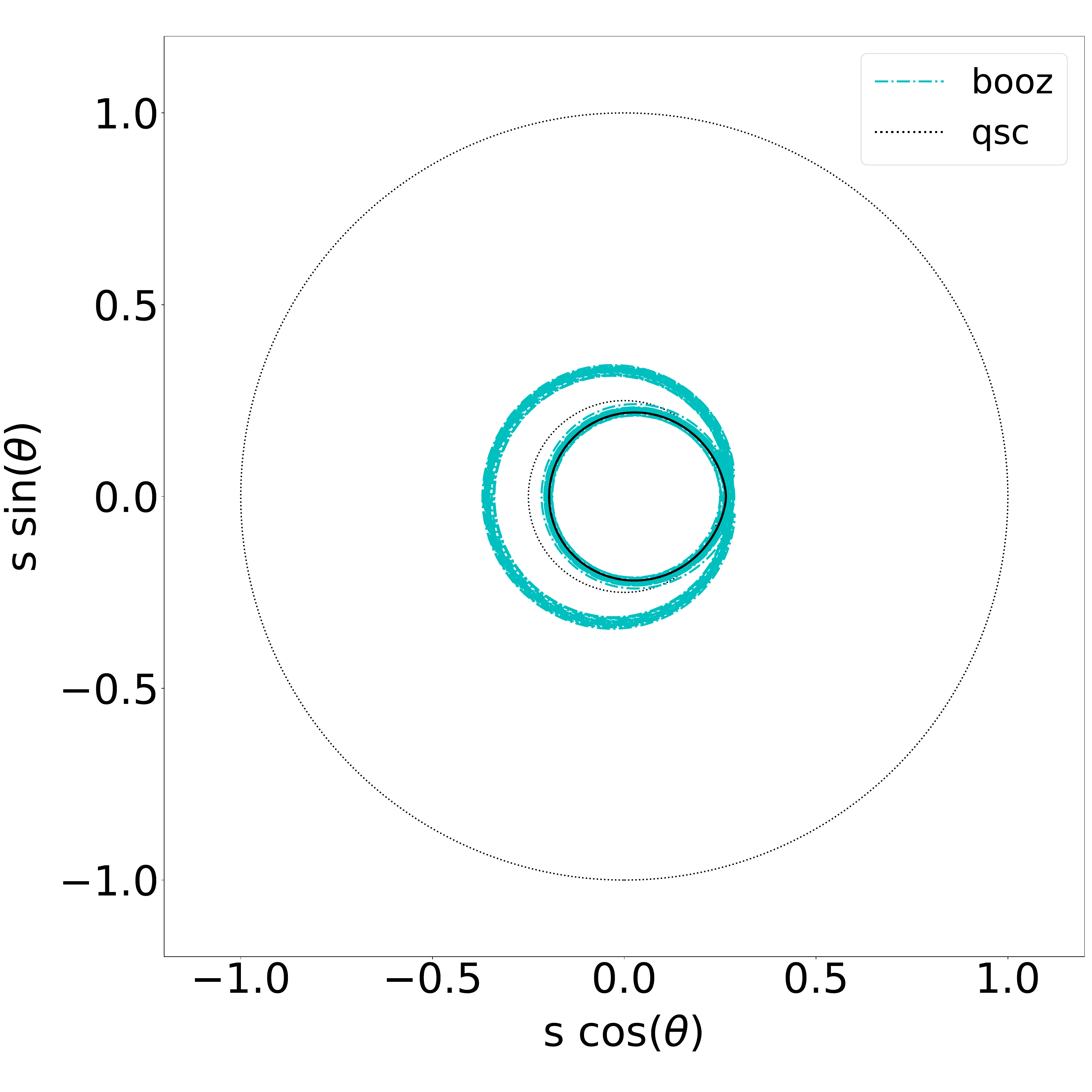}
    \hfill
    \includegraphics[width=0.25\textwidth]{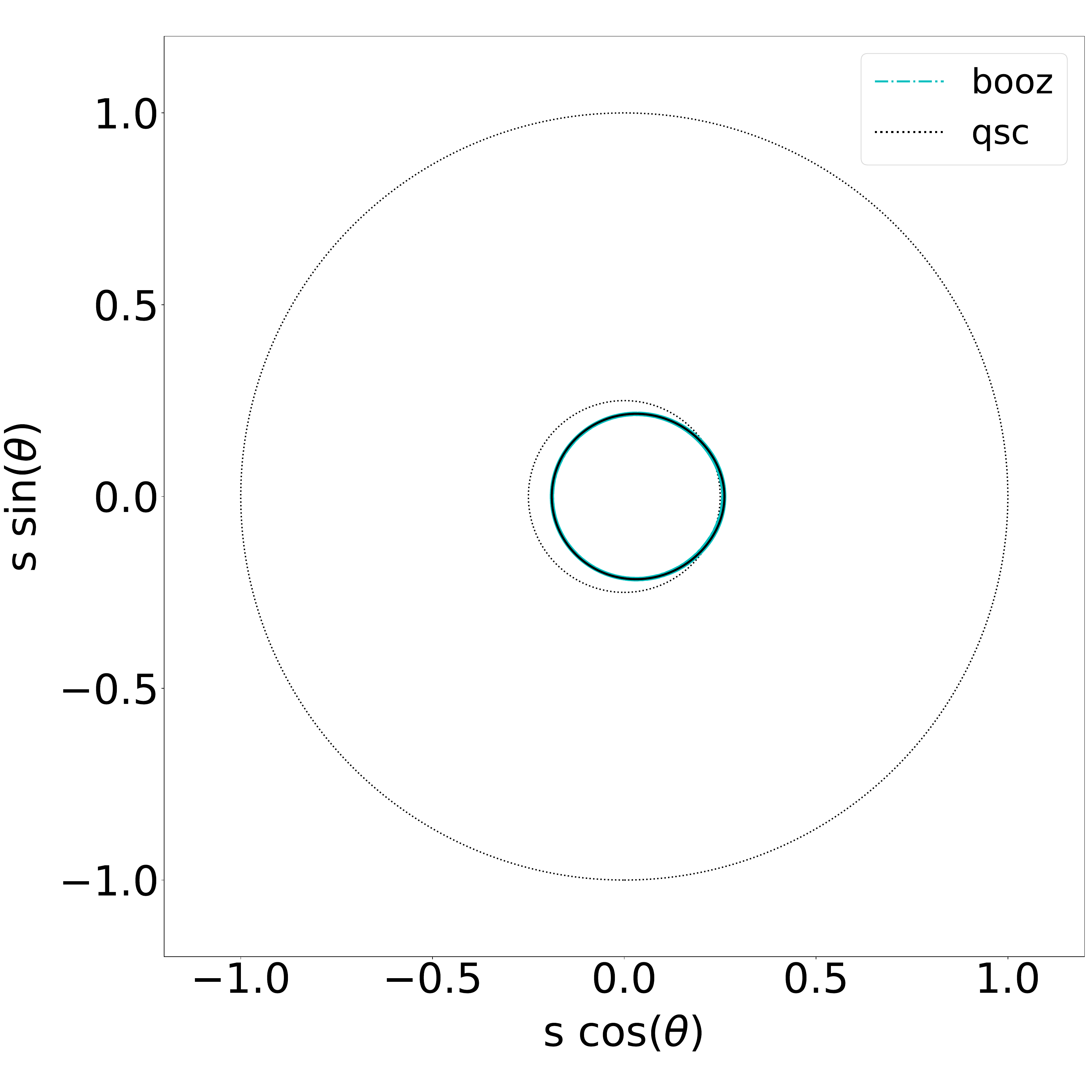}
    \hfill
    \vspace{1em}
    
    \hspace{-1em}
    \includegraphics[width=0.25\textwidth]{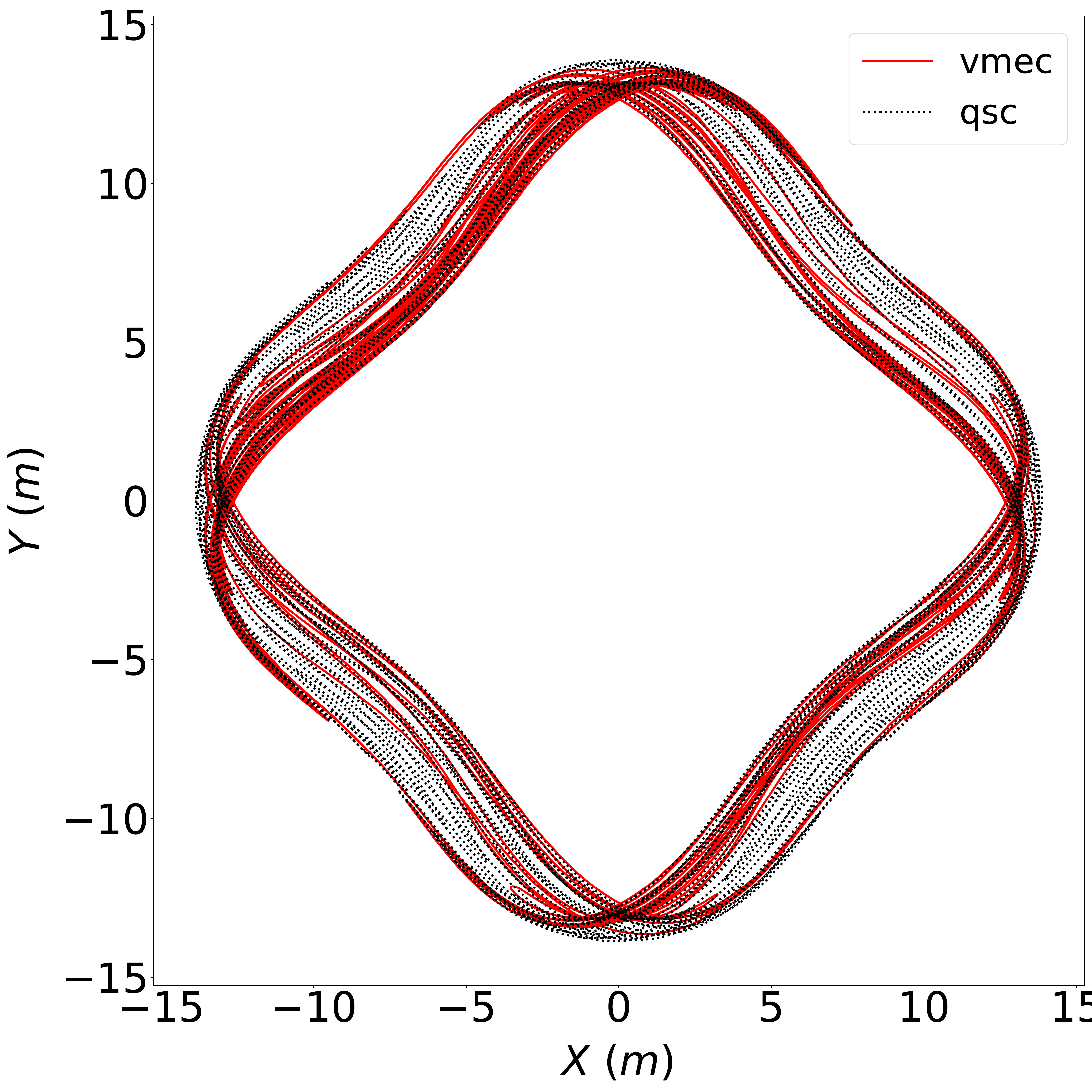}
    \hspace{1.5em}
    \includegraphics[width=0.25\textwidth]{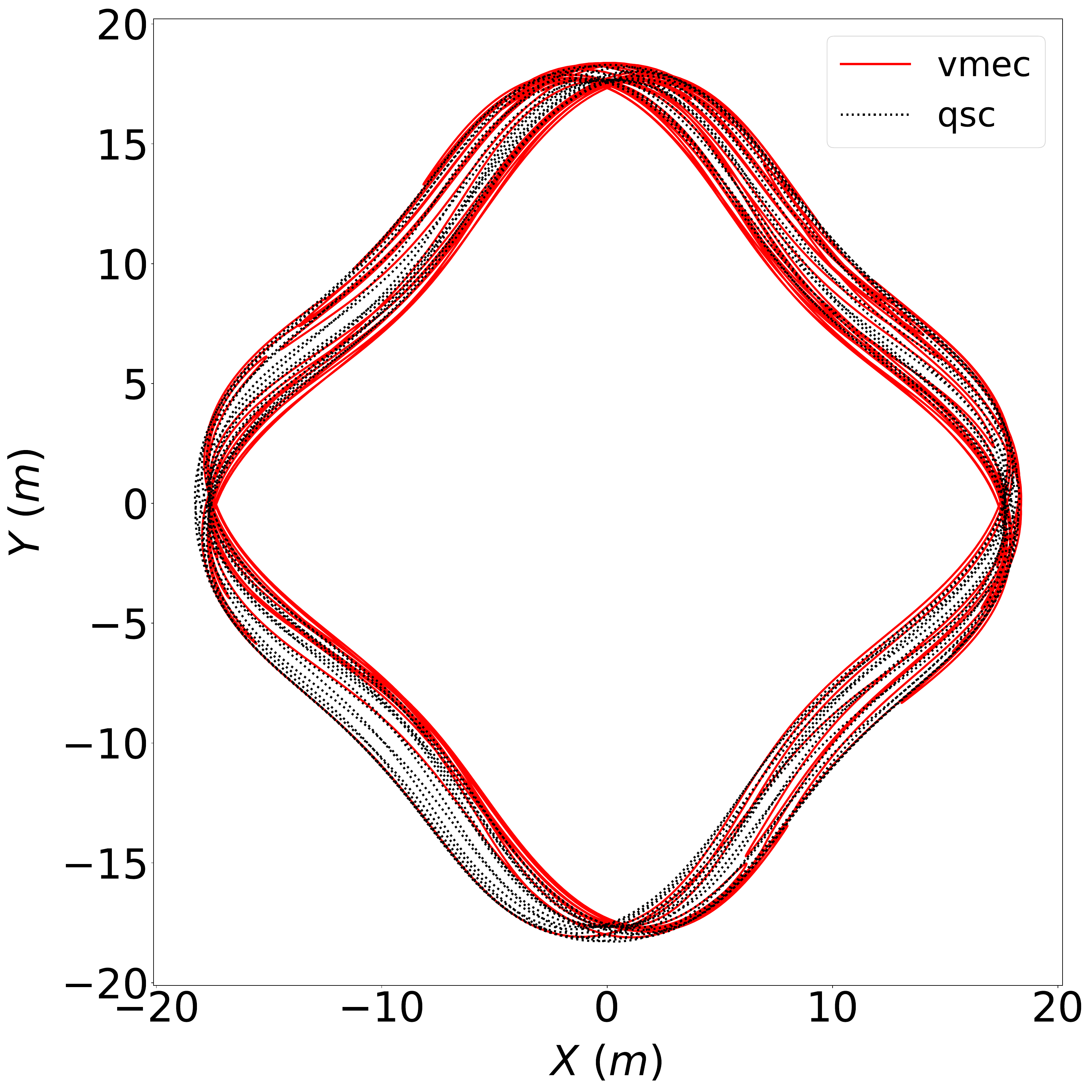}
    \hspace{1.5em}
    \includegraphics[width=0.25\textwidth]{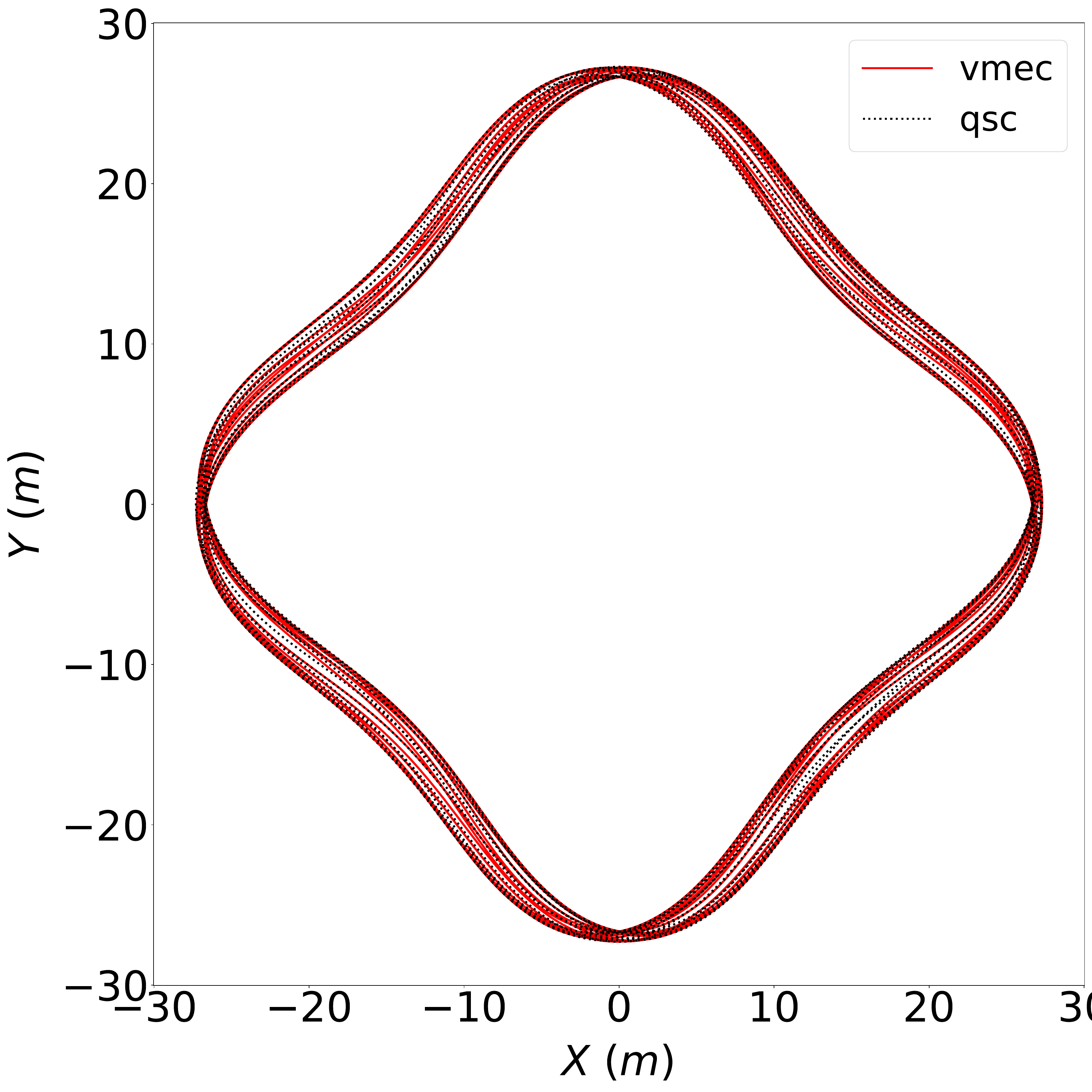}
    \caption{Comparison between alpha particle orbits in \textit{pyQSC} (qsc), \textit{VMEC} (vmec) and \textit{BOOZ\_XFORM} (booz) QH equilibria with initial position $(s, \theta, \varphi) = (0.25, 0.1, 0.1)$ in Boozer coordinates and $\lambda=0.99$. Above: Evolution of the radial position in time. Middle: Normalized poloidal view of the orbit. Below: Top view of the orbit in Cartesian coordinates.  Left: $A=6.8$. Center: $A=9.1$. Right: $A=13.6$.}
    \label{fig:QH-trapped}
\end{figure}

 The orbit widths obtained for the QH stellarator have, in general, smaller amplitudes, which matches the observation in \cite{landremanMagneticFieldsPrecise2022} that the radial excursions are expected to decrease with a higher value of $|\iota - N|$. The lower amplitudes, which can be observed in \cref{fig:QH-passing,fig:QH-trapped}, appear to be accompanied by higher frequencies of oscillation. Additionally, passing orbits such as the ones in \cref{fig:QH-passing} exhibit proportionally higher deviations of the \textit{VMEC} and \textit{BOOZ\_XFORM} orbits from the \textit{pyQSC} ones, although in absolute value they tend to be smaller than their QA counterparts in \cref{fig:QA-passing}. 

\par For the case where $\lambda=0.99$ as depicted in \cref{fig:QH-trapped}, distinct behaviors emerge based on the aspect ratio. When considering a smaller aspect ratio, the orbit in the NAE magnetic field is a purely trapped particle, while the remaining orbits oscillate between trapped and passing states throughout the temporal evolution. In this scenario, the total radial amplitude of the transitioning trajectories exhibits only a marginal increase compared to the purely trapped orbits. For the medium aspect ratio, the particle in an NAE field is now a passing one, while the other particles continue to demonstrate the previously observed behavior, with radial amplitudes exceeding twice that of the passing trajectory. For the case of a larger aspect ratio, as all particles are passing ones, only small deviations are observed between orbits. It is also interesting to note that the banana widths are smaller in the QH case than in the QA case, as expected.

\par Another important aspect is that no signs of trapped particles with an average outward radial drift, $\Delta \psi > 0$, were observed for the orbits in the QH NAE fields. This may in part be caused by the increased quasisymmetry in these fields or an expression of the simplicity of the field, not allowing for locally trapped or detrapping particles. Passing-trapped transitions can be caused by the finite banana width, which is usually a small effect, and by the misalignment of field maxima, which can be attributed to a deviation from perfect quasisymmetry and can cause stochastic diffusion losses of about $10\%$ for particles born at half radius \citep{Beidler2001}. The latter effect is thus more relevant to the QH configurations, where the degree of quasisymmetry achieved is smaller. We note that such deviation is expected to stem mainly from differences in $B$ as the same code, equations and algorithms were used. Despite this fact, the lower transport due to increased rotational transform, an important feature of most QH stellarators, as it provides better confinement of energetic particles for lower timeframes, is partially captured by the NAE.

\subsection{Trapped-Passing Separatrix}
\par  As the alignment of orbits for the different equilibria appears to be appreciably worse for trapped particles than for their passing counterparts, it becomes of interest to try to estimate the threshold between the two of them taking into account their initial coordinates. We attempt exactly that using the first-order NAE analytical formula for the magnetic field. 

\par Defining the energy and magnetic moment as $E= m ( v_\perp ^2 + v_\parallel ^2) / 2$ and $\mu= {m v_\perp ^2}/{2 B(s, \theta, \phi)}$, respectively, and defining $\lambda = E_{\perp_i} / E = \mu B_i / E$, where the subscript $i$ stands for the initial value given for the simulation, we arrive at 
\begin{equation}
    v_\parallel ^2 = v^2 \ \bigg(1 -  \lambda \frac{B(s, \theta, \phi)}{B_i} \bigg).
    \label{eq:v_par}
\end{equation}
For a trapped particle, there is a given critical radial position $(s_c, \theta_c, \varphi_c)$ at which $v_\parallel = 0$, leading to $ B(s_c, \theta_c, \varphi_c)) = B_c = { B_i}/{\lambda}$.
With a first order near axis expansion we can write $B_i = B_0 (1 + a_A \sqrt{s_i} \ \bar{\eta} \cos{\theta_i})$ and $B_c = B_0 (1 + a_A \sqrt{s_c} \ |\bar{\eta}| )$, which we take to be a poloidal maximum of $B$ for this estimation, which leads to $ | \cos{\theta} | =1$. This assumption results in an overestimation of the value of $B_c$, but simplifies the analysis, leading to an estimation of $\lambda$ at which we have the passing-trapped separatrix, $\lambda_s$ depending only on the initial position of the particle and the radial position where the parallel velocity is null, $s_c$.
Taking advantage of the fact that confined orbits perform an oscillatory motion, we may assume $s_c$ should be between $s_i - 2\Delta s$ and $s_i + 2\Delta s$ where $ 2 \Delta s$ is an estimation of the maximum radial peak-to-peak amplitude of orbits. This is done since the starting point can be in the peak or the valley of the oscillation. Finally, we arrive at the following interval for the transition value $\lambda_s$
\begin{align}
    \frac{(1 + a_A \sqrt{s_i} \ \bar{\eta} \cos{\theta_i})}{(1 + a_A \sqrt{s_i + 2 \Delta s} \ |\bar{\eta}| )} \leq  \lambda_s \leq  \frac{(1 + a_A \sqrt{s_i} \ \bar{\eta} \cos{\theta_i})}{(1 + a_A \sqrt{s_i - 2 \Delta s} \ |\bar{\eta}| )}.
    \label{eq:trapped_interval}
\end{align}

\par To demonstrate how effective this interval is at predicting if a particle is trapped or passing we traced particles in the \textit{pyQSC} QA and QH equilibria of aspect ratio $A=9.1$ with linearly spaced initial conditions $\theta$, $\varphi$ and $\lambda$ for three different radial positions $s_i=0.25$, $0.5$ and $0.75$. We then check if the particles have $v_\parallel=0$ at any point of their orbit (trapped) or not (passing), discarding all the lost particles classified as passing orbits. This last step was performed to avoid false negatives, as some orbits could be trapped particles that are lost before achieving $v_\parallel=0$. The interval in \cref{eq:trapped_interval} is depicted in the form of three thresholds, two of them corresponding to the lower and higher bounds on $\lambda_s$ and the third one corresponding to $s_c=s_i$. For the QA stellarator, we take $2 \Delta s$ to be 0.4 and, for the QH, 0.1 as the maximum amplitudes in our examples appear to be of this order.

\par
\begin{figure}
    \centering
    \includegraphics[width=0.42\textwidth]{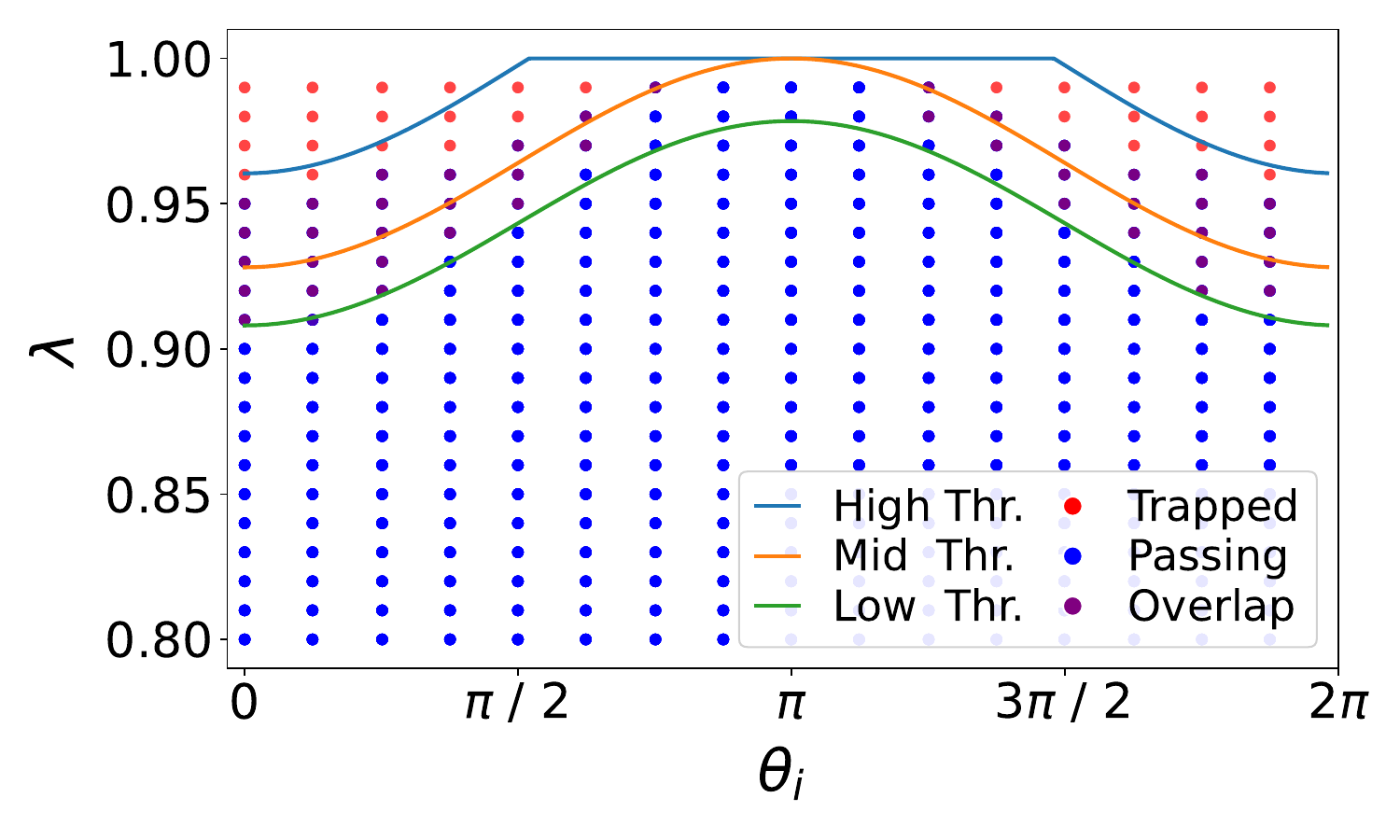}
    \includegraphics[width=0.42\textwidth]{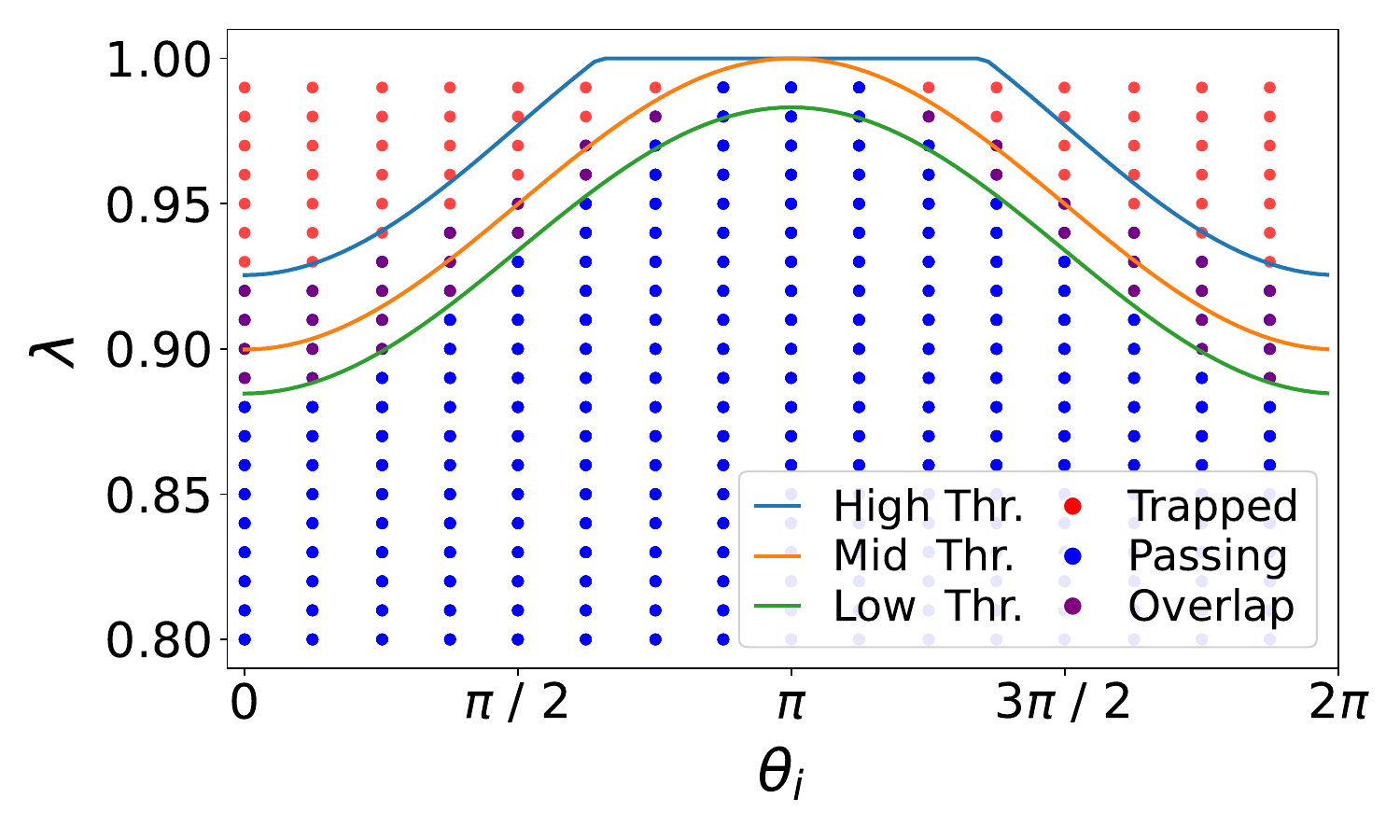}
    \includegraphics[width=0.42\textwidth]{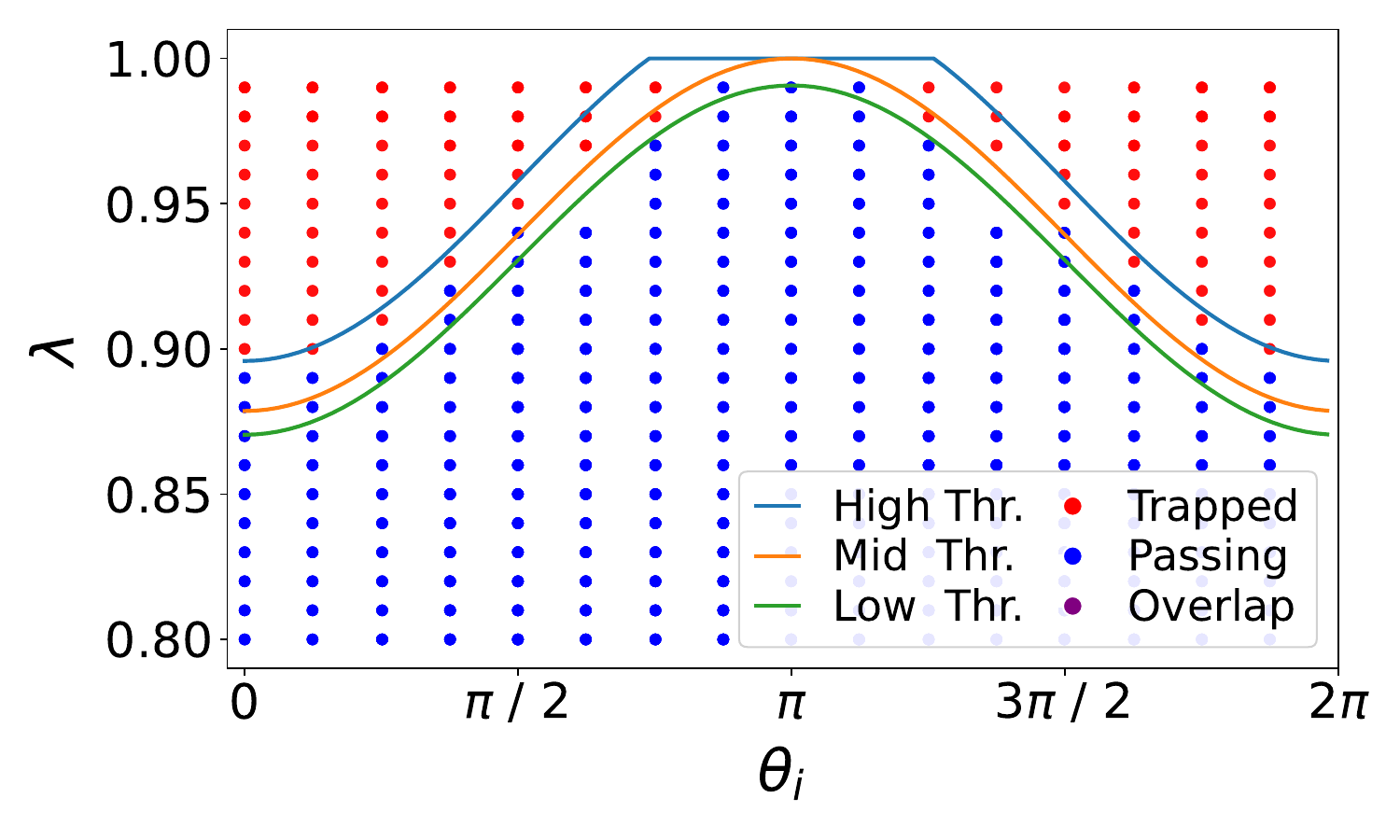}
    \caption{Comparison of estimated interval for trapped to passing separatrix with computed results for a QA stellarator. Trapped (Red) and passing (Blue) particles are plotted as a function of the initial Boozer poloidal coordinate $\theta$ and of $\lambda=\mu B_i/ E$ for three different initial radial positions $s_i$. High and low thresholds correspond to interval bounds, while mid threshold corresponds to an average separatrix. Left: $s_i=0.25$, Right: $s_i=0.5$, Bottom: $s_i=0.75$}
    \label{fig:QA_trapped_passing}
\end{figure}

\begin{figure}
    \centering
    \includegraphics[width=0.42\textwidth]{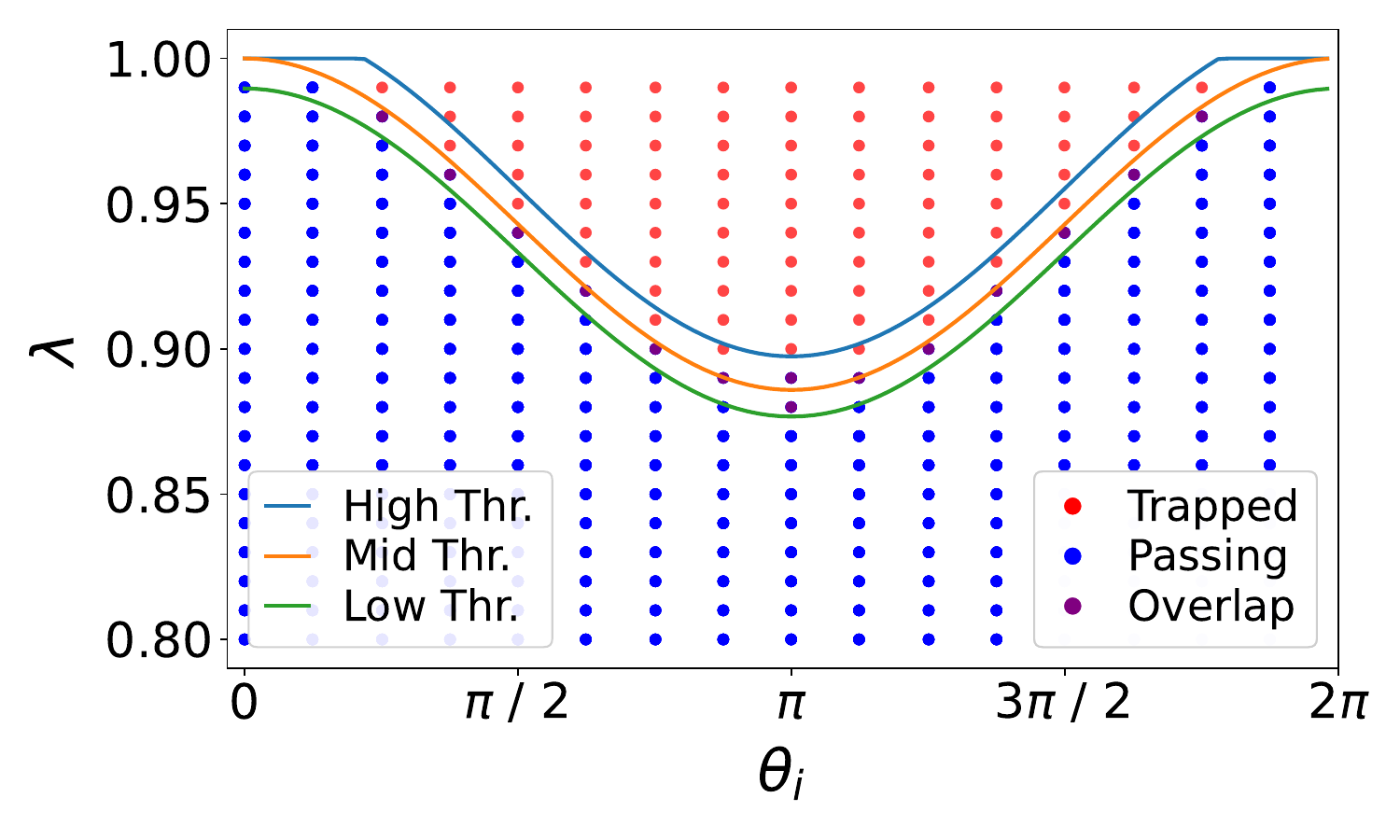}
    \includegraphics[width=0.42\textwidth]{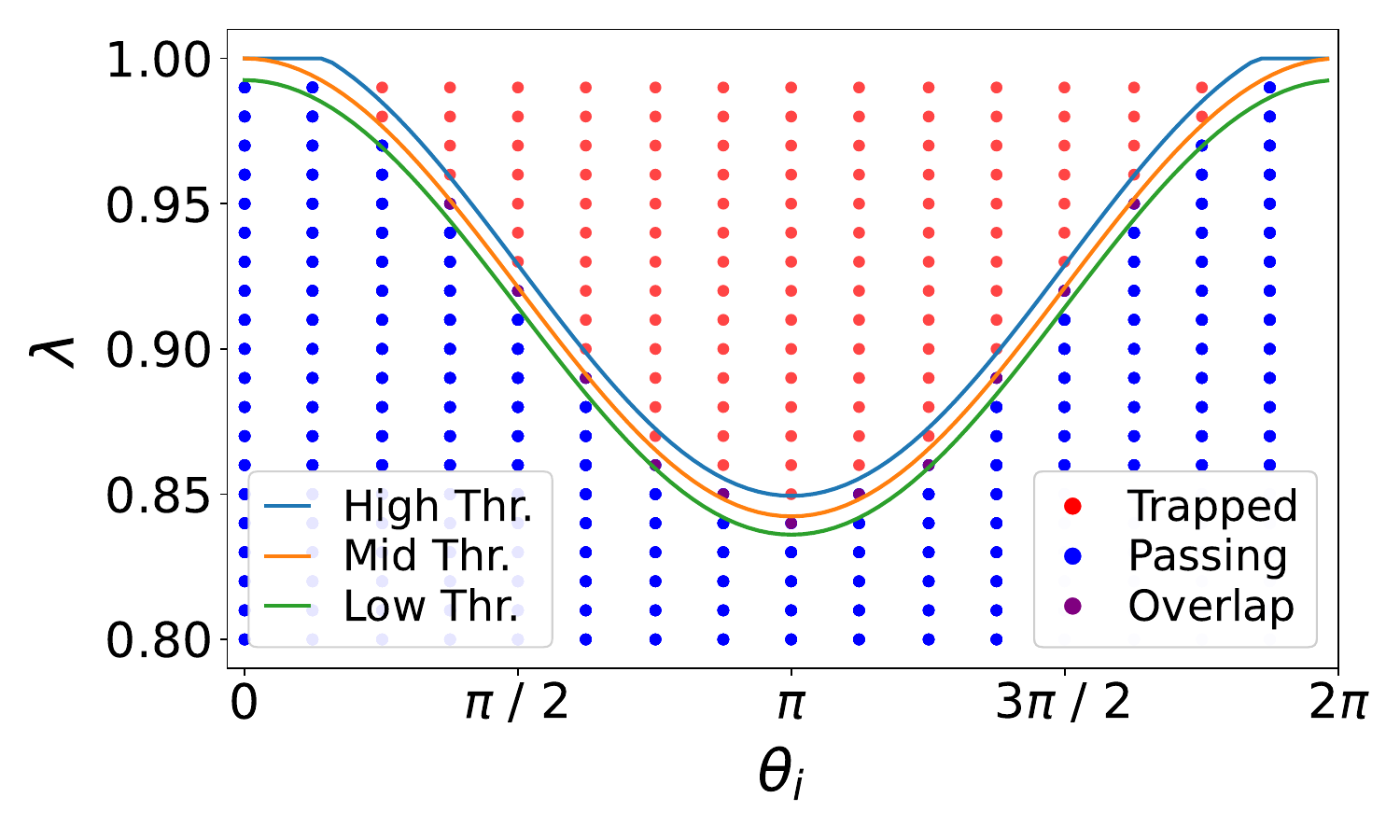}
    \includegraphics[width=0.42\textwidth]{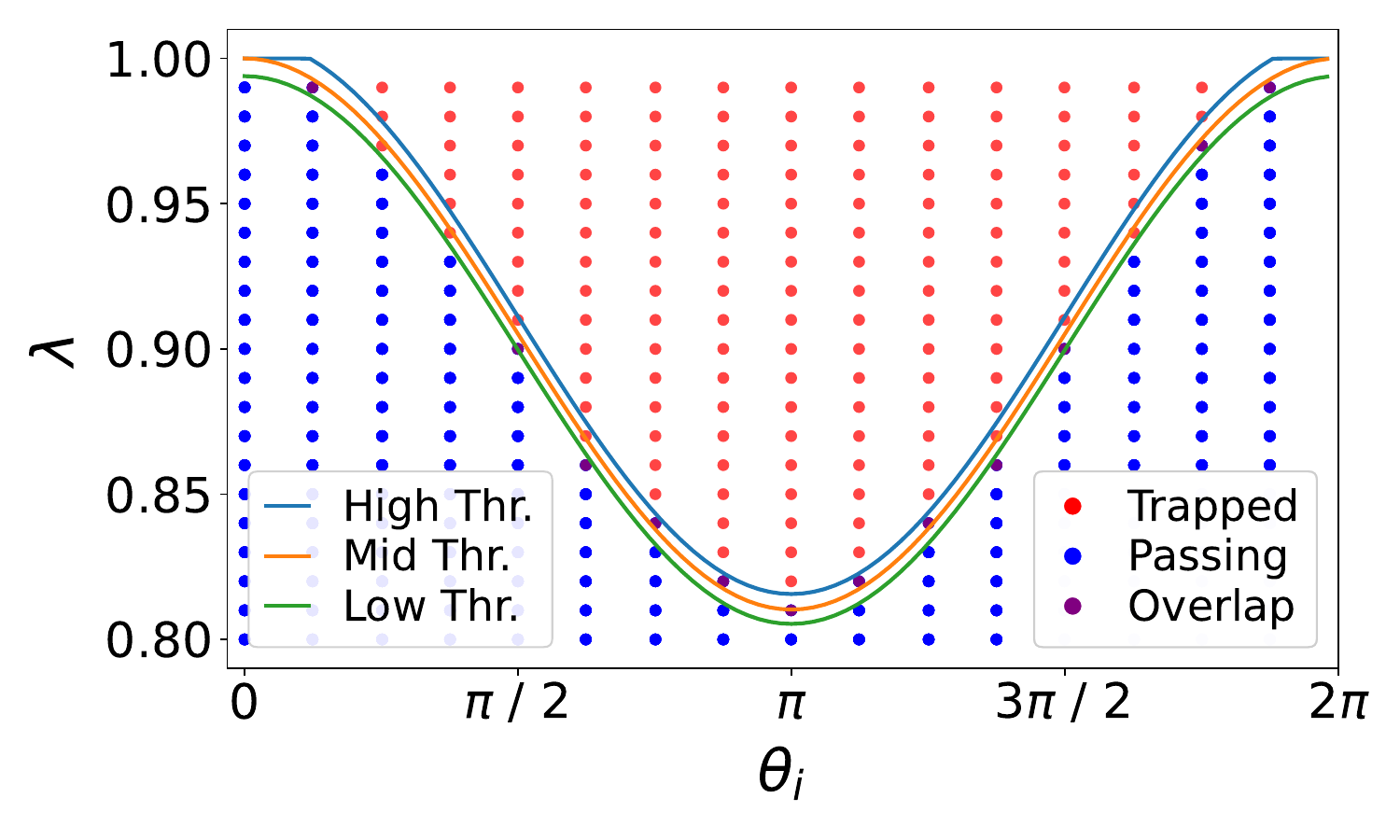}
    \caption{Comparison of estimated interval for trapped to passing separatrix with computed results for a QH stellarator. Trapped (Red) and passing (Blue) particles are plotted as a function of the initial Boozer poloidal coordinate $\theta$ and of $\lambda=\mu B_i/ E$ for three different initial radial positions $s_i$. High and low thresholds correspond to interval bounds, while mid threshold corresponds to an average separatrix. Left: $s_i=0.25$, Right: $s_i=0.5$, Bottom: $s_i=0.75$}
    \label{fig:QH_trapped_passing}
\end{figure}

 The obtained results can be seen in \cref{fig:QA_trapped_passing,fig:QH_trapped_passing}, where the threshold lines match the changes in the trapped area in a quite accurate way both in the QA and the QH stellarators for different radial initial positions. It is also interesting to note that the interval encompasses almost all overlapping dots even when the interval is of a smaller magnitude, such as in the QH case even though it is the product of a few assumptions. That said, the middle threshold can still be useful to have a single predictor of the type of orbit in question, especially in the QH case, where the amount of overlapping points is substantially smaller.

\par In future works, this interval could be used to select \textit{a priori} only passing or trapped particles for analysis. This would be interesting to do because trapped particle orbits are generally unconfined and may exhibit chaotic behaviour, and could be more easily studied with this approach. Furthermore, these particles represent a substantial part of alpha-particle losses and so, computing loss fractions of trapped particles only would be more computationally efficient without considerable information losses. However, as the interval $\Delta s$ was found only \textit{a posteriori} it is still necessary to find a way of estimating this value. An attempt to do exactly this is found in the next subsection.

\subsection{Radial Oscillation Amplitude}

\par Not only is it of interest to have an estimate of the maximum oscillation amplitude $\Delta s$ to complete the above study, but also to have an expression for the dependence of the amplitude with $\lambda$ and $s_i$. To estimate these quantities, we take advantage of the fact that, for a general quasisymmetric stellarator, the conserved momentum $p_\eta$ can be obtained from the Lagrangian in \cref{eq:littleBoozer} through a coordinate transformation from $(s, \theta, \varphi)$ to $(s, \chi, \eta)$. This momentum can be written in the following way

\begin{gather}
    p_\eta = q \frac{N}{M} \psi - q \psi_p + \frac{m v_\parallel}{B(s, \theta, \varphi)} (G + \frac{N}{M}I), 
    \label{eq:lag_chi}
\end{gather}
and is the starting point of our estimation. Taking the first order NAE, we approximate $I = 0$, $\psi_p=\int \iota d\psi = \iota _0 \psi$ and $G = G_0= {L B_0}/(2 \pi)$, where $L=\int_0^{2\pi}d\phi\; \ell'$ is the axis length and $\ell$ is the arc length along the field line. Additionally assuming $\iota_N= \iota_0 - N$ and $M=1$, and using \cref{eq:v_par}, we
find an expression for the radial excursion of the particle in flux surfaces
\begin{gather}
     \psi =  - \frac{p_\eta}{q \ \iota_{N}}  \pm \frac{m v L B_0}{2 \pi q \ \iota_{N} B(s, \theta)}\sqrt{1 - \frac{\lambda B(s, \theta)}{B_i}},
     \label{eq:psi}
\end{gather}
 which depends on the magnetic field $B$, which in turn depends on the radial position $s$ and poloidal angle $\theta$, an implicit equation that would have to be computationally solved. However, we can estimate the amplitude of the orbits by applying the linearization $\Delta \psi / \Delta B = \partial_B \psi$ to \cref{eq:psi} and taking $\partial_B \hspace{0.2em} p_\eta = 0$. For the estimation of the variation of B, we take into account that it depends on $s$ and $\theta$ and thus sum its variation in both coordinates  
\begin{align}
    \Delta B & = \int_0^{\pi} \partial_\theta B \ d\theta + \int_{s_i}^{s_{i}+ \Delta s_{max}} \partial_s B \ ds \notag \\
    & = \bigg(B(s_i, \pi) - B(s_i, 0)\bigg) + \bigg(B(s_i + \Delta {s_{avg}}, \theta_i) - B(s_i, \theta_i)\bigg) \notag \\  
    & = |\bar{\eta}| a_A B_0 \bigg( 2 \sqrt{s_i} +  (\sqrt{s_i + \Delta {s_{max}}} - \sqrt{s_i})  \cos{\theta_i} \bigg).
    \label{eq:delta_B}
\end{align}
Using the expression for $\Delta B$ in \cref{eq:delta_B}, applying the partial derivative of B and making $B(s,\theta) = B_0$ we obtain an estimation of the amplitude of oscillation of the flux surface coordinate $\psi$
which, written as the normalized radial coordinate $s=\psi / \psi_b$, with $\psi_b = B_0 a_A^2 / 2$, becomes
\begin{gather}
    \Delta s =  \frac{ m v L \bar{\eta}}{\pi q \ \iota_{N_0} a_A B_0} \frac{1 - \lambda B_0/ (2 B_i)}{\sqrt{1-\lambda B_0/B_i}} \bigg( 2 \sqrt{s_i} +  (\sqrt{s_i + \Delta {s_{avg}}} - \sqrt{s_i})  \cos{\theta_i} \bigg). 
\end{gather}

\par
\begin{figure}
    \centering
    \includegraphics[width=0.42\textwidth]{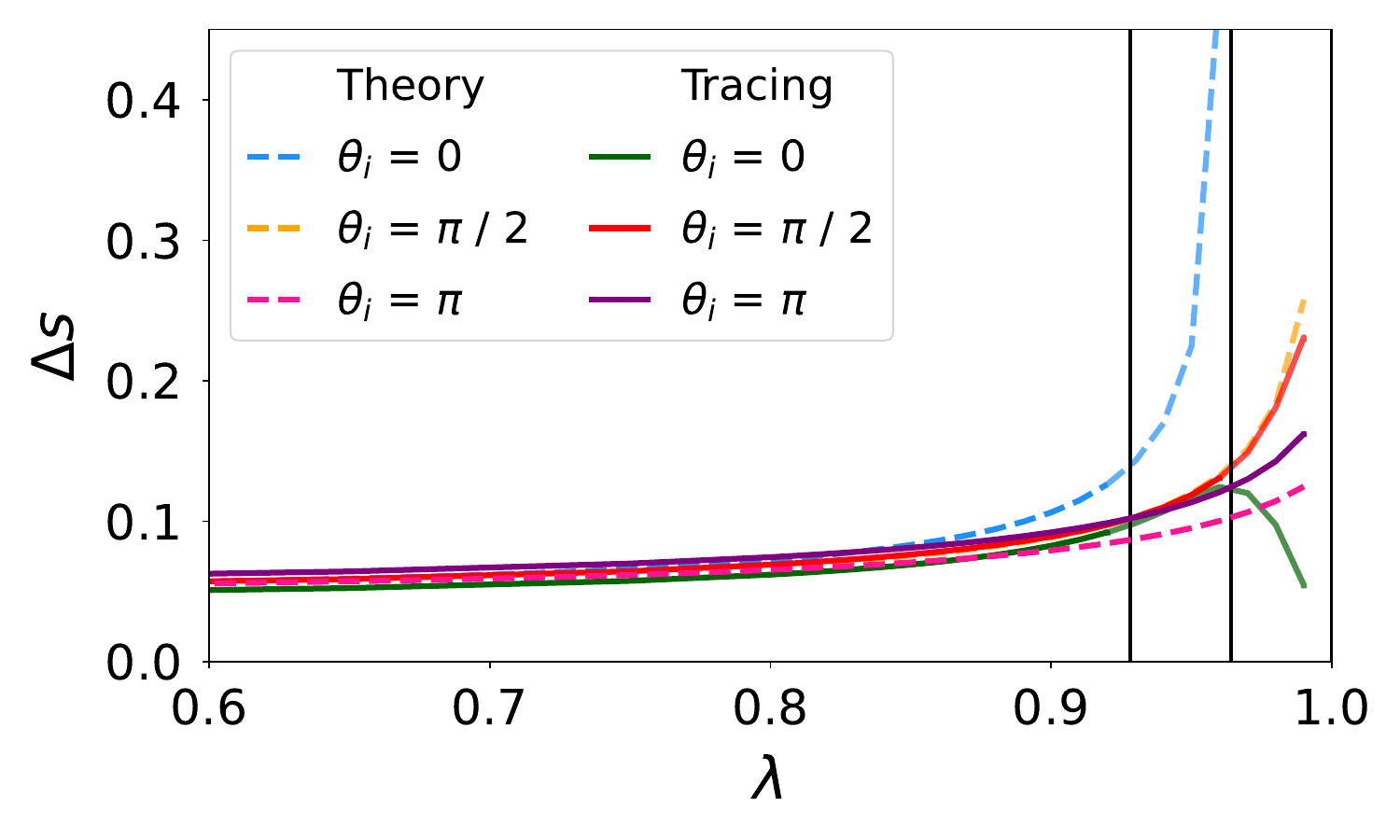}
    \includegraphics[width=0.42\textwidth]{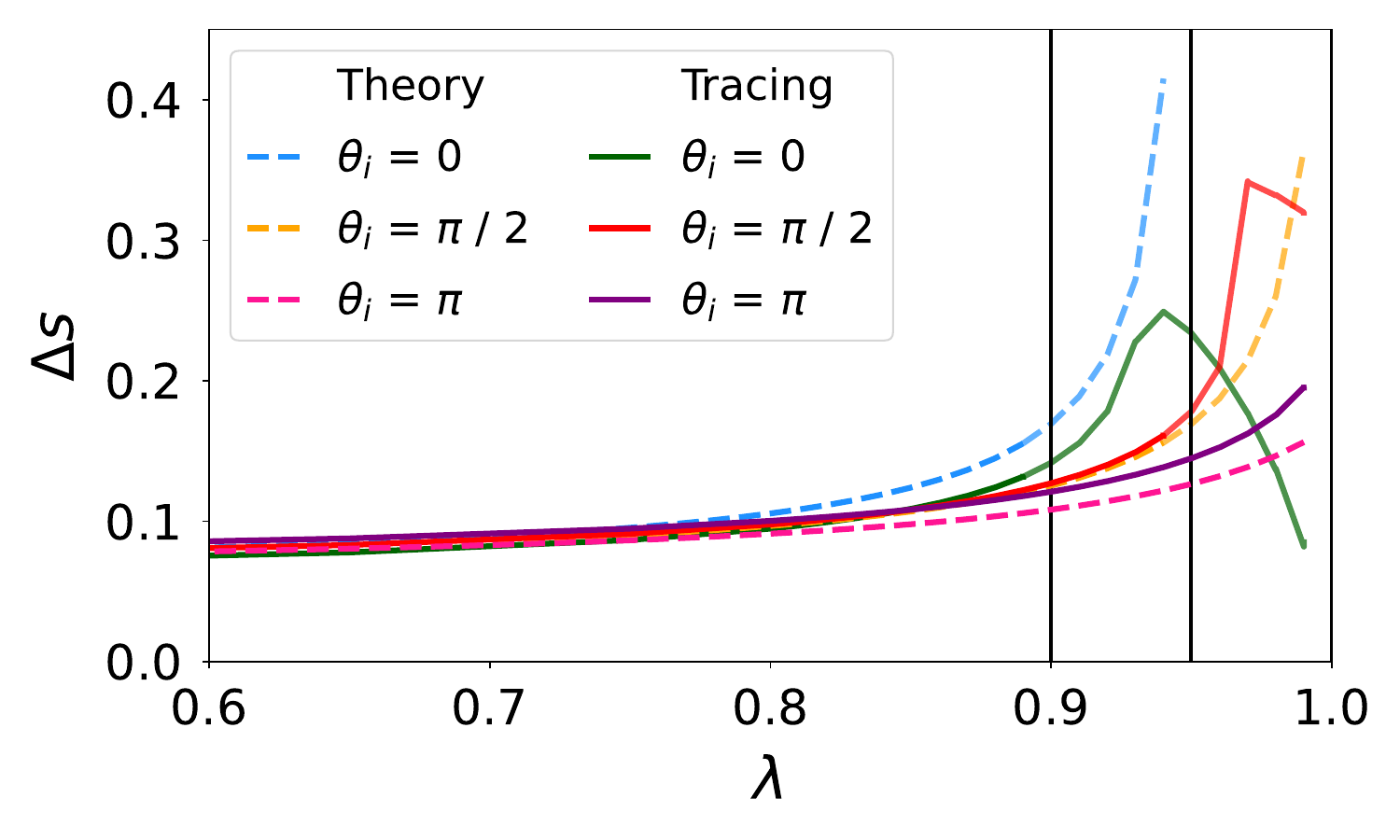}
    \includegraphics[width=0.42\textwidth]{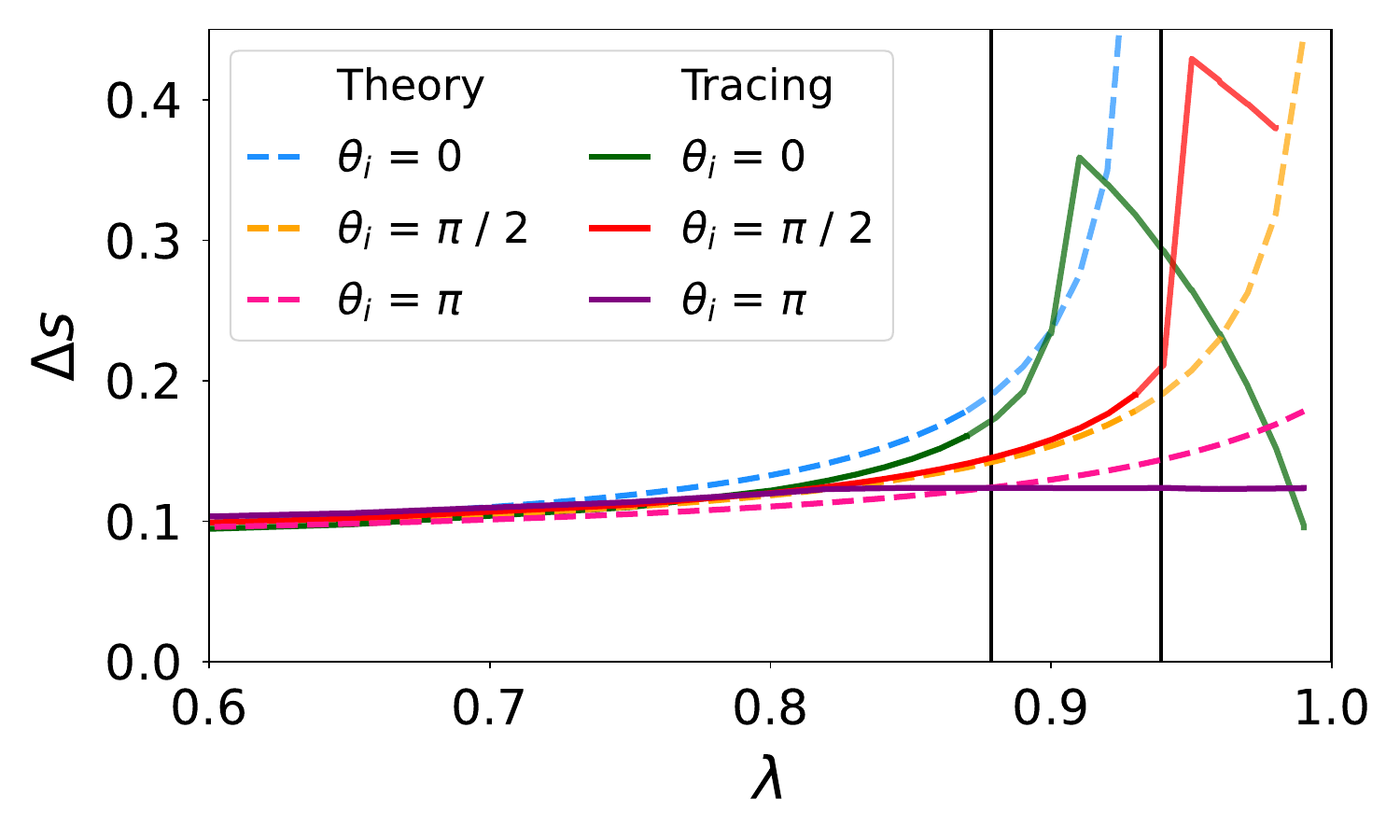}
    \caption{Comparison of computed with estimated orbital radial amplitude $\Delta s$ in a QA configuration as a function of $\lambda= \mu B_i / E$ for different initial poloidal and radial coordinates. Vertical lines indicate the average passing-trapped separatrix for each $\theta_i$, where the two quantities start to diverge.  Left: $s_i=0.25$, Right: $s_i=0.5$, Bottom: $s_i=0.75$}
    \label{fig:QA_amp_est}
\end{figure}

\begin{figure}
    \centering
    \includegraphics[width=0.42\textwidth]{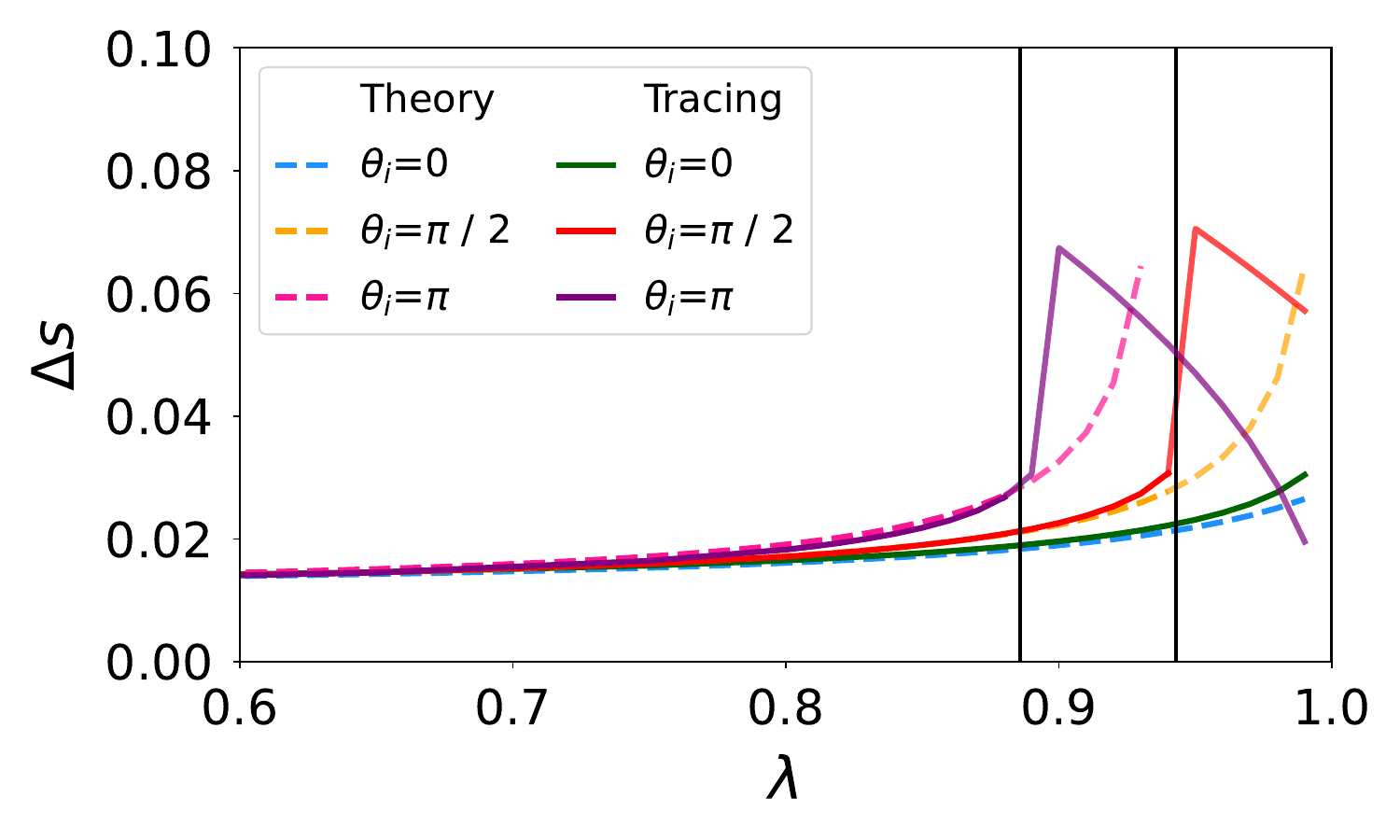}
    \includegraphics[width=0.42\textwidth]{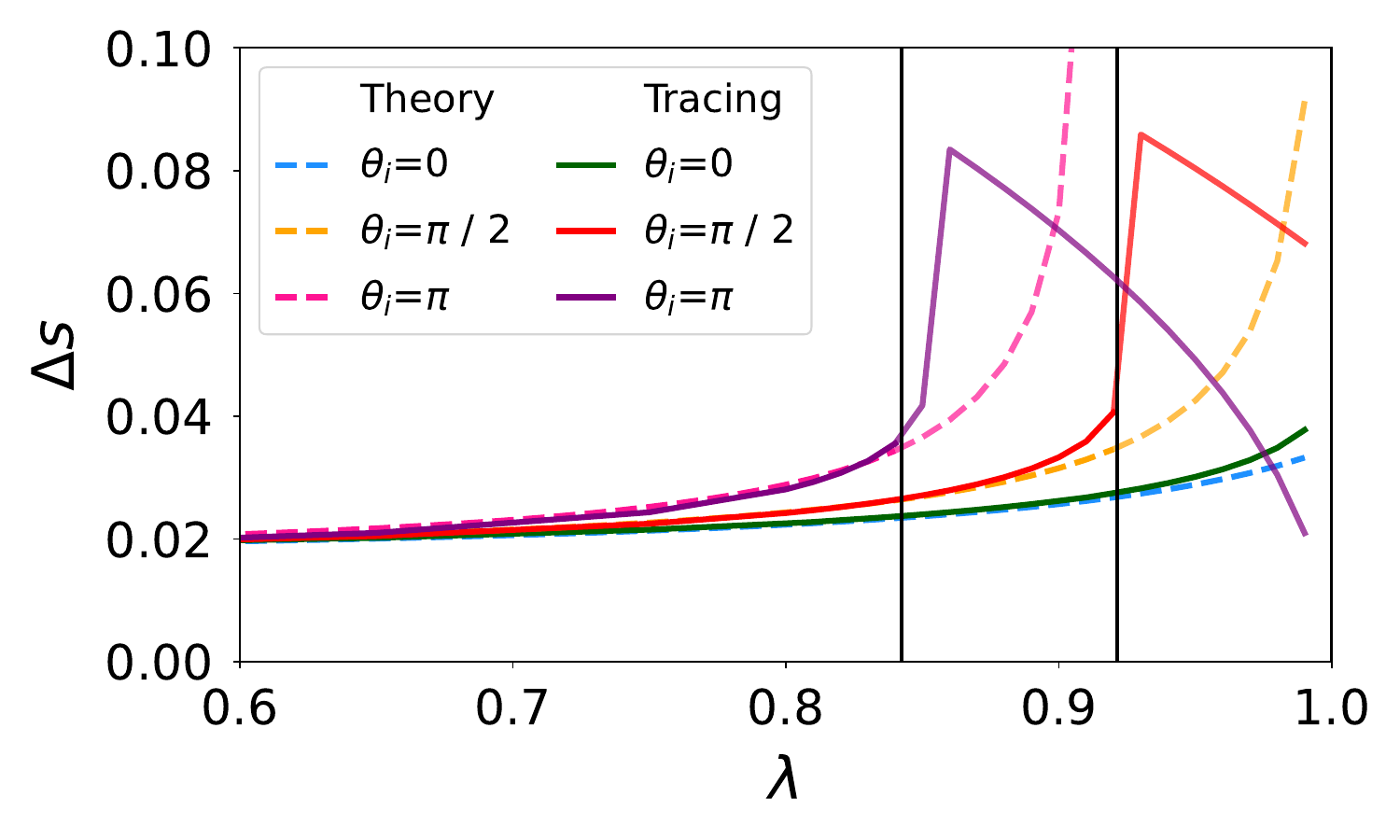}
    \includegraphics[width=0.42\textwidth]{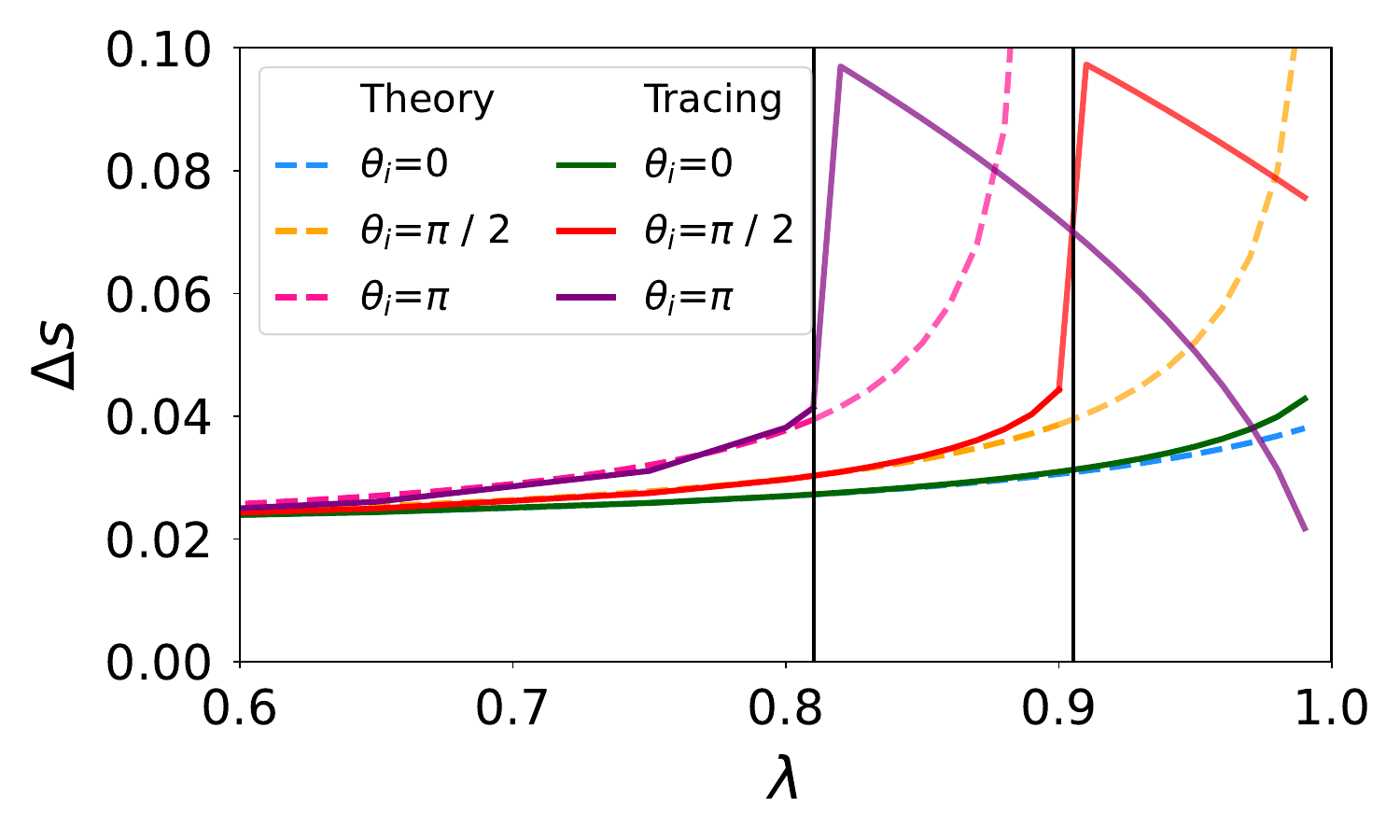}
    \caption{Comparison of computed with estimated orbital radial amplitude $\Delta s$ in a QH configuration as a function of $\lambda= \mu B_i / E$ for different initial poloidal and radial coordinates. Vertical lines indicate the average passing-trapped separatrix for each $\theta_i$, where the two quantities start to diverge.  Left: $s_i=0.25$, Right: $s_i=0.5$, Bottom: $s_i=0.75$}
    \label{fig:QH_amp_est}
\end{figure}

 This expression leads to results that approximate the computed ones up until the passing-trapped separatrix, as can be seen in \cref{fig:QA_amp_est,fig:QH_amp_est}, starting to deviate from it afterward. This appears to be the limit of the applied linearization. To extend the region of applicability of our estimation, it may be possible to go to higher orders on the approximation. The expression appears to capture the differences in amplitude for different stellarator configurations and different initial conditions.

\par Concerning a reasonable estimation of the maximum of the radial amplitude of the motion, it would be necessary to neglect the $\theta$ component of $\Delta B$. However, the maximum value would remain outside this regime so it may not be prudent to use the above expression to obtain the wanted value directly. The value of the proposed expression at the separatrix could instead give an initial guess of the appropriate value. 

\par In conclusion, through examining single particle orbits in different equilibria, we have gained critical insights into the dynamics of particles in NAE and ideal MHD magnetic fields. As we move forward to the analysis of particle ensemble behavior and loss fractions, we expect to improve our understanding of the collective behavior of particles, providing a more holistic view of the confinement process.

\section{Particle Ensembles}

We now change the focus from single particle trajectories to high-energy particle ensembles. The analysis of these groups gives us a more comprehensive look at how well particles are confined and can reveal patterns that might not be evident when examining single-particle dynamics. In this part of our study, we focus on calculating loss fractions, which are the percentage of particles that leave the magnetic equilibrium over a certain timeframe.  These are crucial for assessing how effective the magnetic fields are at keeping particles in place, making them a helpful metric to inform improvements in our confinement techniques.

To determine loss fractions, large groups of particles are initialized for a given duration and the percentage of these particles that cross the last closed flux surface is accounted for. It is important to acknowledge that the real values of loss fractions in an actual device are multifactorial, and the computations performed in this work only take into account some of the relevant physical factors. For a thorough analysis of non-collisional loss fractions, simulations should account for a wide range of random starting conditions and incorporate a density distribution for alpha particles. A realistic account of losses in a given configuration should also take into account that not all of the particles crossing the set boundary would be lost, as some could later reenter the plasma, in addition to taking into account the effects of collisions and turbulence. The former effect was estimated to affect the losses up to $\sim 10 \%$ in the LHD \citep{Miyazawa2014}. However, for a comparison between the losses in different fields, we are less concerned with the realistic global loss values than we are with having similar results between the obtained values, so we compute the loss fractions for different initial radial positions and we have linearly spaced initial poloidal and toroidal \textit{VMEC} coordinates and $\lambda$ in the following ranges: $[0, 2 \pi]$, $[0, 2 \pi/n_{fp}]$ and $[0, 1]$, respectively. For every set of these initial conditions, $s_{v_\parallel}$ is set as -1 and +1, so our number of particles $n_p$ is equal to two times the product of $n_\theta$, $n_\phi$ and $n_\lambda$, where $n_x$ is the number of different possible initial values the initial quantity $x$. To ensure the initial positions matched in the \textit{VMEC} and \textit{pyQSC} equilibria, a root finder was designed to compute the corresponding $\phi_0$ for every $\phi$, with $\varphi$ calculated subsequently. The remaining NAE coordinates were calculated following the reverse of the procedure described in Section \ref{sec:NAE}. This strategy was employed to minimize the arguments passed to the \textit{VMEC} loss fraction algorithm, as this was the computational bottleneck in the workflow. Additionally, as discussed above, the timeframe where the NAE seems to be more relevant is the prompt loss one, and so we will simulate the orbits up until $5 \times 10^{-3}$ s. We note that, for all the results obtained in this work, the computation of loss fractions is two orders of magnitude faster for the \textit{pyQSC} magnetic fields when compared with the \textit{VMEC} calculations, mainly due to the simple closed form solutions available of the first fields..

In the forthcoming discussions, we will connect our earlier findings on individual particle behavior to this larger-scale evaluation. By doing so, we aim to better understand the level of particle confinement in MHD and NAE systems and their differences.

\subsection{Quasi-Axisymmetry}

\par For the calculation of the loss fractions on a QA stellarator we recur to the same baseline equilibrium precise QA, as in the last section, with an aspect ratio of $A= 9.1$. We then trace ensembles of particles with initial radial positions $s_i=0.25$, $0.5$, and $0.75$, facilitating an investigation into the dependence of result correspondence upon the radial distance from the equilibrium axis. 

\par
\begin{figure}
    \centering
    \includegraphics[width=0.42\textwidth]{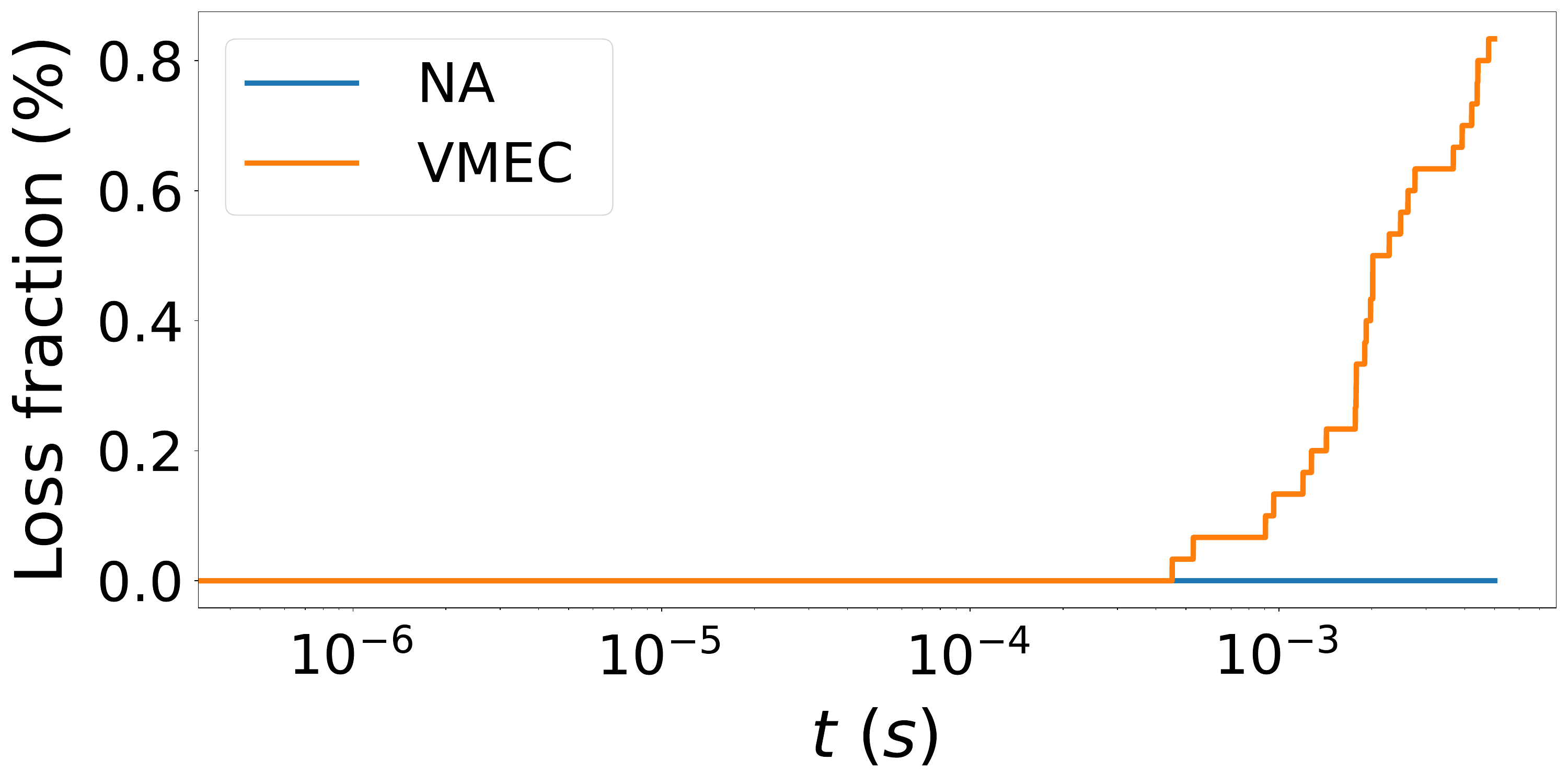}
    \hfill
    \includegraphics[width=0.42\textwidth]{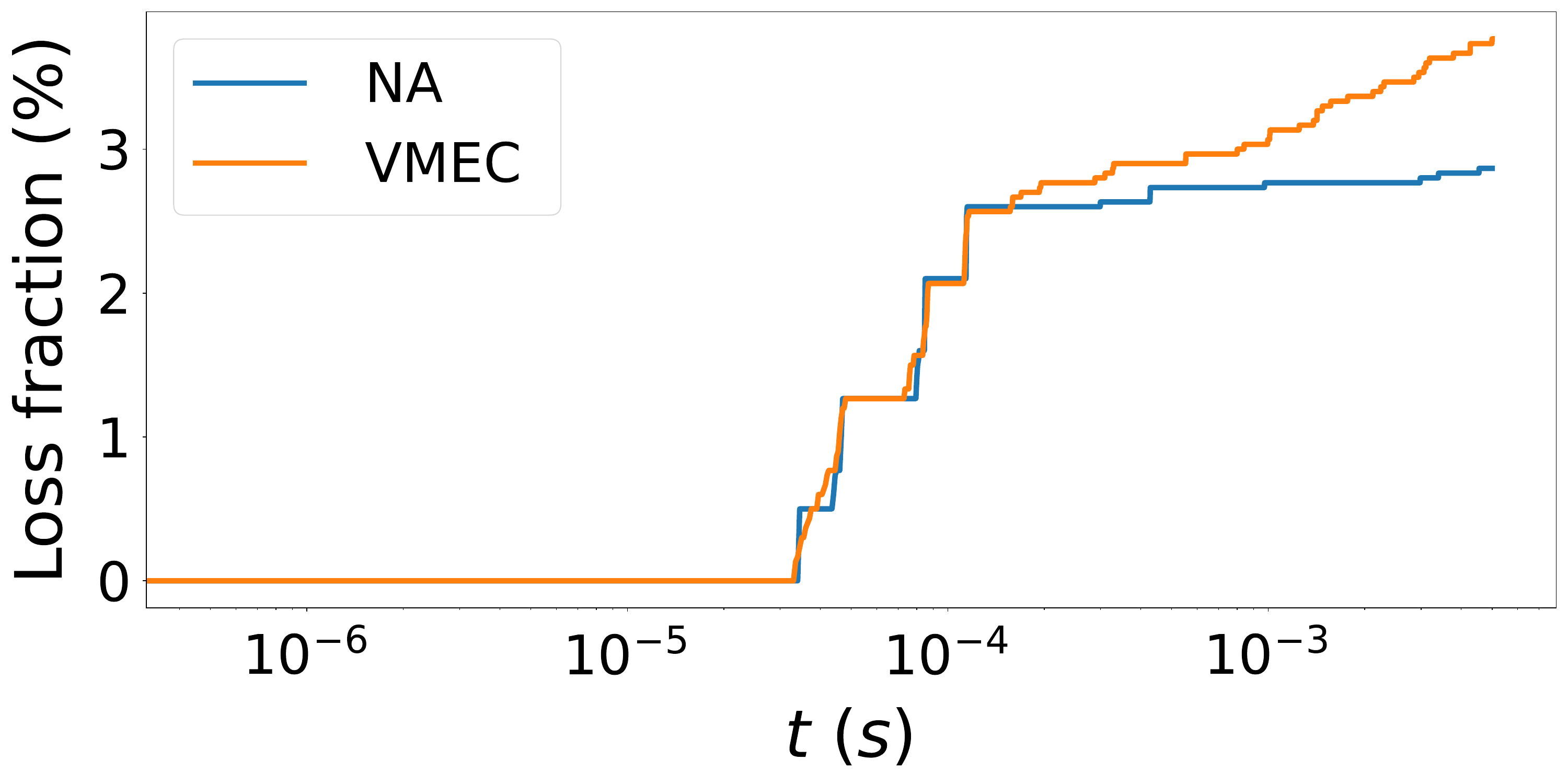}
    \hfill
    \includegraphics[width=0.42\textwidth]{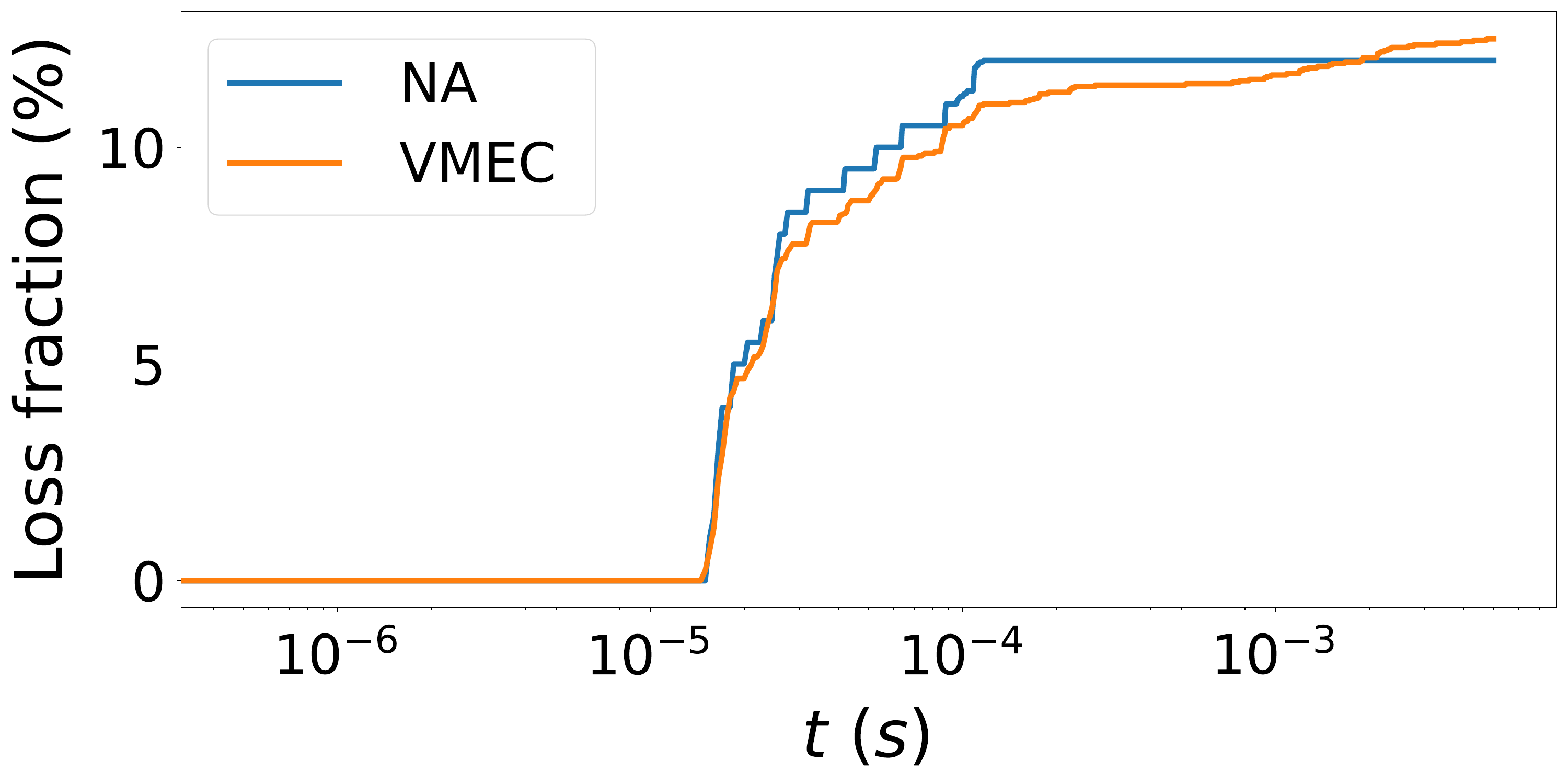}
    \caption{Comparison between alpha particle ensemble loss fractions in \textit{pyQSC} (NA) and \textit{VMEC} (VMEC) with $n_p=9000$ and linearly spaced initial conditions $\theta_i$ and $\phi_i$ in \textit{VMEC} coordinates as well as $\lambda$ for the aspect ratio $A=9.1$ QA equilibria with initial radial positions $s_i=0.25$ (Left), $s_i=0.5$ (Right) and $s_i=0.75$ (Bottom).}
    \label{fig:QA_loss}
\end{figure}

 In \cref{fig:QA_loss} we present the results of the aforementioned ensembles, where we can see an overall agreement over all initial radial positions. The fact that this agreement is present despite differences in the radial oscillations of particles may be attributed to the averaging out of \textit{VMEC} orbits that go slightly above and below the NAE orbits for different values of $\phi$. This conclusion is supported by the fact that increasing the value of $n_\phi$ improves the agreement of the loss fraction. 

\par There is an uncanny agreement for all presented flux surfaces up until the time mark of $t=5 \cdot 10^{-4}$ s, where the values for the two equilibria start to diverge. This is most likely where other effects start being relevant for this stellarator other than first orbit losses. The relative difference in loss fractions is the highest for $s_i=0.25$, where the \textit{pyQSC} equilibrium has no prompt losses, but the percentage on the \textit{VMEC} losses stays below $1 \%$ on the analyzed timeframe. For the ensemble most distant from the axis in \cref{fig:QA_loss}, the amount of lost particles appears to match for the entire measured time, indicating the loss cone particles are the most important loss mechanism at that flux surface. 

\par These results seem to indicate that the computation of loss fractions for initial time intervals could be performed with NAE fields with a low loss in performance and a reduction of the computation time of two orders of magnitude. Each of the $n_p=9000$ ensembles took around four hours to compute in the \textit{VMEC} field and two minutes in an \textit{openMP}-only parallelized code running on eight cores.   

\subsection{Quasi-Helical Symmetry}

\par A similar procedure as in the preceding subsection was followed for the 9.1 aspect ratio QH stellarator used in the single particle tracing. As the orbit amplitude is significantly narrower in this equilibrium, first orbit losses are only seen in its outer regions. For this reason, we present only the results of the ensembles with initial radial positions up from $s_i=0.5$, as results for inner radial starting points do not vary qualitatively, only quantitatively. As the frequency of oscillation in this magnetic field was higher, the divergence between equilibria arose sooner too. Therefore, the integration only goes up to $t=5 \cdot 10^{-4}$ s.

\par
\begin{figure}
    \centering
    \hfill
    \includegraphics[width=0.42\textwidth]{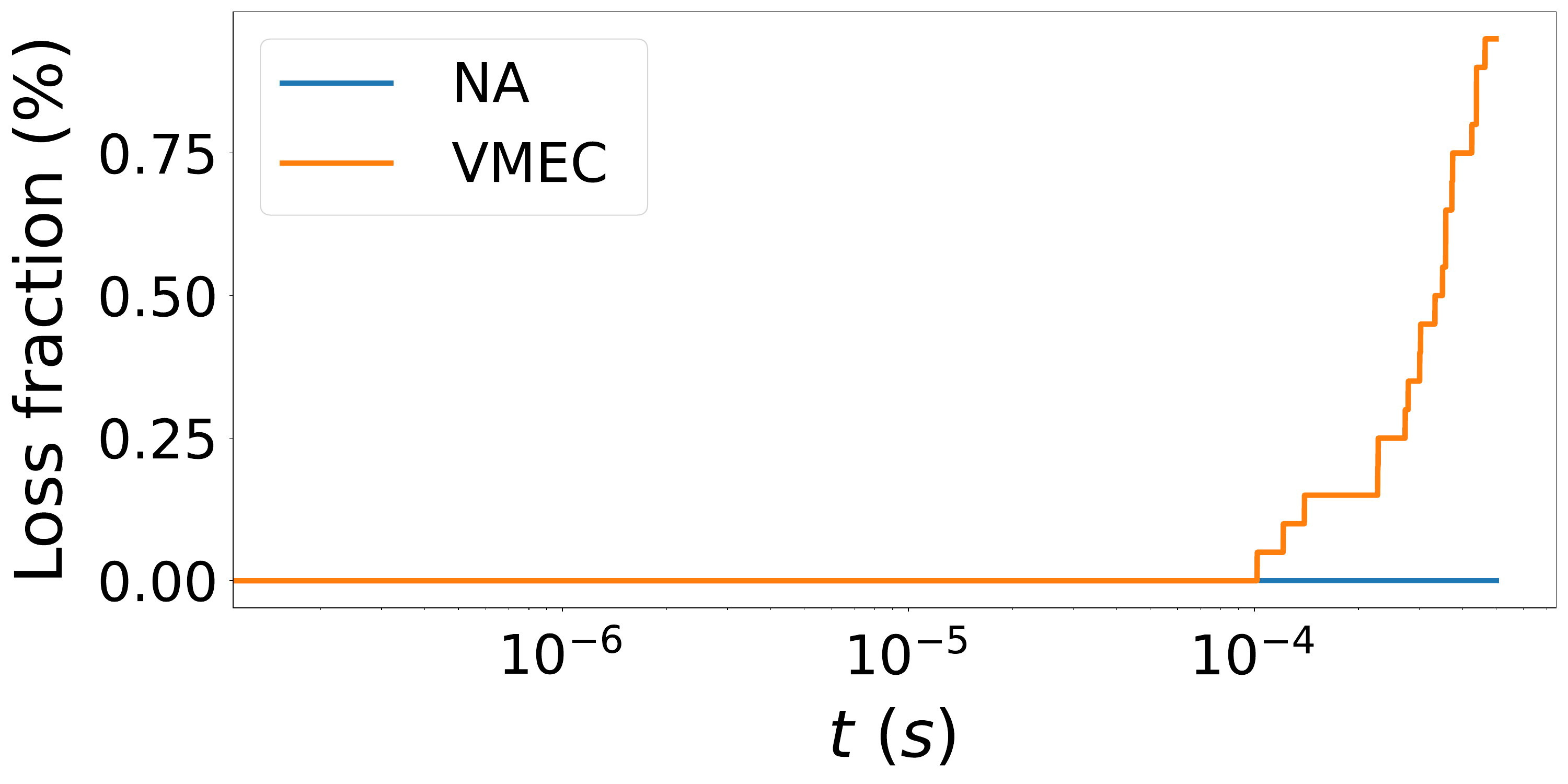}
    \hfill
    \includegraphics[width=0.42\textwidth]{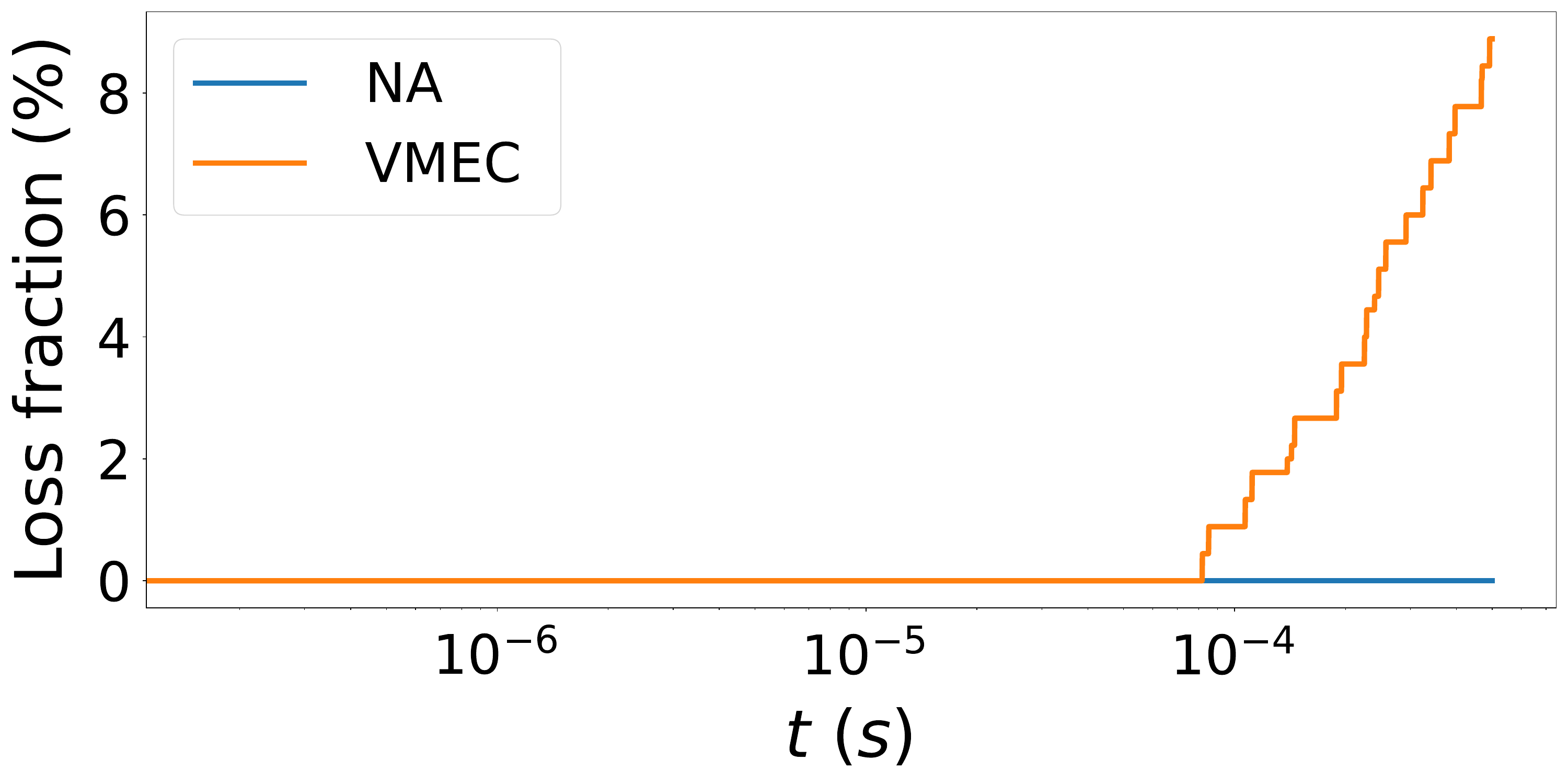}
    \hfill
    \includegraphics[width=0.42\textwidth]{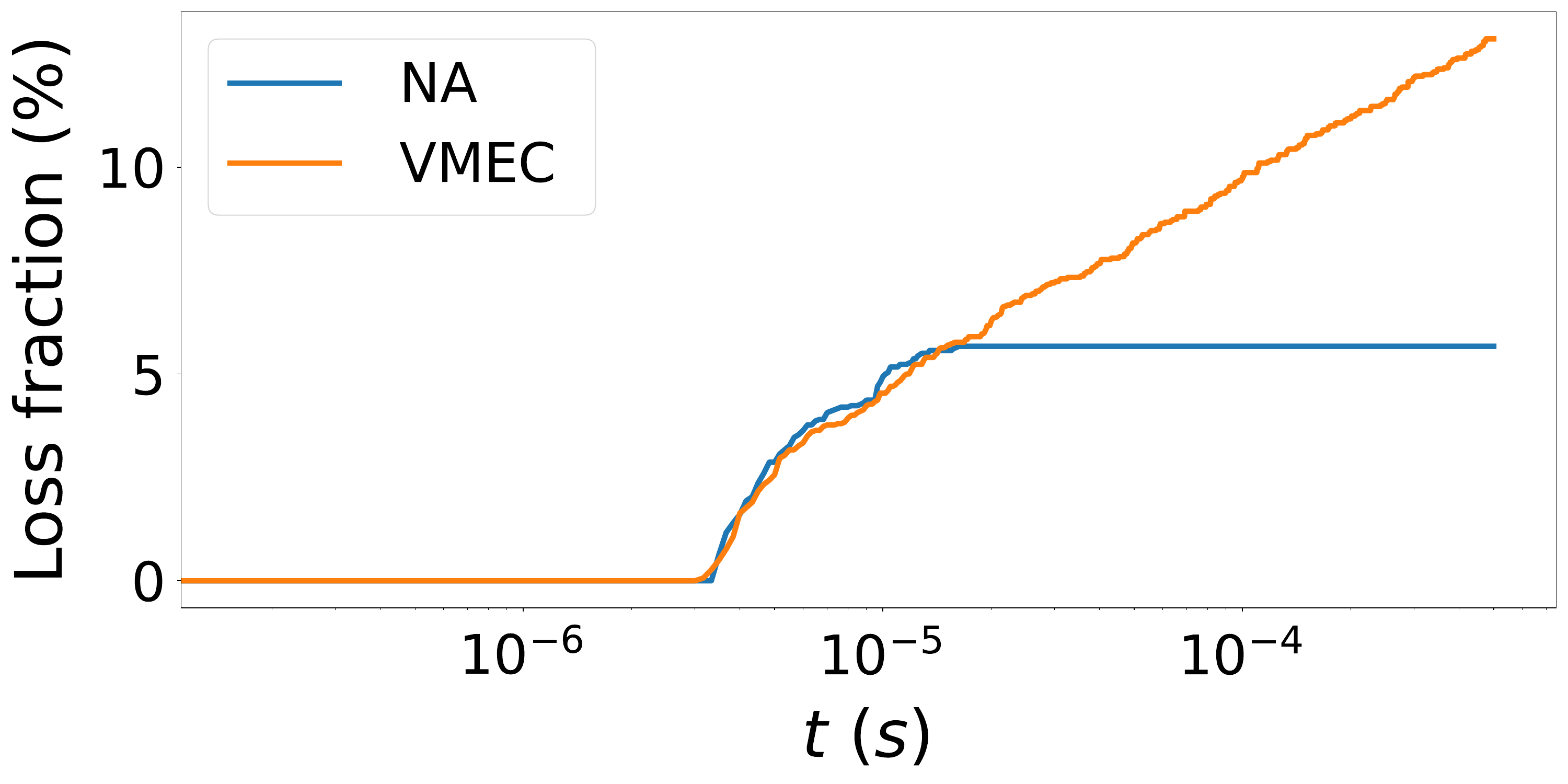}
    \hfill
    \caption{Comparison between alpha particle ensemble loss fractions in \textit{pyQSC} (NA) and \textit{VMEC} (VMEC) with $n_p=9000$ and linearly spaced initial conditions $\theta_i$ and $\phi_i$ in \textit{VMEC} coordinates as well as $\lambda$ for the aspect ratio $A=9.1$ QH equilibria with initial radial positions $s_i=0.5$ (Left), $s_i=0.75$ (Right) and $s_i=0.9$ (Bottom).}
    \label{fig:QH_loss}
\end{figure}

 Although, as shown in \cref{fig:QH_loss}, the behavior of the first few orbits is well captured by the NAE, in the present case, there are no first-orbit losses, and therefore, the loss fraction remains zero until $t=1 \cdot 10^{-4}$ s for the case of $s_i=0.5$ and $0.75$, and later diverges when particles start exhibiting an average radial drift outwards in the \textit{VMEC} equilibrium. In the case of the ensemble generated with initial radial position $s_i=0.9$, there is a region of time where the losses in the \textit{VMEC} magnetic field are emulated by the ones in the \textit{pyQSC} field. However, they start to diverge before the $t=1 \cdot 10^{-4}$ mark. The fact that the convergence is limited in time and space makes the argument for optimization in this case quite more limited. However, the fact that the QH \textit{VMEC} equilibrium exhibited a larger deviation from quasisymmetry may also be responsible for the accelerated divergence between the loss fraction results, as non-vanishing average radial drifts can be a consequence of local loss of symmetry. It can also be a result of particles transitioning from the passing to the trapped regime in \textit{VMEC} without doing so in \textit{pyQSC} fields such as in \cref{fig:QH-trapped}. 

\par Although the effectiveness of the tool explored throughout this work appears to be limited to prompt losses, it is important to relate our findings with the results of \citet{LeViness_2023} and \citet{Velasco_2023}. Where the fact that these point to some correlation between prompt losses and proxies for losses suggests that addressing the mechanisms responsible for these kind of losses may also lead to a reduction of other loss mechanisms that happen at later times. This way, increasing the efficiency for prompt losses optimization can be beneficial beyond the initially expected timeframes. Additionally, we note that the application of the followed methodology is more insightful than simply pursuing QS in a stellarator, providing a direct loss measurement, and more accurate than the aforementioned proxies for initial losses as it includes finite banana width effects. Although lacking in accuracy when compared to direct GC simulations in MHD equilibria \citep{Paul2022}, it benefits from increased computational speed as expressed in this work. 

\section{Conclusions}
\label{section:conclusions}
\par In this work we started by comparing guiding-center orbits computed by the Euler-Lagrange equations of motion implemented in the gyromotion library \textit{gyronimo} with those obtained by the symplectic formulation implemented in the loss-fraction assessment code \textit{SIMPLE}. The explicit methods in \textit{gyronimo} compared well with the implicit symplectic one, showing good levels of numerical agreement. Subsequently, single particle tracing was performed in different aspect ratio reactor scale stellarators. The discrepancy between the NAE fields and the MHD ones such as the \textit{VMEC} and \textit{BOOZ\_XFORM} did not have a noticeable first-order effect on the passing particles independently of the aspect ratio and stellarator, with growing deviations for lower aspect ratios especially in the QH equilibrium. More noticeable disparities were found for the trapped particles, where passing-trapped transitioning particles, radially drifting particles, and other effects were observed for the MHD equilibria in contrast with the NAE ones, where no such phenomena were identified. Once more, these effects were more prominent in the QH magnetic fields. These discrepancies were expected due to the decreased levels of quasisymmetry in the MHD fields. Additionally, effective expressions for estimating the radial amplitude of particles' orbits and the passing-trapped separatrix were found for both QA and QH stellarators in the NAE. 
\par Following the tendencies of the single particle tracing orbits, for each equilibrium studied, the loss fractions of large groups of particles at different initial radial positions agree only for losses of orbits whose radial amplitude is larger than their distance to the edge of the magnetic field. For the QA stellarator, this agreement lasts up until $t=5 \cdot 10^{-4}$ which could be used to do a first optimization of stellarators, or at least eliminate classes of stellarators with too many first orbit losses. Unfortunately, for the QH stellarator, these kinds of losses do not seem to be relevant for the majority of initial positions, as radial amplitudes tend to be smaller, which is concordant with the expectation for a larger value of $\iota$. It is important to emphasize the loss fraction results are obtained up to two orders of magnitude faster for the NAE particle tracing. Although faster performance could be attained in the different tracings by specializing the code to a given field, that is outside the scope of this work.

\par Future work with loss fractions using the NAE should encompass a larger diversity of equilibria, including equilibria generated from relevant MHD fields by fitting them to near-axis fields in order to compare with established results. The implementation of the loss fraction estimations in the optimization of directly constructed equilibria could also be studied as a way of producing a good initial configuration for conventional optimization. 

\section{Acknowledgments}

We thank Matt Landreman for the insightful discussions. We are also grateful the referees of this paper for their thorough analysis and important suggestions to its final form. 

R. J. is supported by the Portuguese FCT-Fundação para a Ciência e Tecnologia, under Grant 2021.02213.CEECIND and DOI  \href{https://doi.org/10.54499/2021.02213.CEECIND/CP1651/CT0004}{10.54499/2021.02213.CEECIND/CP1651/CT0004}.
Benchmarks and loss fraction studies were carried out using the EUROfusion Marconi supercomputer facility.
This work has been carried out within the framework of the EUROfusion Consortium, funded by the European Union via the Euratom Research and Training Programme (Grant Agreement No 101052200 - EUROfusion). Views and opinions expressed are however those of the author(s) only and do not necessarily reflect those of the European Union or the European Commission. Neither the European Union nor the European Commission can be held responsible for them.
IPFN activities were supported by FCT - Fundação para a Ciência e Tecnologia, I.P. by project reference UIDB/50010/2020 and DOI  \href{https://doi.org/10.54499/UIDB/50010/2020}{10.54499/UIDB/50010/2020}, by project reference UIDP/50010/2020 and DOI \href{https://doi.org/10.54499/UIDP/50010/2020}{10.54499/UIDP/50010/2020} and by project reference LA/P/0061/202 and  DOI \href{https://doi.org/10.54499/LA/P/0061/2020}{10.54499/LA/P/0061/2020}.

\section{Declaration of Interests}

\par The authors report no conflict of interest.

\bibliographystyle{jpp}

\bibliography{main}
\end{document}